\documentclass[useAMS,usenatbib]{mn2e}

\usepackage{natbib}

\usepackage{latexsym,graphicx}%,natbib}
\usepackage{color}
\usepackage{fixltx2e}
\usepackage{verbatim} \usepackage{float}
\usepackage{amsmath,amssymb}
\usepackage{times}

\usepackage{setspace}
\usepackage{rotating}
\usepackage{pdflscape}
\usepackage{subfig}

\newcommand\hii{H\,{\sc ii} \,}

\def\apgt{\ {\raise-.5ex\hbox{$\buildrel>\over\sim$}}\ }
\def\aplt{\ {\raise-.5ex\hbox{$\buildrel<\over\sim$}}\ }

\let\oldhat\hat
\renewcommand{\hat}[1]{\oldhat{\mathbf{#1}}}

\title[Excursions of MYSOs in the HR diagram]{On the episodic excursions of massive protostars in the Hertzsprung-Russell diagram}

\author[D. M.-A.~Meyer et al.]
       {D. M.-A.~Meyer$^{1}$\thanks{E-mail: dmameyer.astro@gmail.com},  L.~Haemmerle$^{2}$ and E.~I.~Vorobyov$^{3,4}$ \\ %\textcolor{black}{and B.~Stecklum}$^{5}$ \\
       $^{1}$Astrophysics Group, School of Physics and Astronomy, University of Exeter, Exeter EX4 4QL, United Kingdom \\
       $^{2}$Observatoire de Gen\` eve, Universit\'e de Gen\`eve, chemin des Maillettes 51, CH-1290 Sauverny, Switzerland \\ 
       $^{3}$Research Institute of Physics, Southern Federal University, Stachki 194, Rostov-on-Don, 344090, Russia \\ 
       $^{4}$Department of Astrophysics, The University of Vienna, Vienna, A-1180, Austria \\
       }

\begin{document}

% Date
\date{Received; accepted}

\maketitle

\label{firstpage}

\begin{abstract} 
Massive protostars grow and evolve under the effect of rapid accretion of 
circumstellar gas and dust, falling at high rates ($\ge 10^{-4}$-$10^{-3}\, \rm M_{\odot}\, \rm yr^{-1}$).  
This mass infall has been shown, both numerically and observationally, to be episodically interspersed by 
accretion of dense gaseous clumps migrating through the circumstellar disc to the protostellar surface, 
causing sudden accretion and luminous bursts. 
Using numerical gravito-radiation-hydrodynamics and stellar evolution calculations, we demonstrate that, in addition to the known bloating of 
massive protostars, variable episodic accretion further influences their evolutionary tracks of massive young stellar objects (MYSOs). For each accretion-driven 
flare, they experience rapid excursions toward more luminous, but colder regions of the Hertzsprung-Russell diagram. 
During these excursions, which can occur up to the end of the pre-main-sequence evolution, the photosphere of massive protostars can episodically 
release much less energetic photons and MYSOs surreptitiously adopt the same spectral type as evolved massive 
(supergiants) stars. Each of these evolutionary loop brings the young high-mass stars close to the forbidden Hayashi 
region and \textcolor{black}{might} make their surrounding \hii regions \textcolor{black}{occasionally} fainter, before they recover their quiescent, 
pre-burst surface properties. 
We interpret such cold, intermittent pre-main-sequence stellar evolutionary excursions and \textcolor{black}{the dipping
variability of HII regions} as the signature of the presence of a fragmenting circumstellar accretion disc surrounding the 
MYSOs. \textcolor{black}{We} conjecture that this mechanism might equivalently affect young stars in the intermediate-mass regime. 
\end{abstract}


%%%%%%%%%%%%%%%%%%%%%%%%%%%%%%%%%%%%%%%%%%%%%%%%%%%%%%%%%%%%%%%%%%%%%%%%%%%%%%%%%%%%%%%%%%%
%%%%%%%%%%%%%%%%%%%%%%%%%%%%%%%%%%%%%%%%%%%%%%%%%%%%%%%%%%%%%%%%%%%%%%%%%%%%%%%%%%%%%%%%%%%
%%%%%%%%%%%%%%%%%%%%%%%%%%%%%%%%%%%%%%%%%%%%%%%%%%%%%%%%%%%%%%%%%%%%%%%%%%%%%%%%%%%%%%%%%%%

\section{Introduction}
\label{sect:intro}

\begin{table*}
	\centering
	\caption{
	Characteristics of our models, with initial pre-stellar core mass 
	$M_{\rm c}$, final protostellar age $t_{\rm end}$, final stellar mass $M_{\star}$, details of their corresponding accretion rate histories \textcolor{black}{and of the initial conditions of the stellar evolution calculations}.   
	}
	\begin{tabular}{lccccr}
%	\hline
	\hline
	${\rm {Models}}$             &  $M_{\rm c}$ ($\rm M_{\odot}$)      &    $t_{\rm end}$ ($\rm kyr$)   &   $M_{\star}$ ($\rm M_{\odot}$) & ${\rm {Accretion}\, \rm {rate}}$ &   ${\rm {Method}}$     \\ 
	\hline    
	{\rm Run-100-hydro}     &  $100$                              &    $50.0$                      &  $33.3$     &  Episodic      &    Full hydrodynamical simulation with variable rate   \\  
	{\rm Run-100-constant}  &  $100$                              &    $50.0$                      &  $33.3$     &  Constant      &    Hydrodynamical simulation (collapse) + constant analytic rate (disc)    \\ 
	{\rm Run-100-smoothed}    &  $100$                            &    $50.0$                      &  $33.3$     &  Smoothed      &    Hydrodynamical simulation (collapse) + smoothed burst-free rate (disc) \\
	\textcolor{black}{{\rm Run-100-compact}}    &  \textcolor{black}{$100$}  &  \textcolor{black}{$50.0$}   &  \textcolor{black}{$33.3$}  &  \textcolor{black}{Episodic}   &  \textcolor{black}{As Run-100-hydro, with more compact initial conditions} \\ 
	%\textcolor{black}{{\rm Run-100-large}}    &  \textcolor{black}{$100$}  &  \textcolor{black}{$50.0$}   &  \textcolor{black}{$33.3$}  &  \textcolor{black}{Episodic}   &  \textcolor{black}{As Run-100-hydro with larger stellar embryo} \\ 	
	{\rm Run-60-hydro}      &  $60$                               &    $65.2$                      &  $20.0$     &  Episodic      &    Full hydrodynamical simulation with variable rate   \\  
	\hline    
%	\hline 
	\end{tabular}
\label{tab:models}\\
%\footnotesize{ ${(a)}$This work, ${(b)}$~\citet{meyer_mnras_473_2018} }\\
\end{table*}

% Star formation by accretion
%The evolution of the interstellar medium (ISM) of the Galaxy is governed by the feedback of massive stars onto 
%their direct surroundings~\citep{langer_araa_50_2012}, which strong stellar winds carry important amount 
%of momentum and radiation that enrich the ISM, from their main-sequence~\citep{meyer} to their supernova phase~\citep{meyer_mnras_450_2015}.  
%
Gravitationally-collapsing pre-stellar cores give birth to new young stars, which grow by mass 
accretion of surrounding molecular material. The rates at which protostars 
gain mass have been shown to exhibit such a diversity that the initial picture of isotropic 
collapse~\citep{larson_mnras_145_1969,shu_apj_214_1977} fails to explain the observed spread 
in accretion rates~\citep{vorobyov_apj_704_2009}. Meanwhile, variations 
of the protostellar accretion rate can induce enormous changes in protostellar luminosity, the most extreme 
manifestations of which are the so-called FO-Orionis and the very low luminosity 
objects~\citep{2017A&A...600A..36V}. 
Numerical simulations have demonstrated that this is possible thanks to the presence of a self-gravitating 
circumstellar accretion disc prone to gravitational fragmentation~\citep{vorobyov_apj_633_2005,vorobyov_apj_719_2010,vorobyov_apj_805_2015,machida_mnras_413_2011,zhao_mnras_473_2018,nayakshin_mnras_426_2012,2018arXiv180607675V}.
%, inside which fragmentation mechanisms responsible for variabilities in the accretion rate 
%take place~\citep{clarke_mnras_381_2007,stamatellos_mnras_392_2009,vorobyov_apj_719_2010}. 
%
%The presence of fragmented accretion disc around young stars has been largely confirmed by observations~\citep{}.   

% The same applies to massive stars
The disc fragmentation scenario equivalently applies to star formation in the high-mass regime. 
Numerical simulations predicted the formation of accretion discs around massive young stellar 
objects~\citep{1998MNRAS.298...93B,2002ApJ...569..846Y,peters_apj_711_2010,seifried_mnras_417_2011}, 
together with additional structures such as bipolar cavities filled with ionizing 
radiation generated by the UV feedback of the protostars~\citep{harries_mnras_448_2015,klassen_apj_823_2016,harries_2017}. 
Accretion variability is a direct consequence of asymmetries in the accretion flow~\citep{seifried_mnras_417_2011} and of the 
coupling between the prostellar radiation feedback and its surrounding disc~\citep{peters_apj_711_2010}. 
Such variability is a generic feature of massive star formation in the sense that it is neither 
stopped by the radiation pressure in the bipolar \hii regions~\citep{peters_apj_725_2010} nor 
by disc fragmentation~\citep{meyer_mnras_473_2018}.
\textcolor{black}{ 
Other studies on massive star formation also reported time-variabilities of the disc-to-star mass 
transfer rate~\citep{krumholz_sci_323_2009,rosen_mnras_463_2016}. 
}
In addition, self-gravitating discs around high-mass stars are subject to 
efficient gravitational instabilities, generating heavy spiral arms in which dense gaseous clumps form, 
eventually leading to the formation of multiple hierarchical systems~\citep{krumholz_apj_656_2007,peters_apj_711_2010}. 
These circumstellar clumps can either rapidly migrate onto the stellar surface, trigger increases of the 
protostellar accretion rate which aggravate the variability~\citep{meyer_mnras_473_2018} and produce luminous 
bursts~\citep{meyer_mnras_464_2017} or evolve to secondary low-mass stars which finally end up as low-mass 
spectroscopic companions to the MYSOs~\citep{meyer_mnras_473_2018}. 
\textcolor{black}{
Accretion bursts are a feature of the formation of young massive stellar objects which seems common to 
most massive protostars as it does not depend on the initial properties of their parent pre-stellar core~\citep{2018arXiv181100574A}. 
}
\textcolor{black}{
Moreover, although these eruptive phases represent 
a small fraction ($\sim \%$) of their early formation time, MYSOs can acquire a substential fraction 
of their zero-age-main-sequence (ZAMS) mass via these flaring episods~\citep{2018arXiv181100574A}.
}

From the point of view of observations, (variable) accretion 
flows~\citep{keto_apj_637_2006,stecklum_2017a} and ionized, pulsed, collimated structures~\citep{Cunningham_apj_692_2009,
cesaroni_aa_509_2010,caratti_aa_573_2015,purser_mnras_460_2016,reiter_mnras_470_2017,burns_mnras_467_2017,arXiv180102211B,purser_mnras_475_2018,2018arXiv180311413S} underlined the scaled-up character of massive star formation with respect to low-mass stars~\citep[see also][]{fuente_aa_366_2001,testi_2003,
cesaroni_natur_444_2006,stecklum_2017b}. 
A growing number of observations of (Keplerian) discs around MYSOs have been  
reported~\citep{johnston_apj_813_2015,ilee_mnras_462_2016,forgan_mnras_463_2016,2018arXiv180410622G}, together 
with evidences of a spiral filament feeding the candidate disc MM1-Main~\citep{maud_467_mnras_2017}  
and an infalling gaseous clump in the double-cored system G350.69-0.49~\citep{chen_apj_835_2017}. 
Interestingly, a recent {\it ALMA} view of the massive young object G023.01-00.41 exhibited a clear 
disc-jet association~\citep{2018arXiv180509842S}.  
In addition, some objects revealed the presence of high-mass proto-binary systems within a 
circumbinary disc~\citep{kraus_apj_835_2017}. 
Finally, several MYSOs experienced multi-wavelengths flares~\citep{moscadelli_aa_600_2017,cesaroni_2018} 
in a fashion of the predictions of~\citet{meyer_mnras_464_2017}, among which are the accretion-bursts 
of S255IR NIRS\,3~\citep{fujisawa_atel_2015,stecklum_ATel_2016,caratti_nature_2016} and
from NGC6334I-MM1~\citep{2017arXiv170108637H} experienced accretion-driven outbursts.

\begin{figure*}
%        \centering  
%        \begin{minipage}[b]{ 0.9\textwidth}
%                \includegraphics[width=1.0\textwidth]{./acc_rate_100.pdf}
%        \end{minipage} \\    
%        \begin{minipage}[b]{ 0.9\textwidth}
%                \includegraphics[width=1.0\textwidth]{./acc_rate_60.pdf}
%        \end{minipage}      
%        \centering  
        \begin{minipage}[b]{ 0.95\textwidth}
                \includegraphics[width=1.0\textwidth]{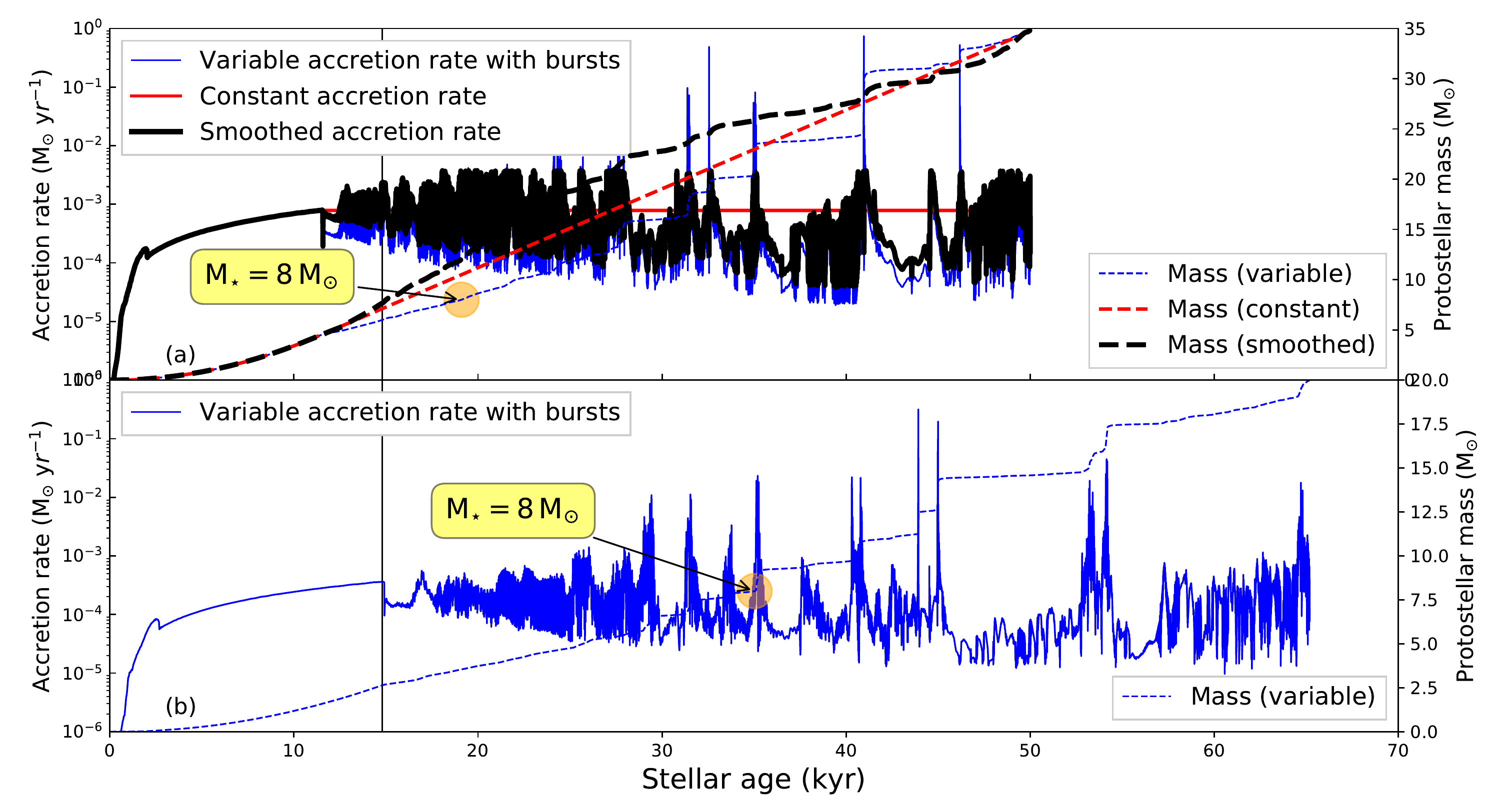}
        \end{minipage} \\                 
        \caption{ 
                 Accretion rate and 
                 protostellar mass evolution in our 
                 models with a pre-stellar core mass of $100\, \rm M_{\odot}$ (a) and 
                 $60\, \rm M_{\odot}$ (b), respectively. 
                 }      
        \label{fig:rate}  
\end{figure*}

% Stellar evolution calculations
Early stellar evolution calculations demonstrated the dominant influence of accretion on the internal  
stellar structure prior to the ZAMS phase~\citep{palla_apj_375_1991}. 
The formation of a radiative barrier that turns the fully convective stellar embryo into a stable 
radiative core and an outer convective shell that causes the star to swell has been shown in the context of young 
intermediate-mass stars for a wide range of different constant accretion 
rates~\citep{palla_apj_392_1992,palla_1993,beech_apjs_95_1994,bernasconi_aa_307_1996,bernasconi_aa_120_1996}. 
A monotonically increasing accretion rate yields similar results~\citep{behrend_aa_373_2001}. 
Calculations with high-rate mass accretion exceeding $10^{-3}\, \rm M_{\odot}\, \rm yr^{-1}$ equivalently 
showed that disc-accreting young massive stellar objects have their pre-main-sequence evolution governed by 
the accretion of circumstellar material, see~\citet{bernasconi_aa_307_1996,norberg_aa_159_2000,behrend_aa_373_2001,hosokawa_apj_691_2009,hosokawa_apj_721_2010}. 
These models reported a rapid swelling of the protostars up to radii $\sim 1000\, \rm R_{\odot}$ produced by the 
so-called luminosity wave~\citep{larson_mnras_157_1972}, an internal redistribution of entropy following the abrupt 
decrease of the opacity in the protostellar interior. 
Interestingly, for the constant accretion rates $\ge\, 10^{-2}\, \rm M_{\odot}\, \rm yr^{-1}$, the protostars evolve to the 
red part of the Hertzsprung-Russell diagram~\citep{hosokawa_apj_721_2010,haemmerle_585_aa_2016}. The authors postulate therein that the \hii region 
generated by those stars can then disappear as their protostellar radius bloats, ultimately leading to lower-luminosity young massive stellar objects. 
%Nevertheless, such high constant rates are not consistent with recent estimates of massive protostars accretion variabilities (). 
%
Simultaneous hydrodynamical and stellar evolution calculations revealed 
that variable-accreting MYSOs can experience a unique loop to the red before recovering their bluer characteristics 
and reach the ZAMS~\citep{kuiper_apj_772_2013}. However, those 2.5-dimensional axisymmetric simulations do not 
account for disc fragmentation physics and its influence on the protostellar variability~\citep{meyer_mnras_464_2017} 
and the corresponding rates were burstsless. 
More complex theoretical studies tackled the problem of the impact of various feedback mechanisms of MYSOs 
onto star formation efficiency, however, without considering their accretion variability~\citep{2018arXiv180401132T}.

% What we do here 
Motivated by the above arguments, we aim at investigating the effects of the repetitive 
accretion events responsible for luminous bursts on the evolution of pre-main-sequence young massive 
stellar objects. With the help of three-dimensional gravito-radiation-hydrodynamics models of 
collapsing rotating massive pre-stellar cores, we first simulate the formation and evolution 
of fragmenting circumstellar accretion discs from which protostars gain mass~\citep{meyer_mnras_473_2018}. 
From the disc models, we extract variable accretion rate histories, interspersed by episodic accretion bursts 
caused by dense gaseous clumps that form in spiral arms and rapidly migrate onto the protostars~\citep{meyer_mnras_464_2017}.  
Finally, we calculate stellar evolution models with the Geneva stellar evolutionary code~\citep{eggenberger_apss_316_2008,haemmerle_phd_2014} 
fed by our accretion rate histories. Using the method developed in the context of constant-accreting  
MYSOs~\citep{haemmerle_585_aa_2016,haemmerle_602_aap_2017} and accretion flows onto large-scale \hii regions~\citep{haemmerle_458_mnras_2016}, 
we calculate the changes in the internal structure and the surface properties of MYSOs experiencing bursts. 
%
%This allows us to make the Hertzsprung-Russell of our young massive stars and to discuss the effects of strong pre-main-sequence disc accretion 
%variabilities on their stellar evolutionary path prior to their zero-age-main-sequence time (ZAMS). 

% Annonce of the plan 
In this paper, we investigate the effects of variable accretion on the evolution of MYSOs.  
We perform numerical gravito-radiation-hydrodynamics simulations and stellar evolution calculations of pre-main-sequence accreting 
massive stellar objects to explore the effects of strong accretion bursts onto their internal as well as their surface properties 
and evolutionary path in the Hertzsprung-Russell diagram. This study is organized as follows. 
In Section~\ref{sect:methods}, we review the methods utilised to perform (i) gravito-radiation-hydrodynamical 
simulations of the monolithic collapse of present-day rotating pre-stellar cores, from which we extract 
accretion rate histories, and (ii) stellar evolutionary calculations of MYSOs which accrete from their 
circumstellar discs at pre-calculated time-variable rates. Our results are presented in Section~\ref{sect:results}, 
\textcolor{black}{
the effects of the initial conditions on the stellar excursions are investigated in Section~\ref{sect:ic},  
}
and our findings further discussed in Section~\ref{sect:discussion}. 
Particularly, we highlight that strong outbursts provoke rapid excursions towards colder regions of the 
Herztsprung-Russel diagram, which typically are not associated with such hot and young stellar objects. 
Finally, we discuss our results and conclude in Section~\ref{sect:cc}.

%%%%%%%%%%%%%%%%%%%%%%%%%%%%%%%%%%%%%%%%%%%%%%%%%%%%%%%%%%%%%%%%%%%%%%%%%%%%%%%%%%%%%%%%%%%
%%%%%%%%%%%%%%%%%%%%%%%%%%%%%%%%%%%%%%%%%%%%%%%%%%%%%%%%%%%%%%%%%%%%%%%%%%%%%%%%%%%%%%%%%%%
%%%%%%%%%%%%%%%%%%%%%%%%%%%%%%%%%%%%%%%%%%%%%%%%%%%%%%%%%%%%%%%%%%%%%%%%%%%%%%%%%%%%%%%%%%%

\section{Modelling}
\label{sect:methods}

In the following paragraphs, we introduce the reader to the method employed to carry out our 
gravito-radiation-hydrodynamical simulations of high-mass star formation, from which we extract time-dependent 
protostellar accretion rate histories. Furthermore, we detail how the outputs of the hydrodynamical 
models are subsequently used as boundary conditions for stellar evolution calculations.

\subsection{Hydrodynamical simulations}
\label{sect:hydro}

Our three-dimensional midplane-symmetric hydrodynamical simulations are initialized with a rigidly-rotating 
spherically symmetric, pre-stellar core of density distribution $\rho(r)\propto r^{\beta_{\rho}}$, 
with $\beta_{\rho}=-3/2$~\citep{butler_apj_754_2012,butler_apj_782_2014} and $r$ is the radial coordinate.  
The inner edge of the core is made of a semi-permeable sink cell centered onto the origin of the 
computational domain and the outer edge of the core, at $R_{\rm c}=0.1\, \rm pc$, is assigned to the outflow boundary conditions. 
We map the domain $[\textcolor{black}{r_{\rm in}},R_{\rm c}]\times[0,\pi/2]\times[0,2\pi]$ with a mesh 
of $N_{\rm r}=128\times\,N_{\rm \theta}=11\times\,N_{\rm \phi}=128$ grid zones, logarithmically expanding  
along the $r$-direction, as a cosine in the polar $\theta$-direction, and uniformly spaced along the azimuthal $\phi$-direction. 
\textcolor{black}{As in~\citet{2018arXiv181100574A}, we use a size of the sink cell of $20\, \rm au$, which is larger} than that of the 
first paper of this series~\citep{meyer_mnras_464_2017} in order 
to reach longer integration times $t_{\rm end}$, while avoiding very restrictive Courant-Friedrich-Levy conditions 
on the time-step within the direct protostellar surroundings.  
We simulate the gravitational collapse of several pre-stellar cores with different initial masses Mc. 
Each core forms a central protostar and a massive circumstellar disc.
The accretion rate from the disc onto the protostar is computed as the rate of mass transport $\dot{M}$ through the sink cell. 
The pre-stellar core temperature is uniformly set to $T_{\rm c}=10\, \rm K$ and its rotational-by-gravitational 
energy ratio is set to a typical value of $\beta=4\, \%$~\citep{meyer_mnras_464_2017}. The models are run 
until the mass of the central star $M_{\star}$ becomes equal to one third that of the initial mass core $M_{\rm c}$. 
The characteristics of our models are summarised in Tab.~\ref{tab:models}.

To solve the evolution of the above described physical system, we numerically integrate the  
equations of gravito-radiation-hydrodynamics with the {\sc pluto} 
code\footnote{http://plutocode.ph.unito.it/}~\citep{mignone_apj_170_2007,migmone_apjs_198_2012}. 
Our method takes into account the direct irradiation of the protostar and radiation transport in 
the accretion disc within the gray approximation using the scheme 
of~\citet{kolb_aa_559_2013}\footnote{http://www.tat.physik.uni-tuebingen.de/~\,pluto/pluto\_radiation/} 
adapted following the prescriptions of~\citet{meyer_mnras_473_2018}, see also equivalent radiation-hydrodynamics 
methods implementations in e.g.~\citet{commercon_aa_529_2011},~\citet{flock_aa_560_2013} and~\citet{bitsch_aa_564_2014}.
The photospheric photons are first ray-traced from the stellar atmosphere and propagate by flux-limited 
diffusion into the circumstellar disc. Such method permits to treat the disc thermodynamics accurately, with 
its central heating together with the outer cooling of the discs, as predicted by~\citet{vaidya_apj_742_2011}.  
%
%Equivalent radiation-hydrodynamics methods have also been presented in 
%e.g.~\citet{commercon_aa_529_2011},~\citet{flock_aa_560_2013} and~\citet{bitsch_aa_564_2014}. 
%
Opacities and calculation of the dust properties are as in~\citet{meyer_mnras_473_2018}, i.e. estimated 
by assuming that disc silicate grains are in equilibrium with the total radiation field. 
The gravitational force includes the gravity of the growing young massive star and the self-gravity of the gas, the latter computed using the prescriptions 
of~\citet{black_apj_199_1975} and the implementation method inspired by~\citet{hirano_sci_357_2017}\footnote{https://shirano.as.utexas.edu/SV.html} 
by time-dependently solving the Poisson equation with the help of the PETSC library\footnote{https://www.mcs.anl.gov/petsc/}. 
We assume that the angular momentum transport in the disc is essentially produced by the spiral arms in 
the discs and we therefore do not include additional turbulent $\alpha$-viscosity~\citep{meyer_mnras_473_2018}.

\subsection{Stellar evolution calculations}
\label{sect:evol}

We used our accretion rate histories to compute the evolutionary tracks of 
our MYSOs in the Hertzsprung-Russel diagram. 
The one-dimensional stellar evolution calculations were performed with the hydrostatic {\sc Geneva} 
code, the original version of which~\citep{eggenberger_apss_316_2008} has recently been updated for disc accretion physics in 
the context of pre-main-sequence massive protostars~\citep{haemmerle_phd_2014}. The code was tested 
in the context of constant-accreting MYSOs~\citep[see details relative to the numerical scheme and 
the implementation method in][]{haemmerle_585_aa_2016} and showed full consistency 
with the original results on high-constant-rate accreting MYSOs of~\citet{hosokawa_apj_721_2010}. 
Accretion is treated within the so-called {\it cold disc accretion} scenario~\citep{palla_apj_392_1992}, which 
assumes that the inner disc region is geometrically thin when the accreted material reached the stellar surface. 
Hence, we follow the mass growth of the hydrostatic core (i.e., the protostar), without considering a spherical 
envelope during the accretion phase~\citep{palla_apj_392_1992}. 
Subsequent theoretical and numerical works demonstrated that such assumption is fully 
reasonable~\citep{vaidya_apj_742_2011,meyer_mnras_473_2018}. Any entropy excess is radiated away 
in direction perpendicular to the disc and it is channeled into the radiatively-driven outflow associated to young massive 
stars~\citep{harries_2017}. The circumstellar material is advected inside the protostar assuming that its 
thermal properties are similar to those of the suface layer of the MYSOs.

Since we assume that most of the energy produced by the accretion shock (not modelled in the 
calculations) is radiated away before it reaches the protostellar surface, no additional entropy from the 
liberation of gravitational energy is added to the surface of the star. Such an assumption is the lower limit 
on the entropy attained by the star during the accretion process, while the upper limit is the 
so-called spherical (or hot) accretion scenario, scenario, in which a fraction of accretion entropy is added 
to the star, see the
sketch in fig.~1 of~\citet{hosokawa_apj_721_2010}. We choose the cold scenario as it has recently been 
used in the context of accreting \hii regions~\citep{haemmerle_458_mnras_2016}. 
Calculations of the stellar structure are performed with the Henyey method within the Lagrangian formulation~\citep{haemmerle_phd_2014}  
at solar metallicity (Z=0.014) using the abundances of~\citet{asplund_ASPC_2005} and~\citet{cunha_apj_647_2006} 
and the deuterium mass fractions of~\citet{norberg_aa_159_2000} and~\citet{behrend_aa_373_2001}.  
The simulations make use of the Schwarzschild criterion for convection, overshooting is 
considered and they are initialised with fully convective stellar embryo because they are the most 
difficult models to bloat~\citep{haemmerle_585_aa_2016,haemmerle_458_mnras_2016}. 
Hence, our method threats the stellar swelling most conservatively, avoiding any artificial effects 
that can lead to excessive swelling. 
To facilitate the stellar evolution calculations, we average the accretion rate histories 
over a time period of $10\, \rm yr$. This excludes the smallest variabilities. However, one can 
justify such an assumption as we know that high accretion rates ($\ge 10^{-2}\, \rm M_{\odot}\, \rm yr^{-1}$) responsible for 
protostellar bloating are exclusively reached during the strongest and longest accretion 
bursts~\citep{meyer_mnras_464_2017,meyer_mnras_473_2018}.

%%%%%%%%%%%%%%%%%%%%%%%%%%%%%%%%%%%%%%%%%%%%%%%%%%%%%%%%%%%%%%%%%%%%%%%%%%%%%%%%%%%%%%%%%%%
%%%%%%%%%%%%%%%%%%%%%%%%%%%%%%%%%%%%%%%%%%%%%%%%%%%%%%%%%%%%%%%%%%%%%%%%%%%%%%%%%%%%%%%%%%%
%%%%%%%%%%%%%%%%%%%%%%%%%%%%%%%%%%%%%%%%%%%%%%%%%%%%%%%%%%%%%%%%%%%%%%%%%%%%%%%%%%%%%%%%%%%

\section{Results}
\label{sect:results}

This section presents the accretion rate histories of our MYSOs and investigates the effects 
of the accretion of dense gaseous clumps on the structure and evolutionary path of high-mass 
protostars in the Hertzsprung-Russell diagram.

\begin{figure}
        \centering
        \begin{minipage}[b]{ 0.49\textwidth}  \centering
                \includegraphics[width=1.0\textwidth]{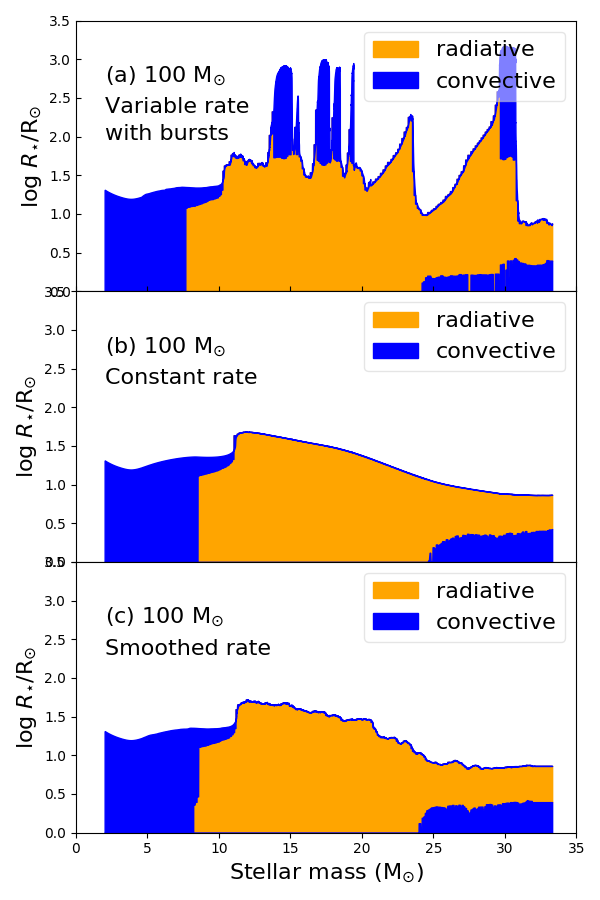}
        \end{minipage}        
        \caption{ 
        		 Kippenhahn diagram our of MYSOs generated with an initial $100\, \rm M_{\odot}$ 
        		 pre-stellar core. It shows the evolution of the internal stellar structure as a  
        		 function of time. The blue and orange regions denote the convective 
        		 and radiative parts of the protostar, respectively. 
                 }      
        \label{fig:structure}  
\end{figure}

\begin{figure*}
        \centering
        \begin{minipage}[b]{ 0.9\textwidth}  \centering
                \includegraphics[width=0.9\textwidth]{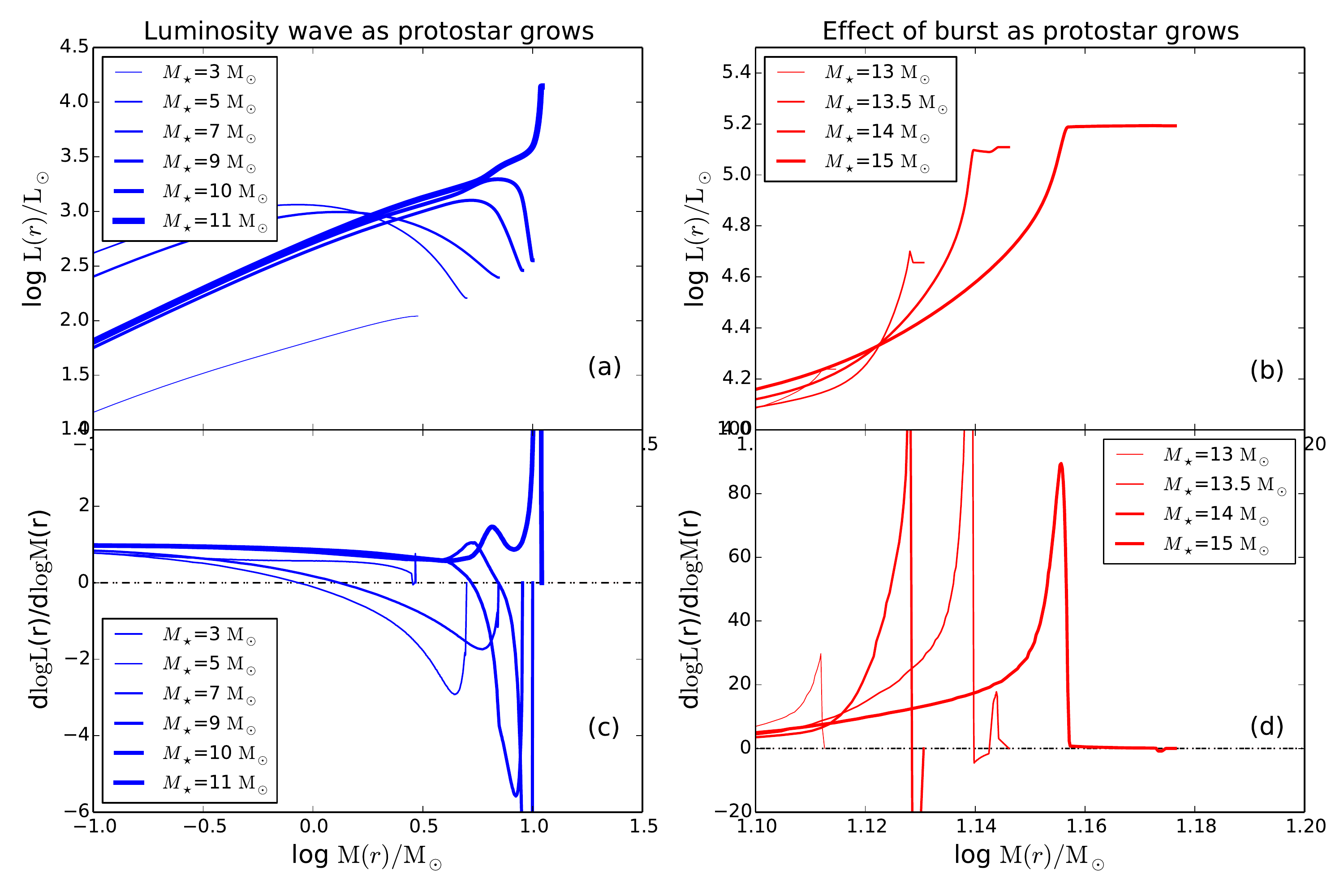}
        \end{minipage}     
%        \centering
%        \begin{minipage}[b]{ 0.475\textwidth}  \centering
%                \includegraphics[width=1.0\textwidth]{./plot_lum_wave_2.pdf}
%        \end{minipage}             
        \caption{ 
        		 Internal luminosity profiles (top panels) and gradient of luminosity profiles (bottom panels) 
        		 in our massive protostellar model Run-100-hydro. 
        		 The panels illustrate the development of the luminosity 
        		 wave (left panels) and the effect of a burst (right panels) onto the structure of a growing MYSO. 
                 }      
        \label{fig:lum_wave}  
\end{figure*}

\begin{figure*}
        \centering
        \begin{minipage}[b]{ 0.8\textwidth}  \centering
                \includegraphics[width=1.0\textwidth]{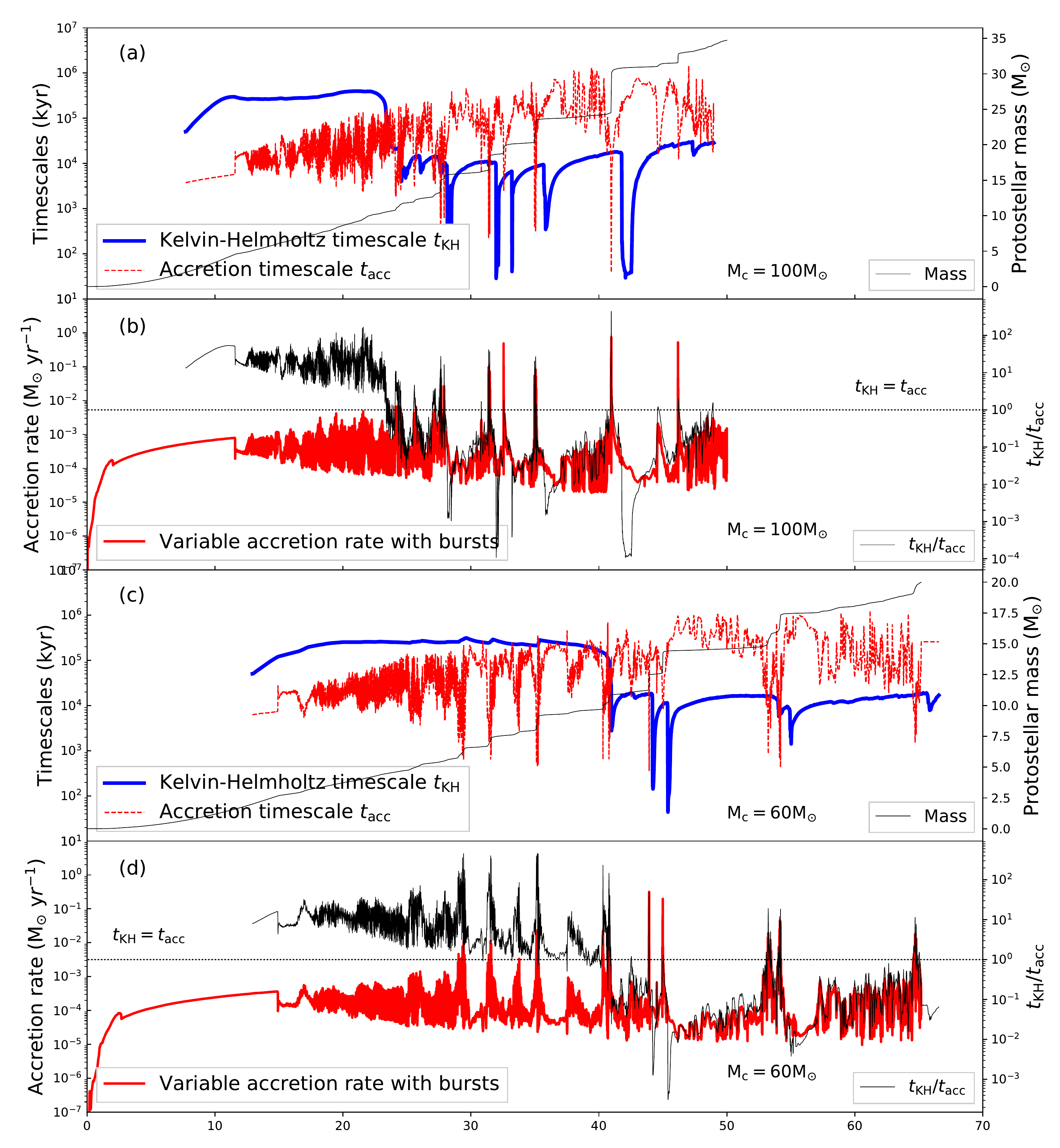}
        \end{minipage}        
        \caption{ 
        		 Characteristic timescales of our bursting protostars accreting at 
        		 variable accretion rates and generated from an initial $100\, \rm M_{\odot}$ 
        		 (a,b) and $60\, \rm M_{\odot}$ (c,d) pre-stellar core, respectively.
                 Panels (a) and (c) plot the Kelvin-Helmholtz timescale $t_{\rm KH}$ (thick solid 
                 blue line, in $\rm kyr$), the accretion timescale $t_{\rm acc}$ (thin dotted red 
                 line, in $\rm kyr$) and the protostellar mass (thin solid black line, in $M_{\odot}$).   
                 Panels (b) and (d) show the accretion rate onto the protostars (thick solid 
                 red line, in $M_{\odot}\, \rm yr^{-1}$), the ratio $t_{\rm KH}/t_{\rm acc}$ 
                 (thin solid black line) and the thin dotted black line corresponds to $t_{\rm KH}/t_{\rm acc}=1$. 
                 }      
        \label{fig:timescale}  
\end{figure*}

\subsection{Accretion rate histories}
\label{sect:rates}

Fig.~\ref{fig:rate} shows the accretion rate histories onto the MYSOs forming during the 
gravitational collapse of $100\, \rm M_{\odot}$ (a) and $60\, \rm M_{\odot}$ (b) pre-stellar cores, respectively. 
The accretion rates (in $\rm M_{\odot}\, \rm yr^{-1}$) are plotted as a function of time (in $\rm kyr$), from 
the beginning of the collapse to the end of the simulation when $M_{\star}=M_{\rm c}/3$. 
The thin black vertical line marks the onset of disc formation when the free-collapsing material 
of the envelope begins to land on a centrifugally balanced circumstellar disc instead of keeping 
on directly falling onto the protostellar surface. From this time instance on, the protostar 
starts gaining mass via accretion from the disc. 
The thick, solid lines represent the accretion rate onto the young high-mass stars while the dashed 
lines are the corresponding protostellar masses (in $\rm M_{\odot}$). 
Furthermore, we indicate by the orange circles the time instance when the MYSOs enter
the high-mass regime ($\rm M_{\star} > 8\, \rm M_{\odot}$). 
We run two distinct hydrodynamical simulation with $M_{\rm c}=100\, \rm M_{\odot}$ (model Run-100-hydro) 
and with $M_{\rm c}=60\, \rm M_{\odot}$ (model Run-60-hydro), and in order to particularly investigate the 
effects of the accretion spikes on the protostellar evolution, we construct additional accretion rate histories 
by modifying the accretion rate of our model Run-100-hydro (our Table~\ref{tab:models}), by keeping constant 
the final mass of $\approx 33.3\, \rm M_{\odot}$ and either (i) imposing a constant rate once the 
disc has formed (model Run-100-constant, see black lines of Fig.~\ref{fig:rate}a) or (ii) filtering out the 
accretion spikes (model Run-100-smoothed, see red lines of Fig.~\ref{fig:rate}a) with a method similar to 
that described in~\citet{vorobyov_apj_805_2015}.

The free-fall collapse of the molecular pre-stellar core produces an initial infall of material through 
the sink cell increasing the accretion rate during the first $\approx 12$-$15\, \rm kyr$, up to reaching the 
canonical value of $\approx 10^{-3}\, \rm M_{\odot}\, \rm yr^{-1}$ at which MYSOs are predicted 
to accrete~\citep{hosokawa_apj_691_2009}. 
Once the disc has formed (see vertical thin line of Fig.~\ref{fig:rate}), variability 
immediately develop in the accretion flow as a direct consequence of important anisotropies 
in the protostellar surroundings~\citep{meyer_mnras_473_2018}. 
The variations amplitude in the accretion rate continues increasing up to being interspersed by 
violent accretion spikes becoming more numerous and more intense as a function of time. 
They are regularly generated by the rapid migration of massive disc fragments forming in its 
outer region by gravitational fragmentation and falling onto the protostellar surface, provoking 
sudden increases of the accretion rate. Those dense gaseous clumps detached from 
spiral arms developing in the self-gravitational discs are responsible for violent accretion-driven 
luminosity bursts~\citep{meyer_mnras_464_2017}. Such mechanism is connected 
to the formation of spectroscopic binary companions to the protostars, as long as the clumps sufficiently
contract in the core and heat up beyond the dissociation temperature of molecular 
hydrogen~\citep{meyer_mnras_473_2018}. 
The protostellar mass evolution reflects the accretion history from the disc in the sense that 
to each strong accretion events ($\dot{M} \ge 10^{-1}\, \rm M_{\odot}\, \rm yr^{-1}$) corresponds 
a step-like increase of the stellar mass~\citep{meyer_mnras_464_2017}. 
The integration time of model Run-60-hydro is longer ($\approx 50\, \rm kyr$) than that of 
Run-100-hydro ($\approx 65.2\, \rm kyr$) because we perform the simulations up to the time 
instance when $M_{\star}=M_{\rm c}/3$. This is longer in the case of the lower mass pre-stellar 
core ($M_{\rm c}=60\, \rm M_{\odot}$) since the mass infall onto the disc is intrinsically smaller.

\subsection{Stellar structure evolution}
\label{sect:structure}

Fig.~\ref{fig:structure} plots the evolution of the stellar structure of our protostars as a function 
of the stellar age for our models with $M_{\rm c}=100\, \rm M_{\odot}$, distinguishing from different 
accretion histories. The figure indicates, in addition to the temporal variations in photospheric 
radius, the convective (blue) and radiative (orange) regions of the protostar, respectively.  
The protostars are initially fully convective (blue regions) with an internal temperature profile too 
cold to ignite Deuterium burning. As its center heats up by gravitational contraction, a radiative core 
forms and grows in mass while rapidly expanding towards the stellar surface once the Deuterium fuel is out. 
The accreting stars further evolve once energetic photons are released out of the radiative core and propagate upwards to 
be absorbed by the still cold and convective enveloppe. During the gradual diminishing of the convective envelope 
thickness (when the protostars of our Run-100-hydro is $\approx 10\, \rm M_{\odot}$, see Fig.~\ref{fig:structure}a) 
concludes the phase transition from radiative to convective stellar interior. 
As the protostellar mass increases, the the so-called luminosity wave mechanism takes place~\citep{larson_mnras_157_1972}. 
We illustrate in Fig.~\ref{fig:lum_wave} the development of the luminosity wave of the growing protostar in our model Run-100-hydro. 
A maximum in the luminosity profile forms (Fig.~\ref{fig:lum_wave}a) and moves outwards until it reaches the stellar surface 
and adopts a strictly monotonically increasing shape, i.e. the luminosity gradient is positive everywhere (Fig.~\ref{fig:lum_wave}c). 
The energy equation of the stellar structure reads as, 
\begin{equation}
	\frac{ dL_\star(r) }{ dM_\star(r) } = -T_{\rm eff} \frac{ ds_\star }{ dt }, 
	\label{eq:structure}
\end{equation}
where $L_\star(r)$, $M_\star(r)$, $T_{\rm eff}$ and $s_\star$ are the internal luminosity, mass, temperature and specific entropy radial profiles. 
It implies that interior to  the luminosity peak (at $r<R_{\rm c}$) the entropy of the star decreases with time, i.e., 
\begin{equation}
	\frac{ dL_\star(r) }{ dM_\star(r)}>0 \Rightarrow \frac{ ds_\star }{ dt }<0 \mbox{ if } r<R_{\rm c},\\
	\label{}
\end{equation}
while exterior to the luminosity peak (at $r>R_{\rm c}$) the entropy of the star increases with time, i.e.,	
\begin{equation}
	\frac{ dL_\star(r) }{ dM_\star(r)}<0 \Rightarrow \frac{ ds_\star }{ dt }>0 \mbox{ if } r>R_{\rm c}. 
	\label{}
\end{equation}
Therefore, the entropy of the surface layers ($r>R_{\rm c}$) 
increase as they absorb energy triggering the stellar bloating, whereas when the $L(r)$ peak reaches the stellar surface at 
$r=R_{\star}$ all the layers lose entropy and the protostar contracts. Each time a violent accretion event 
with an accretion peak of $\approx 10^{-2}$$-$$10^{-1}\, \rm M_{\odot}\, \rm yr^{-1}$ deposits mass to the star, an 
other luminosity maximum sets in the interior (Fig.~\ref{fig:lum_wave}b) and the luminosity wave forms again in 
the upper layers of the protostar (Fig.~\ref{fig:lum_wave}d), provoking a new swelling episod. 
Our model with $M_{\rm c}=60\, \rm M_{\odot}$ gives similar results.

%The convective envelope absorbe the flux so that $L_{\star}$ is maximum at the interface between 
%the convective and radiative regions of the high-mass star, which moves upward as the radius 
%bloats and the envelope recedes. On the whole star is radiative the radius decreases and 
%the star moves black to the blue towards the ZAMS. 
%
%Such wave is further illustrated in the figure of by L.H. [to do]. Such phenomenon has already 
%5been found in the study on accreting \hii region of\citep{haemmerle_458_mnras_2016}, however, 
%its connection to the burst is new. Further deposition of mass onto the radiative surface 
%of the protostars immediately generate additional bloating of their radius, which increase 
%the luminosity (bursts) and correspondingly decrease $T_{\rm eff}$ make the star cooler. 

\begin{figure}
        \centering
        \begin{minipage}[b]{ 0.475\textwidth}  \centering
                \includegraphics[width=1.0\textwidth]{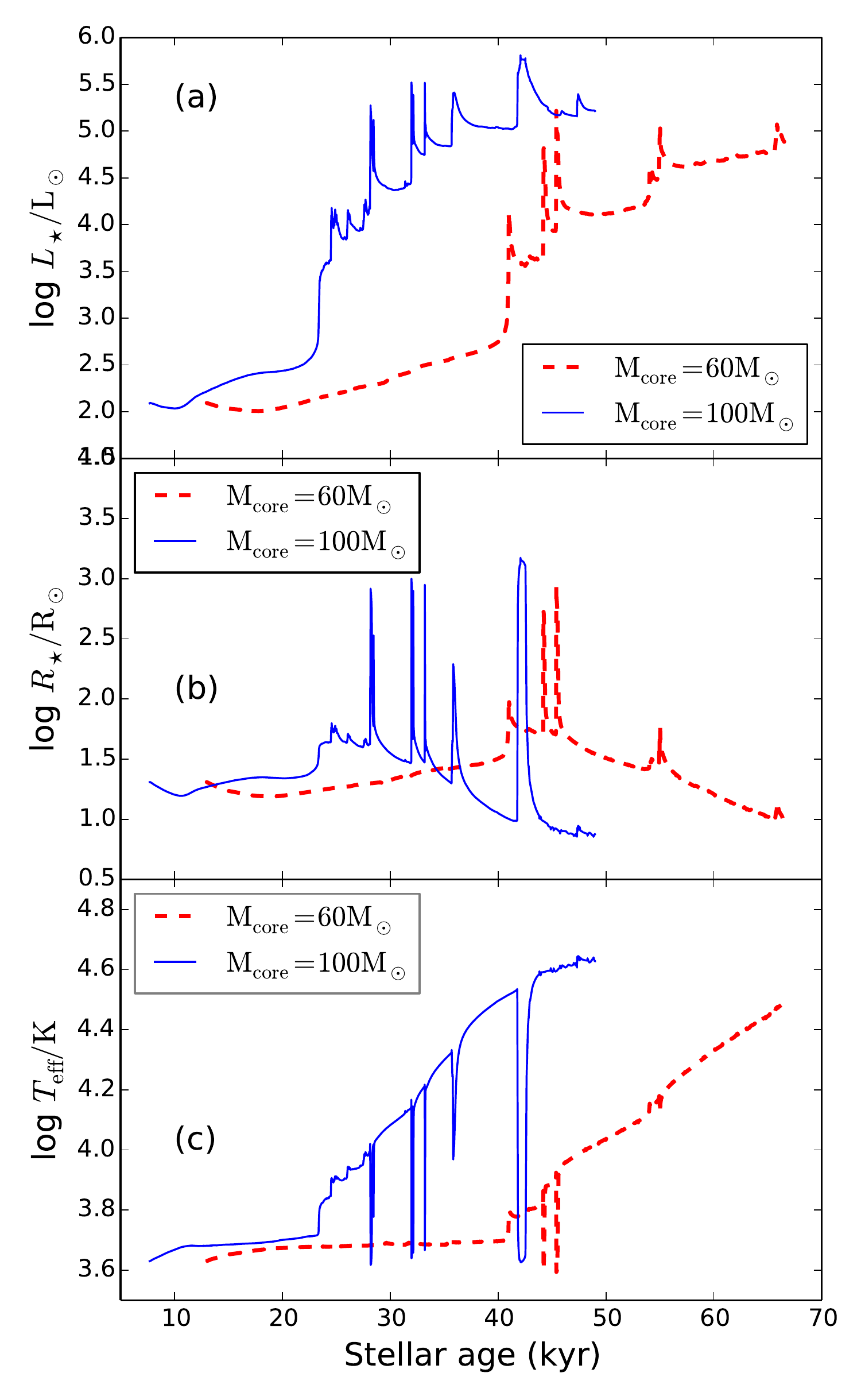}
        \end{minipage}        
        \caption{ 
        		 Evolution as a function of time (in $\rm kyr$) of the stellar surface luminosity (a), 
        		 stellar radius (b) and effective temperature (c) of our MYSOs experiencing 
        		 variable disc accretion interspersed by bursts. The panels distinguish between the models assuming 
        		 an initial $60\, \rm M_{\odot}$ (thick dotted red line) and $100\, \rm M_{\odot}$ 
        		 (thin solid blue line) pre-stellar core, respectively. 
                 }      
        \label{fig:stellar_100}  
\end{figure}

\begin{figure*}
        \centering
        \begin{minipage}[b]{ 0.9\textwidth}  \centering
                \includegraphics[width=1.0\textwidth]{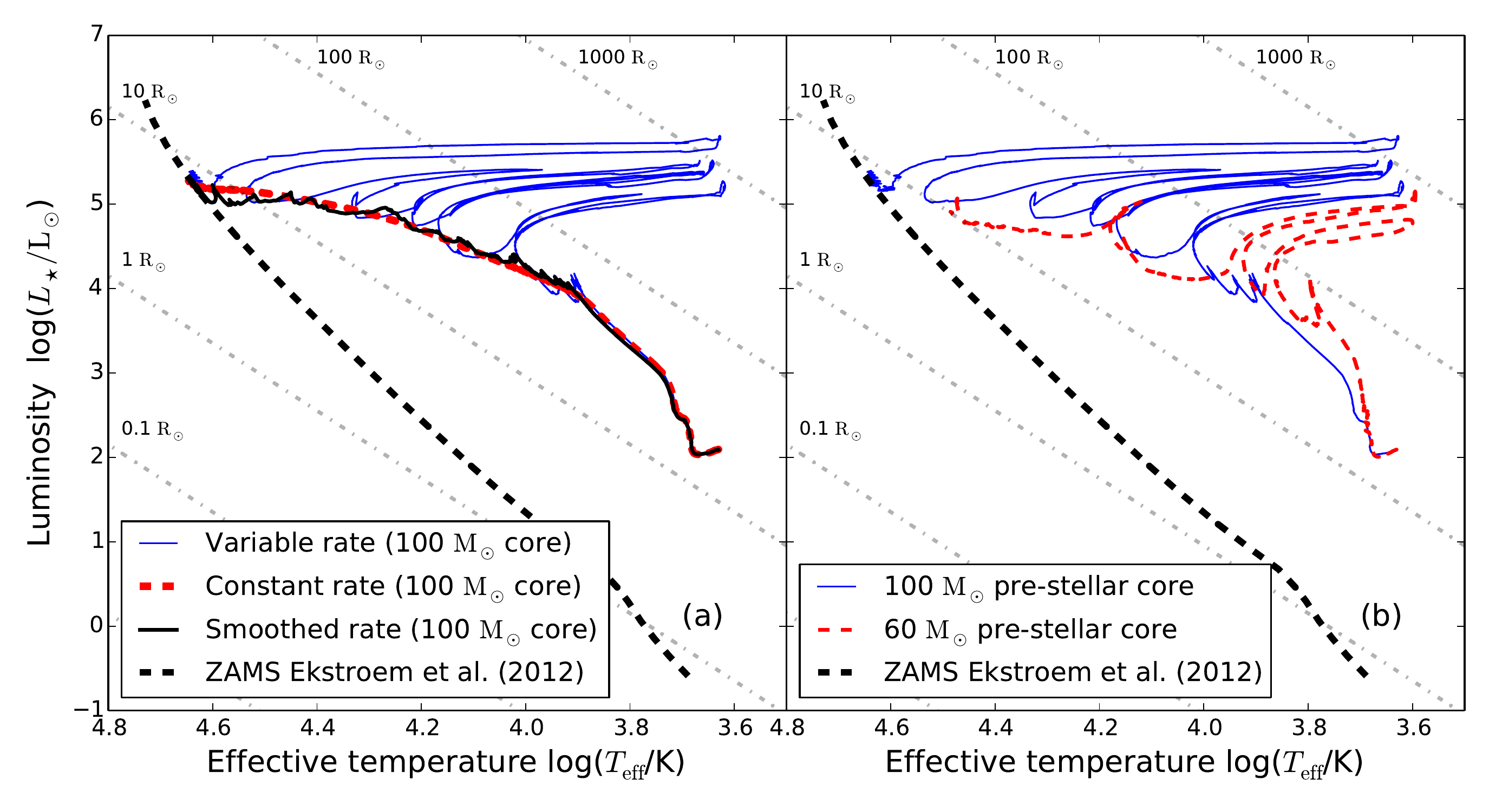}
        \end{minipage}               
        \caption{ 
        		 Hertzsprung-Russell diagram of our proto-stellar models with a similar pre-stellar core mass of 
        		 $100\, \rm M_{\odot}$, but experiencing different initial accretion rate histories (a) 
        		 and evolutionary tracks with pre-stellar cores of different initial masses 
        		 of $60$ and $100\, \rm M_{\odot}$, respectively (b).
        		 Grey dotted solid lines are isoradius and the thick black dashed line is the 
        		 zero-age-main-sequence (ZAMS) track of~\citet{ekstroem_aa_537_2012}.  
                 }      
        \label{fig:hdr}  
\end{figure*}

Fig.~\ref{fig:timescale} presents two characteristic timescales, which are usually used to describe 
the evolution of accreting protostars, and reports their time evolution next to the accretion rate 
history and the protostellar mass evolution. More specifically, the figure plots the so-called accretion timescale, 
\begin{equation}
	t_{\rm acc} = \frac{ M_{\star} }{ \dot{M} }, 
	\label{eq:tacc}
\end{equation}
representing the characteristic timescale of mass growth of the accreting star (thin dashed red lines), and the Kelvin-Helmholtz timescale,
\begin{equation}
	t_{\rm KH} = \frac{G  M_{\star}^{2} }{ R_{\star} L_{\star} }, 
	\label{eq:kh}
\end{equation}
related to the entropy loss through radiation (thick solid blue lines). The protostar begins accreting at          
a variable rate once the disc forms (thick red lines), with $t_{\rm acc}<t_{\rm KH}$ at times when their evolution is 
only governed by disc accretion instead of mass infall from the pre-stellar core. 
During the gravitational collapse and the early disc evolution, the protostars evolve by gaining 
entropy by advection, with negligible radiative loss ($t_{\rm acc}<t_{\rm KH}$, see Fig.~\ref{fig:timescale}b,d). 
Note that our model Run-100-hydro is already a MYSOs when the first burst happens, whereas our model 
Run-60-hydro experiences its first swelling as an intermediate-mass star (thin solid line of Figs.~\ref{fig:timescale}a,c). 
Once the photospheric luminosity has grown enough, the energy loss governs the protostellar evolution 
($t_{\rm acc}>t_{\rm KH}$ because $t_{\rm KH} \propto 1/R_{\star}L_{\star}$ throughout the bursts) 
and \textcolor{black}{$L_{\rm acc}<L_{\star}$}, which means that $L_{\rm acc}$ can be 
neglected (see Section~\ref{sect:excursions}). 
Interestingly, the protostars experience episods with $t_{\rm KH}/t_{\rm acc}>1$ at the same 
time instance of the swelling. The MYSOs intermittently see their evolution ruled by mass 
accretion (the bloating) before recovering a more classical evolution 
governed by energy losses (the unswelling). This repeats each time an accretion burst happens. 
Therefore, each accretion-driven burst corresponds to a swelling episod of the MYSOs followed by a redistribution 
of the internal entropy restoring the internal thermal equilibrium. 
%
%When the protostars enter a bursts phase, $t_{\rm acc}$ decreases inducing a swelling phase, 
%immediately followed by a Kelvin-Helmholtz contraction once the accretion rate recovers its pre-burst, 
%quiescent value. 
%
%Those timescales are global quantities characterising the overall evolution of our MYSOs, so 
%their episodic behaviour reflecting the luminosity-wave-generating accretion of gaseous clumps 
%equivalently affects the protostellar surface properties. 

\subsection{Evolution of the surface properties of the MYSOs}
\label{sect:episodic}

In Fig.~\ref{fig:stellar_100}, we show the evolution of the protostellar iternal photospheric luminosity (a), 
radius (b), and effective temperature (c) as a function of the stellar age of our MYSOs. 
The Figure distinguishes between the variable-accreting models with a \textcolor{black}{$100\, \rm M_{\odot}$} pre-stellar core (thin solid blue lines) 
and with a \textcolor{black}{$60\, \rm M_{\odot}$} pre-stellar core (thick dashed red line), respectively. 
We assume that the MYSOs are black-bodies, therefore, the photospheric luminosity is estimated as,
\begin{equation}
	L_{\star} = 4 \pi R_{\star}^{2} \sigma T_{\rm eff}^{4},
	\label{eq:L}
\end{equation}
where $\sigma$ is the Stefan-Boltzman constant. 
The stellar surface properties does not evolve much during the free-fall gravitational collapse onto the 
MYSOs and the early phase of the disc formation, at times $\le 20\, \rm kyr$. Only a moderate and monotonical 
increase of the stellar radius and corresponding photospheric luminosity occurs, as the deuterieum burning maintains the 
central temperature nearly constant (Fig.~\ref{fig:stellar_100}a). 
When the variations of the accretion rate substantially increase in response to the growing strength of gravitational 
instability and fragmentation in the circumstellar disc, the protostar grows faster, 
and, after the first luminosity wave, its surface becomes radiative so that any episodic deposit of 
mass on it generates an augmentation of the effective temperature and the surface luminosity. 
With one important exception, the time evolution of stellar surface properties is similar to what was 
found in the context of calculations carried out with a constant accretion rate -- the photospheric 
luminosity (Fig. 5a) and effective temperature (Fig. 5c) generally increase, while the stellar 
radius decreases (Fig. 5b) for as long as the star remains in the quiescent accretion phase.
However, this monotonic behaviour is interspersed  with brief excursion events associated with violent accretion bursts, during 
which the MYSOs adopt opposite photospheric properties by becoming bigger and cooler. 
%The brevity of those cool phases corresponds to the time interval $t_{\rm acc}$ of the 
%accretion of the dense gaseous clumps which provoke the accretion spikes of Fig.~\ref{fig:rate}. 
%
Such a behaviour of $R_{\star}$ and $T_{\rm eff}$ is a direct consequence of the changes in the internal 
structure of the MYSOs, which is sensitive to the accretion rate. In the next section, we demonstrate 
how the variable accretion rate with episodic bursts can affect the evolutionary path of MYSOs in 
the Hertzsprung-Russell diagram.

\subsection{Pre-main-sequence excursions in the Hertzsprung-Russell diagram}
\label{sect:excursions}

Fig.~\ref{fig:hdr} shows the evolutionary tracks of our MYSOs in the Hertzsprung-Russell 
diagram. More specifically, the evolution of the central star in the model with a $100\, \rm M_{\odot}$ pre-stellar 
core was calculated using a variable (thin blue 
line), constant (thick dashed red line) and smoothed (thick black line) accretion rates and are plotted in panel 
(a) together with the zero-age-main-sequence (ZAMS) track of~\citet{ekstroem_aa_537_2012}. The tracks 
of the central stars in the models with different pre-stellar core masses of $60$ and $100\, \rm M_{\odot}$ are shown in panel 
(b). In this case, only the variable accretion rate was considered. Grey solid lines are isoradii. 
From the definition of $t_{\rm KH}$ and $t_{\rm acc}$~\citep{hosokawa_apj_691_2009,hosokawa_apj_721_2010} 
it can be shown that, 
\begin{equation}
	\frac{ t_{\rm KH} }{ t_{\rm acc} } \propto \frac{ L_{ \rm acc } }{ L_{ \star } },
	\label{eq:rel_t_L}
\end{equation}
with the accretion luminosity $L_{ \rm acc} \propto G \dot{M} M_{\star} / R_{\star}$, where $G$ is 
the gravitational constant. 
%we can neglect any accretion luminosity when plotting the temperature-luminosity evolution of the 
%MYSOs, see Fig.~\ref{fig:timescale}. 
%
The initial stages of the stellar evolution calculations are similar as we assume identical accretion rates 
during the free-fall collapse. Differences occur at the onset of the disc formation. 
In Fig.~\ref{fig:hdr}a, we directly see that the final stellar mass at the end of the main accretion 
phase is not the key quantity that governs the evolutionary tracks of high-mass protostars 
in the Hertzsprung-Russell diagram. Despite all models are calculated up to reaching a similar 
stellar mass of $33.3\, \rm M_{\odot}$, their track are qualitatively different. Furthermore, 
the track of the Run-100-smoothed model with a smoothed accretion rate (but retaining small-amplitude 
variations) is globally similar to the track of the Run-100-constant model, which assumes a constant 
accretion rate. The small-amplitude variations in the accretion rate only induce tiny 
deviations from the constant accretion track. 
The model with accretion spikes strongly deviates from the constant accretion track. Strong 
accretion bursts generate short but important changes in the pre-main-sequence 
evolutionary track of the MYSOs in the form of evolutionary loops to the red part of the Hertzsprung-Russell 
diagram. These excursions repeat themselves as more and more accretion spikes occur. 
We find similar behaviour for our model with a pre-stellar core mass of $60\, \rm M_{\odot}$ (Fig.~\ref{fig:hdr}b). 
%
%Note that if Run-100-hydro only exhibit strong excursion beyond the $100\, \rm R_{\odot}$ isoradius, 
%Run-60-hydro has both low- and high-amplitude excursions. 

%%%%%%%%%%%%%%%%%%%%%%%%%%%%%%%%%%%%%%%%%%%%%%%%%%%%%%%%%%%%%%%%%%%%%%%%%%%%%%%%%%%%%%%%%%%
%%%%%%%%%%%%%%%%%%%%%%%%%%%%%%%%%%%%%%%%%%%%%%%%%%%%%%%%%%%%%%%%%%%%%%%%%%%%%%%%%%%%%%%%%%%
%%%%%%%%%%%%%%%%%%%%%%%%%%%%%%%%%%%%%%%%%%%%%%%%%%%%%%%%%%%%%%%%%%%%%%%%%%%%%%%%%%%%%%%%%%%

\section{Effect of the initial conditions on the stellar evolutionary tracks of MYSOs}
\label{sect:ic}

\textcolor{black}{
This section explores the effects of different initial conditions of the protostellar seed in 
the evolution calculations on stellar structures, the various protostellar properties 
and on the pre-ZAMS evolutionnary track of the MYSOs in the Hertzsprung-Russell diagram. 
}

\begin{table*}
	\centering
	\caption{
	\textcolor{black}{
	Characteristics of the $2\, \rm M_{\odot}$ protostellar seeds used in our comparison simulations of our models with $100\, \rm M_{\odot}$ core and episodic accretion.   
	}
	}
	\begin{tabular}{lcccr}
%	\hline
	\hline
	${\rm {Models}}$                          &  $R_{\star}$ ($\rm R_{\odot}$)    &    $L_{\star}$ ($\rm L_{\odot}$)   &   $T_{\rm eff}$ ($\rm K$)   & ${\rm {Initial}\, \rm {conditions}}$  \\ 
	\hline    
	\textcolor{black}{\rm Run-100-hydro}       &  $20.4$                           &    $123$                           &  $4270$         &  \textcolor{black}{ Convective embryo with large initial radius mimicing spherical, then disc accretion$^{\star}$ }    \\  
	\textcolor{black}{{\rm Run-100-compact}}    &  $2.9$                            &    $9.7$                           &  $5960$         &  \textcolor{black}{ Radiative embryo with small initial radius mimicing continuous disc accretion$^{\star}$ }    \\ 
	\hline    
%	\hline 
	\end{tabular}
\label{tab:models_com}\\
\footnotesize{ ($\star$) See Section 2.3 of~\citet{haemmerle_458_mnras_2016} }\\
\end{table*}

\subsection{\textcolor{black}{Stellar structures}}
\label{sect:ic_struc}

\textcolor{black}{
The seed core taken as a stellar embryo to initialise the stellar evolution calculations is a 
parameter which both depends on the properties of the pre-stellar core~\citep{vaytet_aa_598_2017,bhandare_aa_618_2018} 
and influences the stellar evolution calculations~\citep{haemmerle_458_mnras_2016}. 
In order to investigate the effects of the bursts on the evolutionary tracks as a function of the 
initial seeds, we carry out a comparison study between two cases using our accretion rate derived from the 
hydrodynamical simulation with a initial $100\, \rm M_{\odot}$ pre-stellar core with variable 
accretion rates (Run-100-hydro and Run-100-compact), see our Table~\ref{tab:models_com}. 
The model Run-100-hydro is initialised with a convective core of $2\, \rm M_{\odot}$ 
of radius $20.4\, \rm R_{\odot}$, temperature $4270\, \rm K$ and luminosity $123\, \rm L_{\odot}$, 
while the model Run-100-compact is started as a fully radiative star of $2\, \rm M_{\odot}$ 
with radius $2.9\, \rm R_{\odot}$, temperature $5960\, \rm K$ and luminosity $9.7\, \rm L_{\odot}$, 
respectively. 
They principally differ by the internal entropy profiles, i.e. the initial convective model 
Run-100-hydro exhibits a flat entropy profile towards the protostellar surface, whereas 
the initial radiative model Run-100-compact has a positive entropy gradient. Therefore, Run-100-hydro  
is adiabatically convective and is larger than Run-100-compact by about an order of magnitude. 
Details about these protostellar seeds are given in~\citet{haemmerle_458_mnras_2016}. 
Particularly, the authors stressed therein that the convective and radiative protostellar embryos have 
properties similar to models of $M_{\star}=2\, \rm M_{\odot}$ built by hot and cold accretion, respectively, 
\textcolor{black}{while the geometry of the accretion is cold for both cases throughout the stellar evolution calculations}. 
Our comparison models therefore investigate the effects of the early accretion geometry on the evolution of the MYSOs. 
}

\begin{figure}
        \centering
        \begin{minipage}[b]{ 0.49\textwidth}  \centering
                \includegraphics[width=1.0\textwidth]{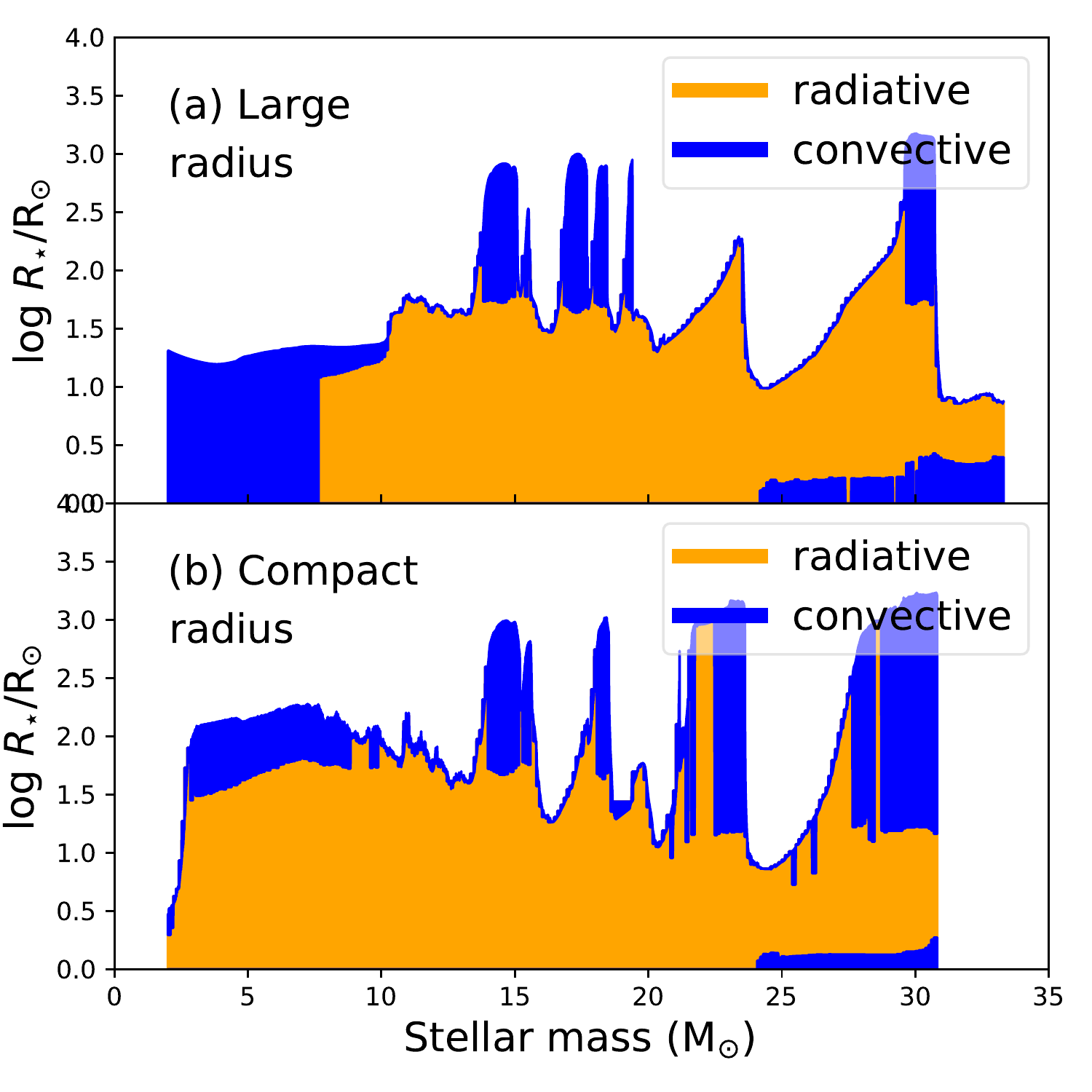}
        \end{minipage}        
        \caption{ 
                 \textcolor{black}{
        		 As Fig.~\ref{fig:structure} for our of MYSOs generated with an initial $100\, \rm M_{\odot}$ pre-stellar 
        		 core, variable accretion rates, large (a) and compact (b) initial protostellar radius. 
                 }      
                 }
        \label{fig:ic_structure}  
\end{figure}

\textcolor{black}{
Fig.~\ref{fig:ic_structure} reports the Kippenhahn diagram two MYSOs calculated with the 
same episodic accretion rate history but different initial conditions (Tab.~\ref{tab:models_com}). 
Despite of differences in the evolution of the outer radius of the protostar, 
notable swelling episods of the MYSO appear at the time instances corresponding 
to the accretion bursts. Although differences are visible, especially (i) during the 
clustered bursts at $15$$-$$20\, \rm M_{\odot}$ in the model Run-100-hydro (a) which 
results in a single inflation of $R_{\star}$ in Run-100-compact (b), and (ii) in the 
burst at $\sim 22.5\, \rm M_{\odot}$ that is more pronounced in the initially 
convective case (b), the major outbursts nevertheless generate similar effects 
regardless of the initial conditions of the calculations. 
Once the protostar reaches $\sim 10\, \rm M_{\odot}$, both simulated MYSOs are 
structured with a convective layer that inflates under the effects of the entropy 
deposition by accretion of gaseous circumstellar clumps and a radiative interior 
which progressively develops a convective core once the protostar is heavy and hot enough 
for H-burning at $\sim 25\, \rm M_{\odot}$. 
The development of a luminosity wave in the outer layer of the MYSO each time a burst 
happens is similar as above pictured in the context of an initially radiative protostellar 
embryo (Fig.~\ref{fig:lum_wave}). 
Note that the radiative model is numerically more difficult to calculate and 
it has been simulated over a slightly smaller time interval. Both initial conditions 
produce similar effects on the evolution of the radius of MYSOs as a response of the 
accretion of circumstellar gaseous clumps. 
}

\subsection{\textcolor{black}{Surface properties and excursions in the Hertzsprung-Russel diagram}}
\label{sect:ic_surface}

\begin{figure}
        \centering
        \begin{minipage}[b]{ 0.49\textwidth}  \centering
                \includegraphics[width=1.0\textwidth]{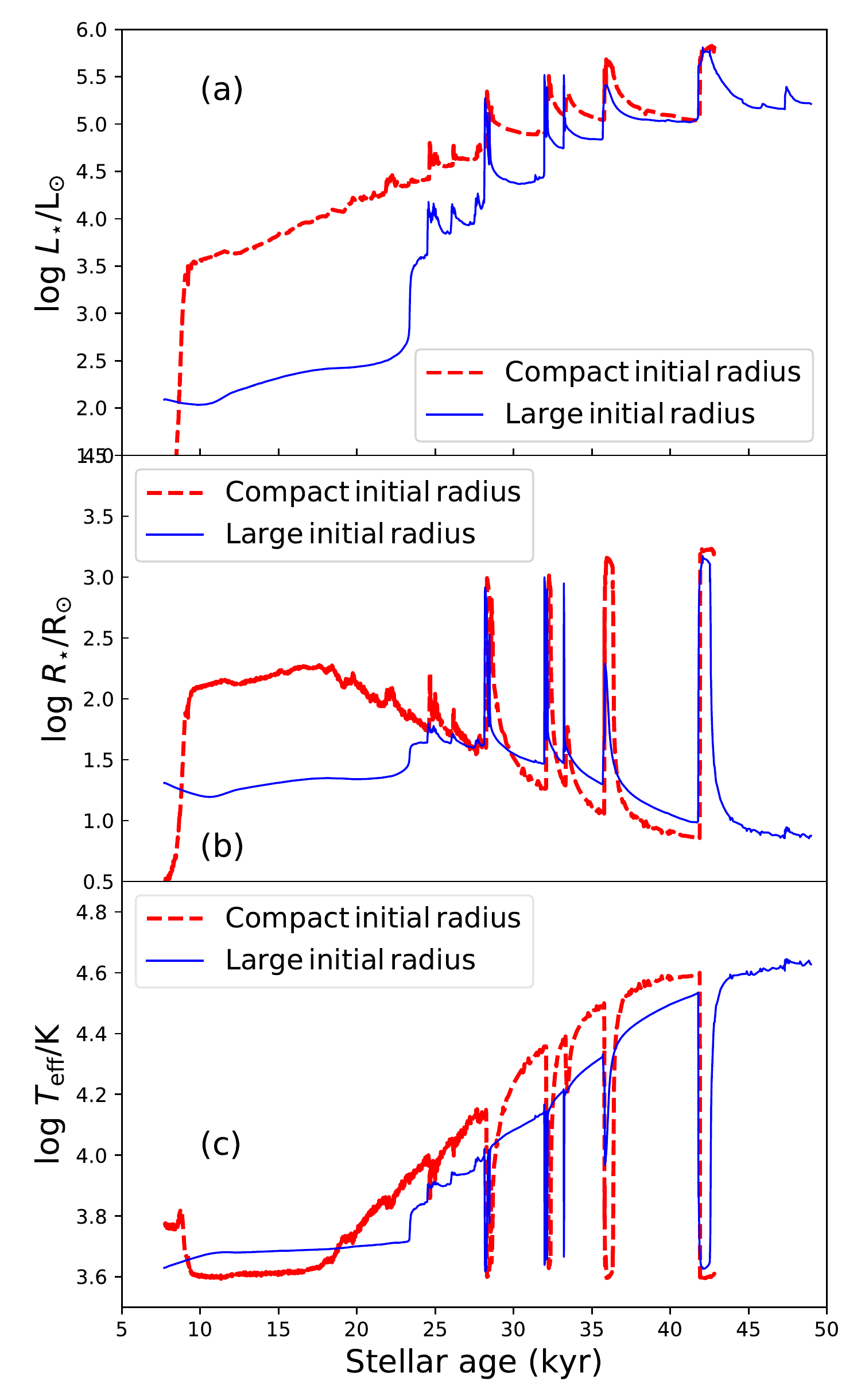}
        \end{minipage}        
        \caption{ 
                 \textcolor{black}{
        		 As Fig.~\ref{fig:stellar_100} with the evolution as a function of time (in $\rm kyr$) 
        		 of the stellar surface luminosity (a), stellar radius (b) and effective temperature (c) 
        		 of our $\rm M_{\rm c}=100\, \rm M_{\odot}$ collapsing pre-stellar cores, considered with 
        		 different initial radii of the protostellar embryo in the stellar evolution calculations, 
        		 respectively. 
        		 }
                 }      
        \label{fig:ic_prop}  
\end{figure}

\textcolor{black}{
Fig.~\ref{fig:ic_prop} plots the evolution of the stellar surface luminosity $L_{\star}$ 
(panel a, in $\rm L_{\odot}$), the stellar radius $R_{\star}$ (panel b, in $\rm R_{\odot}$) and 
(the effective temperature $T_{\rm eff}$ (panel c, in $\rm K$) as 
a function of time (in $\rm kyr$) of our $\rm M_{\rm c}=100\, \rm M_{\odot}$ collapsing 
pre-stellar cores that are considered with large, convective (thin solid blue line) and compact, 
radiative (thick dotted red line) initial conditions, respectively. 
The only difference in the calculations is the past evolution of the stellar embryo of 
$2\, \rm M_{\odot}$ which is mimiced by the initial conditions of the models (Tab.~\ref{tab:models_com}). 
Accretion onto the radiative embryo produces an immediate swelling ($>100\, \rm R_{\odot}$) 
as a response of the deposition of entropy by cold accretion (Fig.~\ref{fig:ic_prop}b), which also 
dimishes the surface temperature (Fig.~\ref{fig:ic_prop}c) and increases the surface luminosity 
(Fig.~\ref{fig:ic_prop}a). 
These differences between the surface properties of the MYSOs persist up to $\approx 20$$-$$25\, \rm kyr$, 
when the series of accretion-driven outbursts begins. As illustrated in Fig.~\ref{fig:ic_structure}, 
the effets of the accretion spikes onto the protostellar properties are qualitatively similar: the 
most important luminosity rises and falls coincide with each other as a function of time and their respective 
offsets are due to slightly shifted values of $R_{\star}$ and $T_{\rm eff}$. 
Note that the initial compact model has a more pronounced swelling during the series of moderate 
bursts at $\approx 20$$-$$25\, \rm kyr$ (Fig.~\ref{fig:ic_prop}b). Once the strong accretion events 
onto the MYSO have started, the initial convective model has baseline values of $R_{\star}$ 
and $T_{\rm eff}$ more compact and hotter than that of the initial radiative calculation (see thick 
red dotted lines of Fig.~\ref{fig:ic_prop}a-c). 
}

\begin{figure}
        \centering
        \begin{minipage}[b]{ 0.48\textwidth}  \centering
                \includegraphics[width=1.0\textwidth]{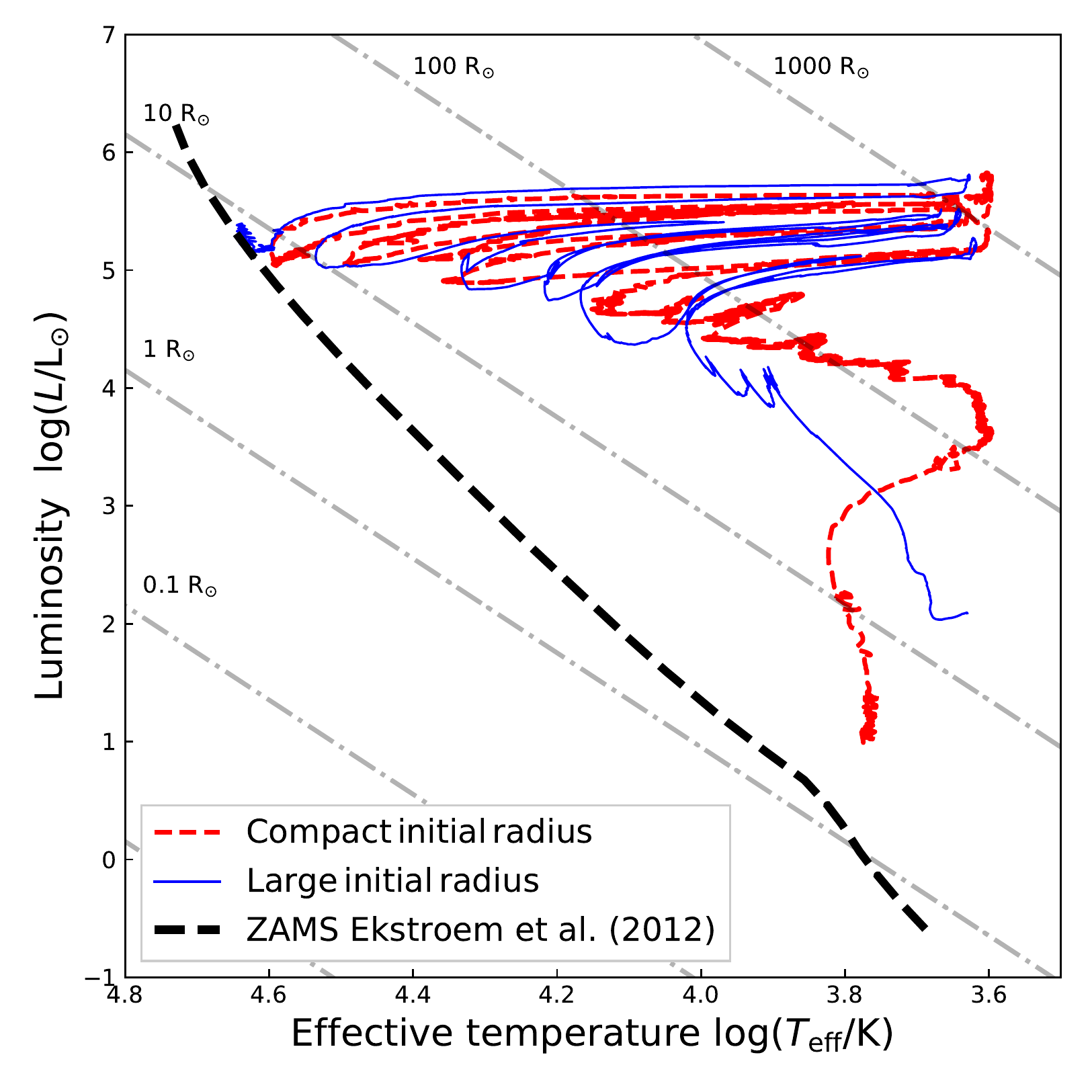}
        \end{minipage}     
        \caption{ 
                 \textcolor{black}{
        		 As Fig.~\ref{fig:hdr} with the evolutionary tracks of our models 
        		 of $M_{\rm c}=100\, \rm M_{\odot}$ collapsing pre-stellar cores 
        		 and episodic accretion rates onto the protostar, considered with 
        		 different initial conditions of the stellar embryo, respectively. 
        		 }
                 }      
        \label{fig:ic_hrd}  
\end{figure}

\textcolor{black}{
Fig.~\ref{fig:ic_hrd} shows the evolutionary tracks in the Hertzsprung-Russell diagram of the models 
Run-100-hydro (thin solid blue line) and Run-100-compact (thick dotted red line), respectively. The 
grey dotted solid lines are isoradii and the thick black dashed line is the zero-age-main-sequence 
(ZAMS) track of~\citet{ekstroem_aa_537_2012}. 
It underlines the initial differences between the two models, especially the rapid 
swelling of the compact model by $\sim 2$ orders of magnitude accompanied by a decrease of 
its temperature and an increase of its luminosity, respectively. It also further illustrates that 
the early bursts are more pronounced in the compact case than it the convective case. 
The protostellar radius of the initially radiative Run-100-compact (thick dotted red line) is 
larger than that of the initially convective model Run-100-hydro in the quiescent phases 
(Fig.~\ref{fig:ic_prop}b), therefore, its evolutionary track is closer to the $100\, \rm R_{\odot}$ 
isoradius. This difference diminishes as a function of time and the two tracks overlap 
each other in the quiescent phase after the third excursion which almost reach the 
$1000\, \rm R_{\odot}$ isoradius occurs. 
Finally, let notice that in both cases, the tracks cross the $1000\, \rm R_{\odot}$ isoradius throughout 
the strongest swelling episods and reach the Hayashi limit. In the initial convective case, when the peak 
of the excursions happens, the track of the MYSO slightly follows vertically the Hayashi track before it 
recovers the values of $R_{\star}$ and $L_{\star}$ corresponding to the quiescent accretion phase. 
Although the properties of the first Larson core depend on the intrinsic pre-stellar core 
characteristics~\citep{vaytet_aa_598_2017,bhandare_aa_618_2018}, our suite of comparison simulations shows that the 
\textcolor{black}{mechanism triggering the excursions} of MYSOs in the Hertzsprung-Russell diagram \textcolor{black}{is} 
independent of their initial conditions.  
This shows that, under our assumptions, the accretion geometry induces little qualitative differences on the 
evolution of the surface properties of episodically-accreting MYSOs. 
}

%%%%%%%%%%%%%%%%%%%%%%%%%%%%%%%%%%%%%%%%%%%%%%%%%%%%%%%%%%%%%%%%%%%%%%%%%%%%%%%%%%%%%%%%%%%
%%%%%%%%%%%%%%%%%%%%%%%%%%%%%%%%%%%%%%%%%%%%%%%%%%%%%%%%%%%%%%%%%%%%%%%%%%%%%%%%%%%%%%%%%%%
%%%%%%%%%%%%%%%%%%%%%%%%%%%%%%%%%%%%%%%%%%%%%%%%%%%%%%%%%%%%%%%%%%%%%%%%%%%%%%%%%%%%%%%%%%%

\section{Discussion}
\label{sect:discussion}

In this section, we present the limitation of our method, compare our outcomes to those of other 
studies assuming burst-free disc accretion histories and extrapolate the findings of our study to 
the formation of intermediate-mass stars, and we discuss the observable implications of our results. 
\textcolor{black}{
Finally, we review alternative explanations for FU-Orionis-like bursts from young stars. 
}

\begin{figure*}
        \centering
        \begin{minipage}[b]{ 0.695\textwidth}  \centering
                \includegraphics[width=1.0\textwidth]{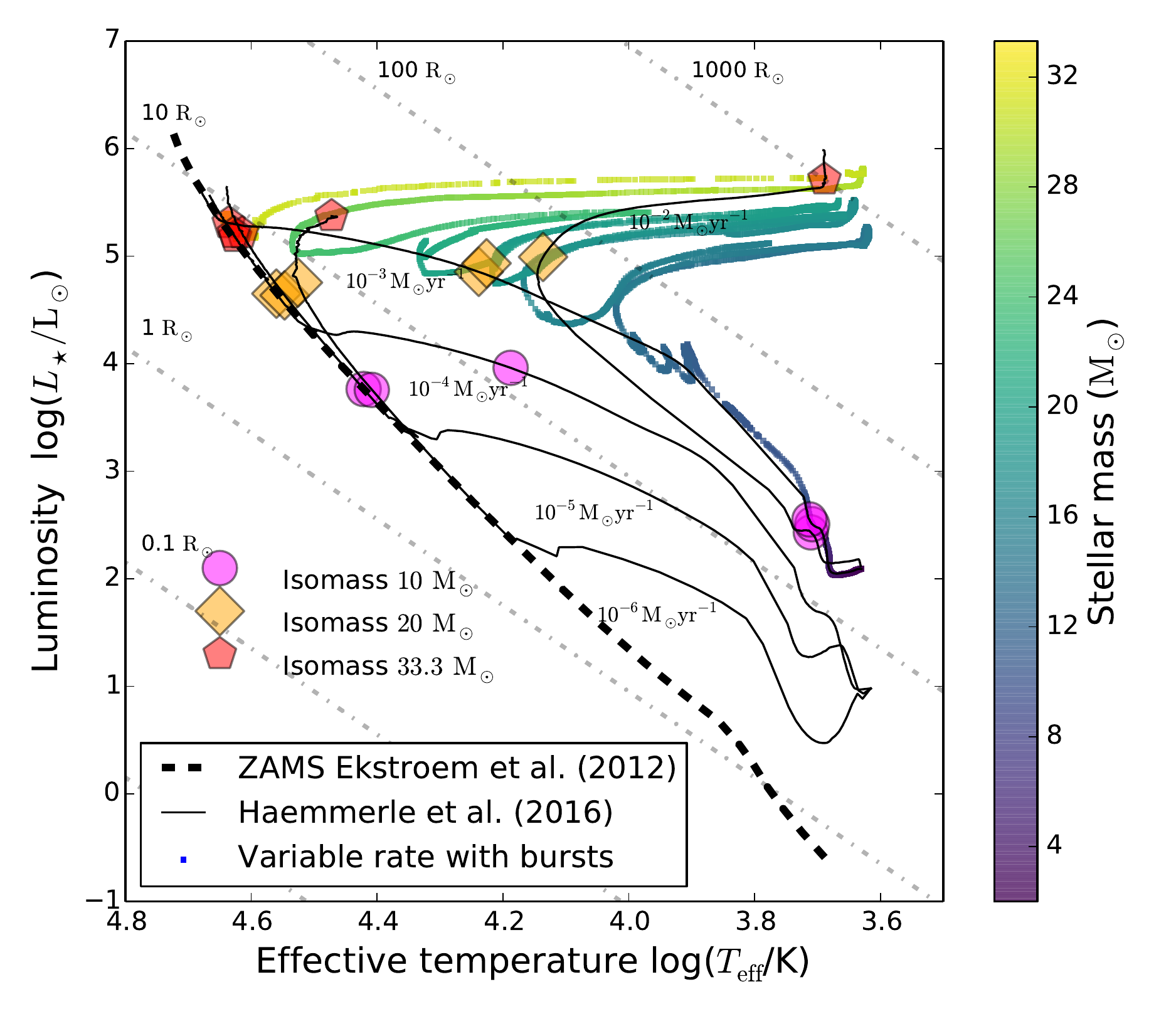}
        \end{minipage} 
%       \centering
%       \begin{minipage}[b]{ 0.44\textwidth}  \centering
%               \includegraphics[width=1.0\textwidth]{./plot_hdr_time.pdf}
%       \end{minipage}     
        \caption{ 
        		 %As Fig.~\ref{fig:hdr1} for our model with $M_{\rm c}=100\, \rm M_{\odot}$ and episodic accretion rate. 
        		 Comparison between our massive protostar with $M_{\rm c}=100\, \rm M_{\odot}$ experiencing variable accretion (thick coloured line) 
        		 and the evolutionary tracks for massive protostars accreting with constant rates (thin black lines) of~\citet{haemmerle_585_aa_2016}. 
        		 The color coding of the track indicate the stellar mass (in $\rm M_{\odot}$) and the coloured dots are 
        		 isomasses, respectively.  
        		 Grey dotted solid lines are isoradius and the thick black dashed line is the 
        		 zero-age-main-sequence (ZAMS) track of~\citet{ekstroem_aa_537_2012}. 
                 }      
        \label{fig:hdr2}  
\end{figure*}

\subsection{Model caveats}
\label{sect:caveats}

The limitations of our method are two-fold. First, the numerical hydrodynamics simulations of disc 
formation are subject to caveats which merit future improvements, whereas the stellar calculations 
are also subject to assumptions potentially calling follow-up betterments. 
In addition to the well-known limitation of disc fragmentation simulations given by the logarithmically-expanding 
radial grid~\citep{meyer_mnras_473_2018}, the sink cell radius, which strongly influences the time-step of the simulations, 
is kept to a value making 
the simulations affordable from the point of view of their numerical cost. Decreasing the inner hole would allow us to better follow  
the migration of the gaseous clumps responsible for the accretion-driven bursts, and therefore make our accretion histories more accurate. 
However, the value used in this paper is kept to a decent value ($r_{\rm in}=20\, \rm au$) that is still smaller 
than that of other studies on disc fragmentation, see e.g. the supermassive protostellar models of~\citet{hosokawa_2015}.

\textcolor{black}{For the sake of completeness,}
future improvements should equivalently include the initial non-sphericity and differential 
rotation of the parent pre-stellar cores~\citep[see, e.g.][]{banerjee_mnras_373_2006} and take into account 
stellar motion in response to the gravitational force of the disc, as massive disc substructures 
can shift notably the center of mass of the system from coordinate center where the protostar resides~\citep{regaly_aa_601_2017}. 
\textcolor{black}{
Nevertheless, despite of the fact that the effects of the stellar inertia on the behaviour of accretion discs has been anaytically 
shown to play a role on the developement of asymetries in their structures~\citep{adam_apj_347_1989}, recent numerical simulations 
demonstrated that wobbling neither prevents disc fragmentation nor reduces the bursts intensities \textcolor{black}{in the 
early formation phase of young high-mass stars}~\citep{2018arXiv181100574A}. 
A set of comparison simulations with and without wobbliofofng is shown in the Fig.~7 of their Section~4 and its illustrates that 
high-magnitudes accretion-driven outbursts develop similarly within the same timescale after the onset of the disc formation which 
followed the free-fall gravitational collapse. Only the time instance and perhaps the long-term occurence of the excursions 
of MYSOs in the Hertzsprung-Russell diagram would change if a moving sink-cell is used in the hydrodynamical simulations. 
}
Our assumptions related to the hydrodynamical simulations are discussed in great detail in~\citet{meyer_mnras_464_2017}.

The manner our stellar evolution calculations treat the accretion of circumstellar material can 
also be improved. Although the so-called cold accretion scenario is a well-established method to include the accretion 
of disc material onto the stellar surface~\citep[see][and references therein]{hosokawa_apj_691_2009}, it is the 
lower limit in terms of for the accretion of entropy~\citep{hosokawa_apj_721_2010}. 
Despite of the fact that high-resolution observations recently demonstrated the disc-plus-jet structure surrounding 
MYSOs, the exact topology of the accretion flow within a few  tens of stellar radii from the protostellar 
surface is unknown and may differ from accretion in the midplane, for example via the formation of accretion 
columns~\citep{romanova_mnras_421_2012}. The deviations of the accretion geometry 
from the cold accretion scenario can be explored within the so-called hot accretion scenario 
or by hybridising the cold and hot accretion scenarios~\citep{vorobyov_aa_605_2017}. 
\citet{haemmerle_458_mnras_2016} showed how the stellar bloating can significantly change as a 
function of the early accretion geometry, regardless of the accretion rate. However, this concerned smaller 
accretion rates than ours and our calculations are performed with the initial convective model ("CV") 
of~\citet{haemmerle_458_mnras_2016}, with a larger initial radius than that of the radiative models 
("RD") for spherical accretion, for which the bloating of the radius is less pronounced.

Our study make use of the methods developed in~\citet{haemmerle_585_aa_2016} and~\citet{haemmerle_458_mnras_2016}. These 
works showed that (i) when a massive protostar accretes at very high constant rates ($\ge 10^{-2}\, \rm M_{\odot}\, \rm yr^{-1}$), its 
corresponding evolutionary track derives towards the red part of the Hertzsprung-Russell diagram~\citep{haemmerle_585_aa_2016}, and (ii) that variabilities in the accretion rate onto 
\hii regions around massive protostar taken from large-scale hydrodycamical simulations~\citep[see][]{peters_apj_711_2010} produce luminosity changes 
reflecting that of the fluctuating accretion rates~\citep{haemmerle_458_mnras_2016}.
Since we focus on the accretion in the vicinity of massive protostars, the accretion rates histories measured from small-scale 
hydrodynamical simulations~\citep{meyer_mnras_464_2017,meyer_mnras_473_2018} are more realistic and we investigate how massive 
protostellar structures reacts under the effect of such accretion variability. 
We show that, high episodic accretion rates ($\ge 10^{-2}$-$10^{-1}\, \rm M_{\odot}\, \rm yr^{-1}$) produce repetitive excursions in 
the Hertzsprung-Russell diagram. It was not the case in~\citet{hosokawa_apj_691_2009} and~\citet{hosokawa_apj_721_2010} 
because their accretion rates were constant and weaker than that ours (up to $10^{-3}\, \rm M_{\odot}\, \rm yr^{-1}$). 
%Because our tracks reach the Hayashi limit, hybrid/hot accretion should principally enhances the swelling of the protostellar radius.
%
Indeed, the entropy accreted is higher than for cold accretion, hence, the radius will be larger according to 
the homology relations, as confirmed for constant rates in~\citet{hosokawa_apj_691_2009,hosokawa_apj_721_2010}, 
and, since the tracks already reached the Hayashi limit, substential differences \textcolor{black}{in the maximal 
values of the prostellar radius should not be expected. 
}

\textcolor{black}{
One can nevertheless wonder whether the mean protostellar radius of the MYOS may be affected by 
hybrid/hot accretion geometries, by making the protostars slightly colder during the phases of 
quiescent accretion and consequently diminishing the amplitudes of the excursions, as shown in 
the context of low-mass star formation~\citep{hosokawa_apj_738_2011}. 
However, the incidence of massive magnetic OB stars is small~\citep{fossati_aa_582_2015,fossati_aa_592_2016} and 
pre-main-sequence massive stars do have strong surface radiation field because of their high effective 
temperatures~\citep{hosokawa_apj_691_2009,hosokawa_apj_721_2010}. Therefore, quiescent accretion processes in most MYSOs should happen via boundary layer mechanisms at the equatorial plane, as investigated by radiation-hydrodynamics simulations~\citep{kee_mnras_479_2018}. 
}
The assumption of cold accretion is therefore appropriate for this \textcolor{black}{first} study on the effects of 
\textcolor{black}{strong, episodic} accretion variability on the stellar evolutionnary tracks of pre-ZAMS massive protostars.

\begin{figure}
        \centering
        \begin{minipage}[b]{ 0.47\textwidth}  \centering
                \includegraphics[width=1.0\textwidth]{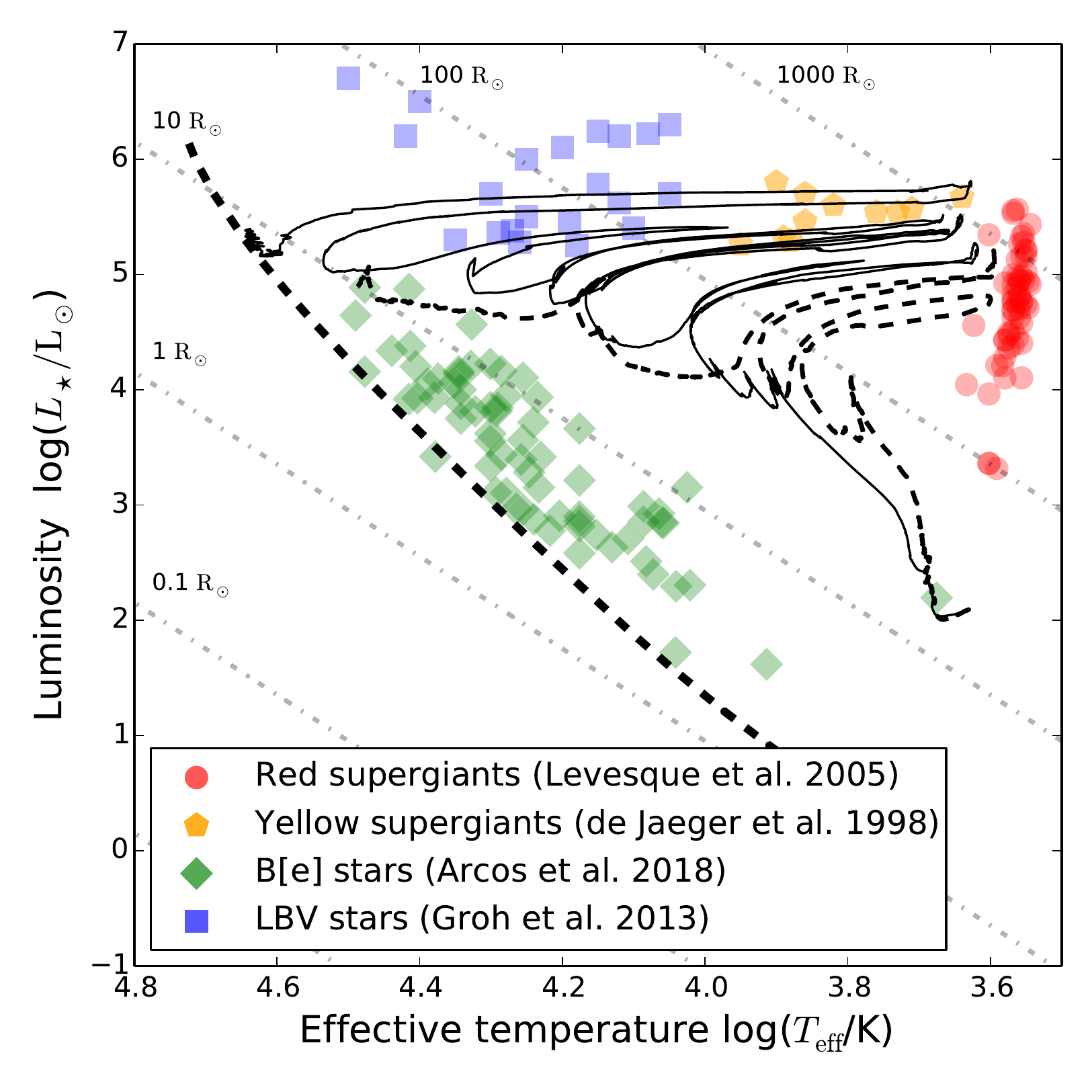}
        \end{minipage}     
        \caption{ 
        		 Comparison between our evolutionary tracks with variable accretion rates 
        		 of $M_{\rm c}=60\, \rm M_{\odot}$ (thick dashed black line) and 
        		 $M_{\rm c}=100\, \rm M_{\odot}$ (thin solid black line) collapsing 
        		 pre-stellar cores, respectively, and observational data. 
        		 Grey dotted-dashed lines are isoradius and the thick black dashed line is 
        		 the zero-age-main-sequence (ZAMS) track of~\citet{ekstroem_aa_537_2012}. 
                 }      
        \label{fig:hdr_data}  
\end{figure}

\subsection{Comparison with other works assuming constant disc accretion rates}
\label{sect:works}

Our work extends previous studies on the modifications brought by disc mass accretion 
onto the evolutionary path of young massive stars in the Hertzsprung-Russell diagram. 
We perform the most spatially-resolved numerical simulations of the inner accretion 
disc around MYSOs that have revealed self-consistent fragmentation of the irradiated 
circumstellar disc into spiral arms interspersed with dense gaseous clumps, which migration 
onto the protostar induces strong variability and bursts in the accretion rate histories. 
These variabilities account for the specifics of the inner disc physics that was informations that were not 
passed to the protostars in previous works on the evolution of massive protostars. 
Indeed, constant accretion rate is considered in early studies~\citep{palla_apj_392_1992,
palla_1993,beech_apjs_95_1994,bernasconi_aa_307_1996,bernasconi_aa_120_1996} 
and in more recent works~\citep{hosokawa_apj_691_2009,hosokawa_apj_721_2010,hosokawa_2015}. 
The latter found that rapid strong accretion produces a strong swelling of the protostars, 
which bloat so that their radii reach $\simeq 100\, \rm R_{\odot}$. This bloating results in inducing a reduction 
of the effective temperature so that the \hii region and reestablishes only when the star contracts 
again and comes back to the bluer part of the Hertzsprung-Russell diagram~\citet{haemmerle_458_mnras_2016}. 
Variabilities produced by the initial gravitational collapse of 
the host pre-stellar core in which the star grows and by the effect of outflows launched 
perpendicularly to the disc have been explored with stellar calculations based on the 2.5D 
axisymmetric gravito-radiation-hydrodynamics models, leading to a single excursion 
to the redder region of the Hertzsprung-Russell diagram~\citep{kuiper_apj_772_2013}.

\begin{figure}
        \centering
        \begin{minipage}[b]{ 0.495\textwidth}  \centering
                \includegraphics[width=1.0\textwidth]{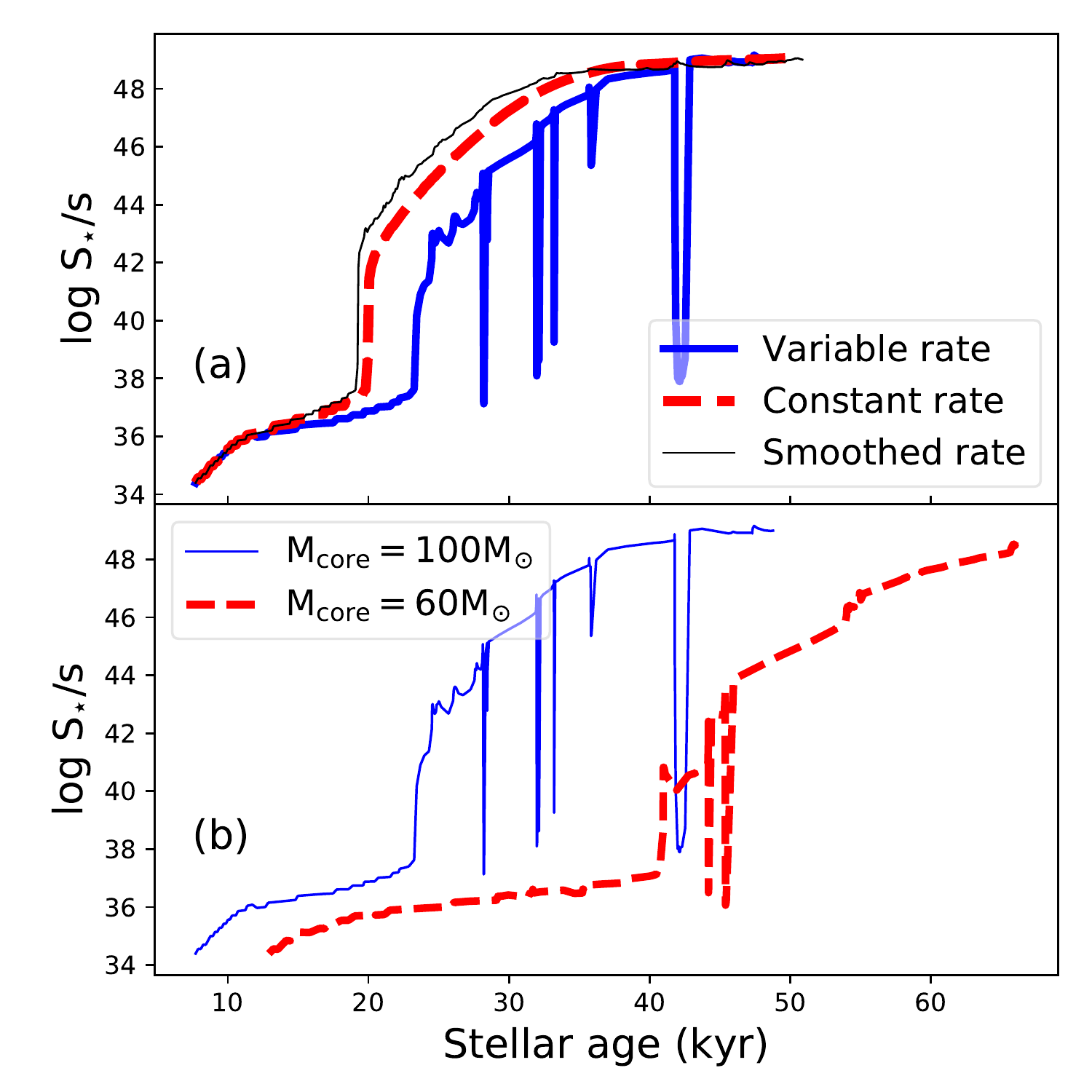}
        \end{minipage}     
        \caption{ 
        		 Evolution of the number of surface ionizing photons $S_{\star}$ per second 
        		 produced by our young massive stars as a function of the protostellar age, 
        		 for our models with an initial $100\, \rm M_{\odot}$ assuming different accretion 
        		 histories (a) and generated by different initial pre-stellar core masses but 
        		 experiencing variable accretion rate (b). 
                 }      
        \label{fig:hdr3}  
\end{figure}

Our study investigates the effects of strong accretion-driven bursts predicted to happen 
in the context of massive protostars~\citep{meyer_mnras_464_2017}. 
We show that the swelling of the MYSOs, \textcolor{black}{occasionally} going to the cold part of the 
Hertzsprung-Russell diagram~\citep{kuiper_apj_772_2013}, can be generalised in a  
wider context, as a successive series of evolutionary loops to the red region of the 
Hertzsprung-Russell diagram, before the protostars converge to the ZAMS. 
Fig.~\ref{fig:hdr2} compares our models with the tracks of~\citet{haemmerle_585_aa_2016} 
by plotting our evolutionary tracks with a color coding informing on the stellar mass 
(in $\rm M_{\odot}$).  
In general, our MYSOs follow the track of a massive prototar accreting at a constant 
rate of $10^{-3}\, \rm M_{\odot}\, \rm yr^{-1}$~\citet{haemmerle_585_aa_2016}, but, in addition, our tracks exhibit 
excursions to the red part of the Hertzsprung-Russell diagram caused by repetitive accretion bursts 
that modify their surface properties. 
Our tracks are therefore consistent in terms of final mass and location on the ZAMS~\citep{ekstroem_aa_537_2012}, 
modulo the excursions which deviate from the constant-accretion solutions. Particularly, we recover 
the non-monotoneous evolution of the radius and effective temperature of two-dimensional present-day models of disc-accreting 
MYSOs noted by~\citet{kuiper_apj_772_2013}, but reveal their episodic nature together with the intermittency of their ionized 
flux, as it is known to exist in the context of primordial supermassive protostars gaining mass from a fragmented disc~\citep{hosokawa_2015}.

\begin{figure*}
        \centering
        \begin{minipage}[b]{ 0.8\textwidth}  \centering
                \includegraphics[width=1.0\textwidth]{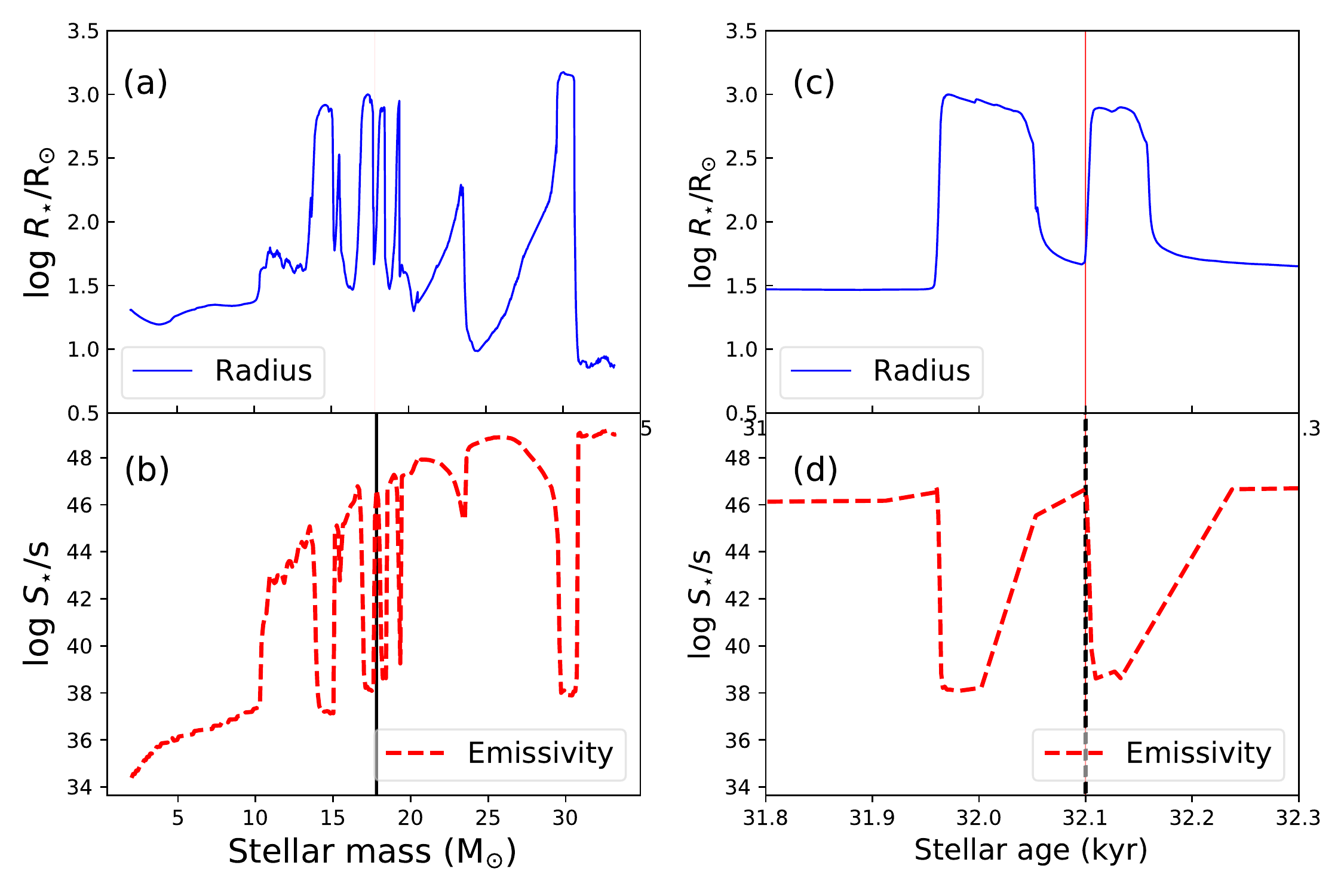}
        \end{minipage}     
        \caption{ 
                 \textcolor{black}{
        		 Correlation between the evolution of the protostellar radius and the 
        		 number of surface UV ionizing photons $S_{\star}$ in our model Run-100-hydro. 
        		 The protostellar radius and the emissivity is shown as a function of the mass 
        		 of the MYSO (a,b) and as a function of the stellar age for two of its main 
        		 bursts-producing excursions in the Herztsprung-Russel diagram (c,d). 
        		 The vertical thick dashed black line marks the beginning of a bloating episod 
        		 of the protostar. A similar figure illustrating the intermittency of the \hii 
        		 region of supermassive protostars can be found in~\citet{hosokawa_2015}.  
        		 }
                 }      
        \label{fig:yso_photons_flux}  
\end{figure*}

\subsection{Observational implications}
\label{sect:obs}

Fig.~\ref{fig:hdr_data} compares the excursions of our protostars in the 
Hertzsprung-Russell diagram with observational data of various high-mass stars. The figure 
clearly illustrates that the properties of young massive stars during the bursts may be similar 
to those of evolved, supergiant massive stars with high-luminosity, large radii and cold effective 
temperature. One can note that the model with $M_{\rm c}=60\, \rm M_{\odot}$ 
(dashed black line) has surface characteristics at the peak of the burst 
that are similar to the characteristics of a red supergiant~\citep{levesque_apj_628_2005}. 
On the other hand, the $M_{\rm c}=60\, \rm M_{\odot}$n model has surface characteristics 
similar to Be stars~\citep{arcos_mnras_474_2018} when converging to the ZAMS in the post burst phase.
At the same time, the model with a more massive pre-stellar 
core ($M_{\rm c}=100\, \rm M_{\odot}$) crosses the luminous blue variables (LBV) region~\citep{groh_aa_558_2013} 
attains the characteristics a yellow supergiant (YSG) star at the peak of the burst~\citep{jaeger_arv_8_1998}. 
To avoid confusion with evolved massive stars, one should use unique spectral signature of young massive 
protostars, such as infrared excess and line emission typically associated to accretion discs.
%
%However, one should not confound bursting protostars with episodic surface properties of 
%evolved massive stars because of their unique spectral signature including multi-wavelengths features such 
%as infrared excess and line emission typically associated to accretion disc makes our MYSOs 
%easily distinguishable from an evolved massive star when experiencing bursts. 

Most important observational implication of our results is the intermittent character 
of the \hii regions associated to massive protostars accreting from a fragmented 
circumstellar disc. 
We estimate its impact on the \textcolor{black}{ionization feedback} with the stellar properties 
obtained from our stellar evolution calculations by computing the number of UV-ionizing photons 
per unit time $S_{\star}$ as the integral of the blackbody spectrum above the ionizing 
energy threshold~\citep{haemmerle_458_mnras_2016}, i.e. 
\begin{equation}
	S_{\star} = 4 \pi R_{\star}^{2} \int_{h\nu>13.6\, \rm eV} \frac{F_{\nu}}{h\nu} d \nu,
	\label{eq:photons}
\end{equation}
with, 
\begin{equation}
	F_{\nu} = \frac{ 2 \pi (h \nu)^{3} }{ c^{2} h^{2}  } \frac{ 1   }{ e^{ \frac{ h\nu}{k_{\rm B}T_{\rm eff}} } -1  },
	\label{eq:flux}
\end{equation}
where $h$, $\nu$, $c$ and $k_{\rm B}$ are the Planck constant, photon frequency, speed of light and Boltzman 
constant, respectively. We report its evolution during the pre-main-sequence phase for our protostellar 
models in Fig.~\ref{fig:hdr3}. 
The number of photons gradually increases up to the time instance of the disc formation at 
$10$-$20\, \rm kyr$ (Fig.~\ref{fig:hdr3}a) after the beginning of the gravitational collapse. 
The accretion variability breaks the strict monotonic time-evolution of $S_{\star}$ at 
$\approx 25\, \rm kyr$ (Run-100-hydro with variable accretion, thick blue line of Fig.~\ref{fig:hdr3}a) 
and at the time of each burst (equivalently each spectroscopic excursions), $S_{\star}$ sharply decreases 
by up to $\approx 8-9$ orders of magnitude, \textcolor{black}{inducing dippering in the variability of the} \hii regions of MYSOs. Such a phenomenon is 
only a consequence of the bursts, as our models with constant and smoothed accretion histories do not 
exhibit sharp decreases in $S_{\star}$ (thin solid black and thick dotted lines of Fig.~\ref{fig:hdr3}a). A similar behaviour 
is found for our model Run-60-hydro (\ref{fig:hdr3}b). 
The \hii regions become fainter, which makes them much more difficult to detect on timescales corresponding 
to those of the bursts. 
\textcolor{black}{
Fig.~\ref{fig:yso_photons_flux} further highlights the correlation between evolution of the protostellar 
radius and number of ionizing photons released by the protostellar surface $S_{\star}$ as a function of the 
growing mass of the MYSOs. The time interval corresponding to the bloating phases is of the order of the 
collisional recombinaison timescale in the plasma ($\sim 100\, \rm yr$) derived for the intermittent \hii 
regions around supermassive stars in~\citet{hosokawa_2015}. 
}

\subsection{Implication for intermediate-mass star formation}
\label{sect:obs}

Both theoretical and observational works have recently highlighted a possible similarity  
between the star forming processes in different mass regimes. First evidence came form 
the direct observation of \hii regions piercing opaque pre-stellar clouds in high-mass 
star forming regions~\citep{fuente_aa_366_2001,testi_2003,cesaroni_aa_509_2010}. Then, 
the observation of accretion flow~\citep{keto_apj_637_2006} and Keplerian disc structure 
surrounding protostars of various mass~\citep{johnston_apj_813_2015,ilee_mnras_462_2016,forgan_mnras_463_2016,2018arXiv180410622G} 
strengthened that picture. 
The numerical proof of disc fragmentation around MYSOs and triggering of FU-Orionis-like 
events supported the scaled-up character of high-mass star formation with respect to that of low-mass 
and primordial stars~\citep{vorobyov_apj_805_2015,hosokawa_2015,meyer_mnras_464_2017}. 
Moreover, the intermittent character of \hii regions from young primordial stars has 
been demonstrated with the help of gravito-radiation-hydrodynamics models of the same 
kind as ours~\cite{2016ApJ...824..119H}. By coupling the outcomes of the hydrodynamical 
results to stellar evolution calculations, they derive the intermittent properties of the 
UV feedback that is channelled through the radiation-driven cavity perpendicular to the 
accretion disc of young supermassive stars.

Our results extends such phenomenon to young massive stars, and we speculate that similar mechanisms may 
be at work in the intermediate-mass systems as they equivalently 
generate ionizing photons~\citep{haemmerle_458_mnras_2016} and form accretion disks and jets~\citep{kessel_aa_337_1998,fuente_aa_366_2001,torrelles_mnras_442_2014,2017arXiv170604657R}. 
Young intermediate-mass stars indeed have all necessary prerequisites, i.e. circumstellar disc and \hii 
regions~\citep{lumsen_mnras_424_2012,fontani_mnras_423_2012,menu_aa_581_2015,zakhozhay_mnras_477_2018}, 
to experience disc fragmentation and accretion variability. They may episodically produce FU-Orionis-type 
bursts which will subsequently make the UV feedback that fills their radiation-driven bubbles intermittent. 
Such a prediction is supported by various observational results, including amongst others 
the variability of the eruptive intermediate-mass stellar object IRAS 18507+0121~\citep{nikoghosyan_aa_603_2017} 
and the ionised outflow of the intermediate-mass stellar object IRAS 05373+2349 VLA 2~\citep{brown_mnras_463_2016}. 
Additionally, we interpret the spectroscopic excursions of massive 
protostars in the Hertzsprung-Russell diagram triggered by FU-Orionis bursts as a direct consequence 
of the inward migration of a gaseous clump in their fragmented accretion discs onto the stellar 
surface, and, consequently, consider the intermittency of their \hii regions as a possible signature 
of disc gravitational fragmentation.

\subsection{Alternative explanations for bursting young stellar objects}
\label{sect:alternative}

\textcolor{black}{
ALMA views of bursting objects such as protostars with EXor revealed that their accretion discs seem not to be Toomre-instable, 
which suggests that outbursts should be trigered by mechanisms different than gravitational instabilities followed 
by inward migration of clumps in the disc~\citep{cieza_mnras_474_2018}. 
Although the massive accretion discs of MYSOs implies that the fragmentation scenario is likely to 
happen~\citep{kratter_mnras_373_2006,kratter_araa_54_2016}, various other mechanisms have indeed been proposed to 
explain luminosity rises from young low-mass stellar objects, particularly in the low-mass regime of star formation. 
The work of~\citet{cieza_mnras_474_2018} lists the principal alternatives for the generation of bursts without the 
classical picture of migrating disc gaseous clumps, i.e. (i) coupling between magneto-rotational and gravitational 
instabilities, (ii) thermal-viscous instability, (iii) instabilities induced by planets or companions and (iv) the 
infall clumpiness mechanism. 
}

\textcolor{black}{
Although these mechanisms are very different, they all consist in providing a manner to suddenly/episodically increase 
the accretion rate onto young stars. 
First, a cyclic magnetohydrodynamical instability in the inner $\sim 1\, \rm au$ of protoplanetary discs is proposed in~\citet{armitage_mnras_324_2001}. 
It is further demonstrated in~\citet{zhu_apj_694_2009} that the magneto-rotational instability or the gravitational instability alone can 
not be responsible for the radial mass transport over the overall disc and produce high fluctuations of the accretion rate 
on young protostars that is required for FUor bursts. However, these two instabilities can couple together in the innermost au 
of the disc and induce accretion variabilities compatible with the observed infrared spectra of FU-Orionis objects~\citep{zhu_apj_701_2009}. 
Secondly, a thermal ionisation instability at the inner region of accretion discs around low-mass protostars has been proposed 
to explain episodic increases of the disc-star mass transferts associated to luminous flashes from ionized gas. This model has 
successfully been compared to the major observational characteristcs of FU-Orionis objects~\citep{clarke_mnras_442_1990,bell_apj_427_1994,bell_apj_444_1995}. 
Additionally, the thermal viscous ionisation instability scenario has been extended to a wider mechanism, which assumes that 
it naturally develops away from the disc edge by the presence of an embedded planet~\citep{lodato_mnras_353_2004}. 
Similarly, instable mass transferts in a binary system modelled with Lagrangian methods gave accretion rate and 
luminosity rises consistent with that of FU-Orionis protostars~\citep{bonnell_apj_401_1992}. 
Last, the infall clumpiness scenario consists in assuming than the dense material which falls onto the 
young stars is directly formed by gas from converging, clumpy filamentary flows in the collapsing turbulent molecular 
clouds in which stellar clusters form~\citep{padoan_apj_797_2014}. Such gravito-turbulent, large-scale models do not 
resolve small-scale structures as accretion discs and spot the forming stars with sink particles~\citep{federrath_apj_713_2010}, 
however, it gives results consistent with the disc fragmentation scenario~\citep{vorobyov_apj_633_2005,vorobyov_apj_805_2015}. 
}

\textcolor{black}{
These alterlative mechanisms can all explain the production of flares from young stellar objects, with or without the 
presence of an accretion disc around them, and they potentially can be applied to the high-mass regime of star formation. 
The strong ionization feedback of MYSOs~\citep{vaidya_apj_742_2011}, the efficient gravitational instability in their surrounding discs~\citep{meyer_mnras_473_2018} and the 
possible therein magneto-rotational instability~\citep{kratter_apj_681_2008} make the model of~\citet{armitage_mnras_324_2001} and~\citet{zhu_apj_694_2009} 
applicable in the context of massive protostars. 
The thermal viscous ionisation instability scenario is conceivable in objects such as the high-mass proto-binary 
IRAS17216-3801~\citep{kraus_apj_835_2017} and the infall clumpiness scenario applicable to regions such as the 
massive collapsing and high-mass star-forming filament IRDC 18223~\citep{beuther_aa_584_2015}. 
}

%%%%%%%%%%%%%%%%%%%%%%%%%%%%%%%%%%%%%%%%%%%%%%%%%%%%%%%%%%%%%%%%%%%%%%%%%%%%%%%%%%%%%%%%%%%
%%%%%%%%%%%%%%%%%%%%%%%%%%%%%%%%%%%%%%%%%%%%%%%%%%%%%%%%%%%%%%%%%%%%%%%%%%%%%%%%%%%%%%%%%%%
%%%%%%%%%%%%%%%%%%%%%%%%%%%%%%%%%%%%%%%%%%%%%%%%%%%%%%%%%%%%%%%%%%%%%%%%%%%%%%%%%%%%%%%%%%%

%\section{Discussion and conclusion}
\section{Conclusion}
\label{sect:cc}

% What we have done
Our study explores for the first time the effects of a strongly variable protostellar accretion rate history, 
including including strong accretion-driven luminosity bursts~\citep{meyer_mnras_464_2017}, on the 
internal structure and evolutionary path of pre-main-sequence, massive young stellar objects (MYSOs). 
We model growing massive protostars by performing three-dimensional gravito-radiation-hydrodynamics 
simulations of the formation and evolution of their circumstellar discs, unstable to gravitational 
instability, from which the protostars gain mass~\citep{meyer_mnras_473_2018}. Direct stellar irradiation feedback and 
appropriate disc thermodynamics are taken account, in addition to a sub-au spatial resolution of the 
inner region of the self-gravitating circumstellar discs. 
Gaseous clumps produced in the fragmented discs episodically migrate towards the protostar  
and produce brief, but violent increases of the accretion rate onto the MYSOs, subsequently responsible 
for luminous outbursts via the mechanism described in~\citet{meyer_mnras_464_2017}. 
We post-process our accretion rate histories by using them as inputs when feeding 
a stellar evolutionary code including the physics of pre-main-sequence disc accretion at high rates 
($\ge 10^{-3}\, \rm M_{\odot}\, \rm yr^{-1}$) within the so-called {\it cold accretion} 
formalism~\citep{hosokawa_apj_691_2009,hosokawa_apj_721_2010,haemmerle_585_aa_2016,haemmerle_602_aap_2017}. 
The internal and surface stellar properties are self-consistently calculated, together with the evolutionary 
track of the protostars in the Hertzsprung-Russell. 
Our models differ by the initial mass of the collapsing pre-stellar cores, taken to be $60$ and 
$100\, \rm M_{\odot}$, respectively.

% Our results and opening
The protostars are initially fully convective and highly opaque, but soon develop a radiative core. 
At the outer boundary of the radiative core the internal luminosity peaks and moves outwards to the 
surface as a luminosity wave following the decrease of the internal opacity~\citep{larson_mnras_157_1972}. 
At the moment of the strong increase of the accretion rate ($\ge 10^{-2}\, \rm M_{\odot}\, \rm yr^{-1}$), 
the MYSOs bloat as a consequence of the redistribution of entropy, thus reestablishing the internal thermal equilibrium. 
The photospheric luminosity first sharply augments as a response of the increase of the accretion rate, 
before gradually decreasing up to recovering its quiescent, pre-burst value. In addition to the swelling of the stellar 
radius that accompanies the burst, the effective temperature decreases during the bursts. 
This found decrease in the effective temperature is in stark contrast to what was found for 
low-mass stars~\citep{vorobyov_aa_605_2017}. In the latter case, the effective temperature increases 
and the protostars migrate to the left portion of the Hertzsprung-Russell diagram. 
Our models recover the principal feature of the protostars accreting at rates $\ge 10^{-2}\, \rm M_{\odot}\, \rm yr^{-1}$  
extensively studied in~\citet{hosokawa_apj_721_2010} and~\citet{haemmerle_585_aa_2016}, i.e. their evolutionary track 
reach the forbidden Hayashi region. The decrease of the accretion rate which follows each accretion 
spike brings back the evolutionary track of our massive protostars to the standard ZAMS~\citep{ekstroem_aa_537_2012}, 
as previously calculated in the models of~\citet{kuiper_apj_772_2013}. 
\textcolor{black}{
Importantly, this phenomenon is \textcolor{black}{qualitatively} independent of the choice of 
the initial stellar radius considered in the stellar evolution calculations. 
}
Our simulations with accretion histories derived from three-dimensional simulations show that this 
mechanism is repetitive. Consequently, while gradually gaining mass, the protostellar radius episodically 
swells and the MYSOs experience multiple pre-main-sequence spectroscopic excursions towards the colder regions of 
the Hertzsprung-Russell diagram. Each additional evolutionary loop to the red corresponds to an ongoing 
accretion burst inducing intermittent changes in the surface entropy of the protostar.

% Last part
Our work demonstrates that the successive excursions of massive protostars in the Herztsprung-Russel are only 
possible during strong accretion bursts, and that mild disc-induced variability is insufficient to trigger them. 
Accretion bursts make MYSOs appear colder and much fainter than during their quiescent accretion phase, 
eventually crossing the luminous-blue-variable and yellow supergiant sectors of the Hertzsprung-Russell diagram 
to reach the red supergiants region, particularly if the initial the pre-stellar core mass is sufficiently large 
($\ge 100\, \rm M_{\odot}$). 
\textcolor{black}{
To each high-magnitude accretion bursts~\citep{meyer_mnras_473_2018} corresponds an excursion assuming that none 
of them produce a close/spectroscopic binary companion to the massive 
protostar~\citep{chini_424_mnras_2012,2013A&A...550A..27M}.
}
The changes in the stellar structure cause a decrease the  
effective temperature while increasing the stellar radius, which greatly affects number of ionizing UV photons 
released by the MYSOs and induces intermittencies in their surrounding \hii region, as was previously noticed 
in the context of primordial stars~\citep{2016ApJ...824..119H}. 
%
%i.e. in regions of the Hertzsprung-Russell diagram which were 
%so far not associated to such typically hot, UV-feedback producing stars. 
%
\textcolor{black}{
The present study highlights the scaled-up character of star-forming processes, as models for the 
evolution of brown dwarfs and low-mass stars also revealed excursions in the Herztsprung-Rusell as a response 
to strong variable disc accretion~\citep{baraffe_apj_702_2009,baraffe_aa_597_2017,vorobyov_aa_605_2017}. 
}
Finally, we conjecture that such mechanism should equivalently affect star formation in the 
intermediate-mass regime and constitute a typical feature of hot, UV-feedback producing young stars. 
%
%These excursions should be investigated in the context of disc-mediated burst-producing MYSOs such as the ones monitored within 
%S255IR NIRS\,3~\citep{caratti_nature_2016} and NGC6334I-MM1~\citep{2017arXiv170108637H,2018arXiv180102141H}. 
%
%Future stellar evolution calculations of variable-accreting MYSOs should also consider the intrinsic stellar 
%rotation, e.g. within the formalism developed in~\citet{haemmerle_aa_602_2017}. 

%%%%%%%%%%%%%%%%%%%%%%%%%%%%%%%%%%%%%%%%%%%%%%%%%%%%%%%%%%%%%%%%%%%%%%%%%%%%%%%%%%%%%%%%%%%
%%%%%%%%%%%%%%%%%%%%%%%%%%%%%%%%%%%%%%%%%%%%%%%%%%%%%%%%%%%%%%%%%%%%%%%%%%%%%%%%%%%%%%%%%%%
%%%%%%%%%%%%%%%%%%%%%%%%%%%%%%%%%%%%%%%%%%%%%%%%%%%%%%%%%%%%%%%%%%%%%%%%%%%%%%%%%%%%%%%%%%%

\section*{Acknowledgements}

\textcolor{black}{
The authors thank the anonymous referee for useful advices 
and suggestions which greatly improved the manuscript. 
}
D.~M.-A.~Meyer thanks N.~Castro-Ramirez for kindly sharing his knowledge on observational data  
\textcolor{black}{and B.~Stecklum for insightful remarks}. 
%
%D.~M.-A.~Meyer is funded from the European Research Council (ERC) under the European 
%Unions Horizon 2020 research and innovation programme (grant agreement No 682393). 
%
This work was sponsored by the Swiss National Science Foundation (project number 200020-172505). 
E.~I.~Vorobyov acknowledges acknowledges support form the Russian Ministry of Education 
and Science grant 3.5602.2017.
    
%%%%%%%%%%%%%%%%%%%%%%%%%%%%%%%%%%%%%%%%%%%%%%%%%%%%%%%%%%%%%%%%%%%%%%%%%%%%%%%%%%%%%%%%%%%
%%%%%%%%%%%%%%%%%%%%%%%%%%%%%%%%%%%%%%%%%%%%%%%%%%%%%%%%%%%%%%%%%%%%%%%%%%%%%%%%%%%%%%%%%%%
%%%%%%%%%%%%%%%%%%%%%%%%%%%%%%%%%%%%%%%%%%%%%%%%%%%%%%%%%%%%%%%%%%%%%%%%%%%%%%%%%%%%%%%%%%%

% Bibliography style for the bibtex
\bibliographystyle{mn2e}

\footnotesize{
% Create the reference Section using BibTeX:
\bibliography{grid}

\begin{thebibliography}{}

\bibitem[\protect\citeauthoryear{{Abdo}, {Ackermann} \& {Ajello}}{{Abdo}
  et~al.}{2010}]{abdo_apj_722_2010}
{Abdo} A.~A.,  {Ackermann} M.,    {Ajello} M.,  2010, \apj, 722, 1303

\bibitem[\protect\citeauthoryear{{Arnaud}}{{Arnaud}}{1996}]{arnaud_aspc_101_19%
96}
{Arnaud} K.~A.,  1996, in {Jacoby} G.~H.,  {Barnes} J.,  eds, Astronomical Data
  Analysis Software and Systems V Vol.~101 of Astronomical Society of the
  Pacific Conference Series, {XSPEC: The First Ten Years}.
p.~17

\bibitem[\protect\citeauthoryear{{Aschenbach} \& {Leahy}}{{Aschenbach} \&
  {Leahy}}{1999}]{aschenbach_aa_341_1999}
{Aschenbach} B.,  {Leahy} D.~A.,  1999, \aap, 341, 602

\bibitem[\protect\citeauthoryear{{Asplund}, {Grevesse}, {Sauval} \&
  {Scott}}{{Asplund} et~al.}{2009}]{asplund_araa_47_2009}
{Asplund} M.,  {Grevesse} N.,  {Sauval} A.~J.,    {Scott} P.,  2009, \araa, 47,
  481

\bibitem[\protect\citeauthoryear{{Baade}}{{Baade}}{1938}]{baade_apj_88_1938}
{Baade} W.,  1938, \apj, 88, 285

\bibitem[\protect\citeauthoryear{{Badenes}, {Maoz} \& {Draine}}{{Badenes}
  et~al.}{2010}]{badenes_mnras_407_2010}
{Badenes} C.,  {Maoz} D.,    {Draine} B.~T.,  2010, \mnras, 407, 1301

\bibitem[\protect\citeauthoryear{{Baranov}, {Krasnobaev} \&
  {Kulikovskii}}{{Baranov} et~al.}{1971}]{baranov_sphd_15_1971}
{Baranov} V.~B.,  {Krasnobaev} K.~V.,    {Kulikovskii} A.~G.,  1971, Soviet
  Physics Doklady, 15, 791

\bibitem[\protect\citeauthoryear{{Bedogni} \& {D'Ercole}}{{Bedogni} \&
  {D'Ercole}}{1988}]{bedogni_190_aa_1988}
{Bedogni} R.,  {D'Ercole} A.,  1988, \aap, 190, 320

\bibitem[\protect\citeauthoryear{{Blaauw}}{{Blaauw}}{1993}]{blau1993ASPC...35.%
.207B}
{Blaauw} A.,  1993, in {Cassinelli} J.~P.,  {Churchwell} E.~B.,  eds, Massive
  Stars: Their Lives in the Interstellar Medium Vol.~35 of Astronomical Society
  of the Pacific Conference Series, {Massive Runaway Stars}.
p.~207

\bibitem[\protect\citeauthoryear{{Blandford}, {Kennel}, {McKee} \&
  {Ostriker}}{{Blandford} et~al.}{1983}]{blandford_301_natur_1983}
{Blandford} R.~D.,  {Kennel} C.~F.,  {McKee} C.~F.,    {Ostriker} J.~P.,  1983,
  \nat, 301, 586

\bibitem[\protect\citeauthoryear{{Blondin} \& {Koerwer}}{{Blondin} \&
  {Koerwer}}{1998}]{blondin_na_57_1998}
{Blondin} J.~M.,  {Koerwer} J.~F.,  1998, \na, 3, 571

\bibitem[\protect\citeauthoryear{{Blondin}, {Lundqvist} \&
  {Chevalier}}{{Blondin} et~al.}{1996}]{blondin_apj_472_1996}
{Blondin} J.~M.,  {Lundqvist} P.,    {Chevalier} R.~A.,  1996, \apj, 472, 257

\bibitem[\protect\citeauthoryear{{Borkowski}, {Blondin} \&
  {Sarazin}}{{Borkowski} et~al.}{1992}]{borkowski_apj_400_1992}
{Borkowski} K.~J.,  {Blondin} J.~M.,    {Sarazin} C.~L.,  1992, \apj, 400, 222

\bibitem[\protect\citeauthoryear{{Brighenti} \& {D'Ercole}}{{Brighenti} \&
  {D'Ercole}}{1994}]{brighenti_mnras_270_1994}
{Brighenti} F.,  {D'Ercole} A.,  1994, \mnras, 270, 65

\bibitem[\protect\citeauthoryear{{Brighenti} \& {D'Ercole}}{{Brighenti} \&
  {D'Ercole}}{1995a}]{brighenti_mnras_277_1995}
{Brighenti} F.,  {D'Ercole} A.,  1995a, \mnras, 277, 53

\bibitem[\protect\citeauthoryear{{Brighenti} \& {D'Ercole}}{{Brighenti} \&
  {D'Ercole}}{1995b}]{brighenti_mnras_273_1995}
{Brighenti} F.,  {D'Ercole} A.,  1995b, \mnras, 273, 443

\bibitem[\protect\citeauthoryear{{Brott}, {de Mink}, {Cantiello}, {Langer}, {de
  Koter}, {Evans}, {Hunter}, {Trundle} \& {Vink}}{{Brott}
  et~al.}{2011}]{brott_aa_530_2011a}
{Brott} I.,  {de Mink} S.~E.,  {Cantiello} M.,  {Langer} N.,  {de Koter} A.,
  {Evans} C.~J.,  {Hunter} I.,  {Trundle} C.,    {Vink} J.~S.,  2011, \aap,
  530, A115

\bibitem[\protect\citeauthoryear{{Bucciantini}, {Blondin}, {Del Zanna} \&
  {Amato}}{{Bucciantini} et~al.}{2003}]{bucciantini_aa_405_2003}
{Bucciantini} N.,  {Blondin} J.~M.,  {Del Zanna} L.,    {Amato} E.,  2003,
  \aap, 405, 617

\bibitem[\protect\citeauthoryear{{Chevalier}}{{Chevalier}}{1982}]{chevalier_ap%
j_258_1982}
{Chevalier} R.~A.,  1982, \apj, 258, 790

\bibitem[\protect\citeauthoryear{{Chevalier} \& {Liang}}{{Chevalier} \&
  {Liang}}{1989}]{chevalier_apj_344_1989}
{Chevalier} R.~A.,  {Liang} E.~P.,  1989, \apj, 344, 332

\bibitem[\protect\citeauthoryear{{Chiotellis}, {Schure} \& {Vink}}{{Chiotellis}
  et~al.}{2012}]{chiotellis_aa_537_2012}
{Chiotellis} A.,  {Schure} K.~M.,    {Vink} J.,  2012, \aap, 537, A139

\bibitem[\protect\citeauthoryear{{Chita}, {Langer}, {van Marle},
  {Garc{\'{\i}}a-Segura} \& {Heger}}{{Chita} et~al.}{2008}]{chita_aa_488_2008}
{Chita} S.~M.,  {Langer} N.,  {van Marle} A.~J.,  {Garc{\'{\i}}a-Segura} G.,
  {Heger} A.,  2008, \aap, 488, L37

\bibitem[\protect\citeauthoryear{{Ciotti} \& {D'Ercole}}{{Ciotti} \&
  {D'Ercole}}{1989}]{ciotti_aa_215_1989}
{Ciotti} L.,  {D'Ercole} A.,  1989, \aap, 215, 347

\bibitem[\protect\citeauthoryear{{Comer\'{o}n} \& {Kaper}}{{Comer\'{o}n} \&
  {Kaper}}{1998}]{comeron_aa_338_1998}
{Comer\'{o}n} F.,  {Kaper} L.,  1998, \aap, 338, 273

\bibitem[\protect\citeauthoryear{{Cowie} \& {McKee}}{{Cowie} \&
  {McKee}}{1977}]{cowie_apj_211_1977}
{Cowie} L.~L.,  {McKee} C.~F.,  1977, \apj, 211, 135

\bibitem[\protect\citeauthoryear{{Cox}, {Gull} \& {Green}}{{Cox}
  et~al.}{1991}]{cox_mnras_250_1991}
{Cox} C.~I.,  {Gull} S.~F.,    {Green} D.~A.,  1991, \mnras, 250, 750

\bibitem[\protect\citeauthoryear{{Cox}, {Kerschbaum}, {van Marle}, {Decin},
  {Ladjal} \& {Mayer}}{{Cox} et~al.}{2012}]{cox_aa_537_2012}
{Cox} N.~L.~J.,  {Kerschbaum} F.,  {van Marle} A.~J.,  {Decin} L.,  {Ladjal}
  D.,    {Mayer} A.,  2012, \aap, 543, C1

\bibitem[\protect\citeauthoryear{{Decin}, { }, {Royer}, {Van Marle},
  {Vandenbussche}, {Ladjal}, {Kerschbaum}, {Ottensamer}, {Barlow}, {Blommaert},
  {Gomez}, {Groenewegen}, {Lim}, {Swinyard}, {Waelkens} \& {Tielens}}{{Decin}
  et~al.}{2012}]{decin_aa_548_2012}
{Decin} L.,  { } N.~L.~J.,  {Royer} P.,  {Van Marle} A.~J.,  {Vandenbussche}
  B.,  {Ladjal} D.,  {Kerschbaum} F.,  {Ottensamer} R.,  {Barlow} M.~J.,
  {Blommaert} J.~A.~D.~L.,  {Gomez} H.~L.,  {Groenewegen} M.~A.~T.,  {Lim} T.,
  {Swinyard} B.~M.,  {Waelkens} C.,    {Tielens} A.~G.~G.~M.,  2012, \aap, 548,
  A113

\bibitem[\protect\citeauthoryear{{Dgani}, {van Buren} \&
  {Noriega-Crespo}}{{Dgani} et~al.}{1996}]{dgani_apj_461_1996}
{Dgani} R.,  {van Buren} D.,    {Noriega-Crespo} A.,  1996, \apj, 461, 927

\bibitem[\protect\citeauthoryear{{Dopita}}{{Dopita}}{1973}]{dopita_aa_29_1973}
{Dopita} M.~A.,  1973, \aap, 29, 387

\bibitem[\protect\citeauthoryear{{Dwarkadas}}{{Dwarkadas}}{2005}]{dwarkadas_ap%
j_630_2005}
{Dwarkadas} V.~V.,  2005, \apj, 630, 892

\bibitem[\protect\citeauthoryear{{Dwarkadas}}{{Dwarkadas}}{2007}]{dwarkadas_ap%
j_667_2007}
{Dwarkadas} V.~V.,  2007, \apj, 667, 226

\bibitem[\protect\citeauthoryear{{Eldridge}, {Langer} \& {Tout}}{{Eldridge}
  et~al.}{2011}]{eldridge_mnras_414_2011}
{Eldridge} J.~J.,  {Langer} N.,    {Tout} C.~A.,  2011, \mnras, 414, 3501

\bibitem[\protect\citeauthoryear{{Ferreira} \& {de Jager}}{{Ferreira} \& {de
  Jager}}{2008}]{ferreira_478_aa_2008}
{Ferreira} S.~E.~S.,  {de Jager} O.~C.,  2008, \aap, 478, 17

\bibitem[\protect\citeauthoryear{{Filippenko}}{{Filippenko}}{1997}]{filippenko%
_araa_35_1997}
{Filippenko} A.~V.,  1997, \araa, 35, 309

\bibitem[\protect\citeauthoryear{{Frail}, {Goss}, {Reynoso}, {Giacani}, {Green}
  \& {Otrupcek}}{{Frail} et~al.}{1996}]{frail_aj_111_1996}
{Frail} D.~A.,  {Goss} W.~M.,  {Reynoso} E.~M.,  {Giacani} E.~B.,  {Green}
  A.~J.,    {Otrupcek} R.,  1996, \aj, 111, 1651

\bibitem[\protect\citeauthoryear{{Gaensler}}{{Gaensler}}{1998}]{gaensler_apj_4%
93_1998}
{Gaensler} B.~M.,  1998, \apj, 493, 781

\bibitem[\protect\citeauthoryear{{Gaensler}}{{Gaensler}}{1999}]{gaensler_phd_1%
999}
{Gaensler} B.~M.,  1999, PhD thesis, University of Sydney

\bibitem[\protect\citeauthoryear{{Garcia-Segura}, {Langer} \& {Mac
  Low}}{{Garcia-Segura} et~al.}{1996}]{garciasegura_1996_aa_316}
{Garcia-Segura} G.,  {Langer} N.,    {Mac Low} M.-M.,  1996, \aap, 316, 133

\bibitem[\protect\citeauthoryear{{Gies}}{{Gies}}{1987}]{gies_apjs_64_1987}
{Gies} D.~R.,  1987, \apjs, 64, 545

\bibitem[\protect\citeauthoryear{{Gonz{\'a}lez-Casanova}, {De Colle},
  {Ramirez-Ruiz} \& {Lopez}}{{Gonz{\'a}lez-Casanova}
  et~al.}{2014}]{gonzalezcasanova_apj_781_2014}
{Gonz{\'a}lez-Casanova} D.~F.,  {De Colle} F.,  {Ramirez-Ruiz} E.,    {Lopez}
  L.~A.,  2014, \apjl, 781, L26

\bibitem[\protect\citeauthoryear{{Green}}{{Green}}{2009}]{green_cat_2009}
{Green} D.~A.,  2009, VizieR Online Data Catalog, 7253, 0

\bibitem[\protect\citeauthoryear{{Green} \& {Stephenson}}{{Green} \&
  {Stephenson}}{2003}]{green_lnp_598_2003}
{Green} D.~A.,  {Stephenson} F.~R.,  2003, in {Weiler} K.,  ed., Supernovae and
  Gamma-Ray Bursters Vol.~598 of Lecture Notes in Physics, Berlin Springer
  Verlag, {Historical Supernovae}.
pp 7--19

\bibitem[\protect\citeauthoryear{{Gvaramadze}, {Menten}, {Kniazev}, {Langer},
  {Mackey}, {Kraus}, {Meyer} \& {Kami{\'n}ski}}{{Gvaramadze}
  et~al.}{2014}]{Gvaramadze_2013}
{Gvaramadze} V.~V.,  {Menten} K.~M.,  {Kniazev} A.~Y.,  {Langer} N.,  {Mackey}
  J.,  {Kraus} A.,  {Meyer} D.~M.-A.,    {Kami{\'n}ski} T.,  2014, \mnras, 437,
  843

\bibitem[\protect\citeauthoryear{{Hobbs}, {Lorimer}, {Lyne} \&
  {Kramer}}{{Hobbs} et~al.}{2005}]{hibbs_360_mnras_2005}
{Hobbs} G.,  {Lorimer} D.~R.,  {Lyne} A.~G.,    {Kramer} M.,  2005, \mnras,
  360, 974

\bibitem[\protect\citeauthoryear{{Huthoff} \& {Kaper}}{{Huthoff} \&
  {Kaper}}{2002}]{huthoff_aa_383_2002}
{Huthoff} F.,  {Kaper} L.,  2002, \aap, 383, 999

\bibitem[\protect\citeauthoryear{{Kane}, {Drake} \& {Remington}}{{Kane}
  et~al.}{1999}]{kane_apj_511_1999}
{Kane} J.,  {Drake} R.~P.,    {Remington} B.~A.,  1999, \apj, 511, 335

\bibitem[\protect\citeauthoryear{{Katsuda}, {Tsunemi}, {Mori}, {Uchida},
  {Petre}, {Yamada} \& {Tamagawa}}{{Katsuda}
  et~al.}{2012}]{katsuda_apj_754_2012}
{Katsuda} S.,  {Tsunemi} H.,  {Mori} K.,  {Uchida} H.,  {Petre} R.,  {Yamada}
  S.,    {Tamagawa} T.,  2012, \apjl, 754, L7

\bibitem[\protect\citeauthoryear{Kaufman \& Kaufman}{Kaufman \&
  Kaufman}{2009}]{kaufman_EAS2009}
Kaufman M.,  Kaufman M.,  2009, EAS Publications Series, 34, 151

\bibitem[\protect\citeauthoryear{{Kothes}, {Fedotov}, {Foster} \&
  {Uyan{\i}ker}}{{Kothes} et~al.}{2006}]{kothes_aa_457_2006}
{Kothes} R.,  {Fedotov} K.,  {Foster} T.~J.,    {Uyan{\i}ker} B.,  2006, \aap,
  457, 1081

\bibitem[\protect\citeauthoryear{{Langer}}{{Langer}}{2012}]{langer_araa_50_201%
2}
{Langer} N.,  2012, \araa, 50, 107

\bibitem[\protect\citeauthoryear{{Langer}, {Garc{\'{\i}}a-Segura} \& {Mac
  Low}}{{Langer} et~al.}{1999}]{langer_ApJ_520_1999}
{Langer} N.,  {Garc{\'{\i}}a-Segura} G.,    {Mac Low} M.-M.,  1999, \apjl, 520,
  L49

\bibitem[\protect\citeauthoryear{{Lockett}, {Gauthier} \& {Elitzur}}{{Lockett}
  et~al.}{1999}]{lockett_apj_511_1999}
{Lockett} P.,  {Gauthier} E.,    {Elitzur} M.,  1999, \apj, 511, 235

\bibitem[\protect\citeauthoryear{{Lodders}}{{Lodders}}{2003}]{lodders_apj_591_%
2003}
{Lodders} K.,  2003, \apj, 591, 1220

\bibitem[\protect\citeauthoryear{{Lyne} \& {Lorimer}}{{Lyne} \&
  {Lorimer}}{1994}]{lyne_natur_369_1994}
{Lyne} A.~G.,  {Lorimer} D.~R.,  1994, \nat, 369, 127

\bibitem[\protect\citeauthoryear{{MacDonald} \& {Bailey}}{{MacDonald} \&
  {Bailey}}{1981}]{macdomald_mnras_197_1981}
{MacDonald} J.,  {Bailey} M.~E.,  1981, \mnras, 197, 995

\bibitem[\protect\citeauthoryear{{Mackey}, {Gvaramadze}, {Mohamed} \&
  {Langer}}{{Mackey} et~al.}{2014}]{mackey_sept_2014}
{Mackey} J.,  {Gvaramadze} V.~V.,  {Mohamed} S.,    {Langer} N.,  2014, ArXiv
  e-prints

\bibitem[\protect\citeauthoryear{{Mackey}, {Mohamed}, {Neilson}, {Langer} \&
  {Meyer}}{{Mackey} et~al.}{2012}]{mackey_apjlett_751_2012}
{Mackey} J.,  {Mohamed} S.,  {Neilson} H.~R.,  {Langer} N.,    {Meyer}
  D.~M.-A.,  2012, \apjl, 751, L10

\bibitem[\protect\citeauthoryear{{Manchester}}{{Manchester}}{1987}]{manchester%
_aa_171_1987}
{Manchester} R.~N.,  1987, \aap, 171, 205

\bibitem[\protect\citeauthoryear{{Medina}, {Raymond}, {Edgar}, {Caldwell},
  {Fesen} \& {Milisavljevic}}{{Medina} et~al.}{2014}]{medina_791_apj_2014}
{Medina} A.~A.,  {Raymond} J.~C.,  {Edgar} R.~J.,  {Caldwell} N.,  {Fesen}
  R.~A.,    {Milisavljevic} D.,  2014, \apj, 791, 30

\bibitem[\protect\citeauthoryear{{Meyer}, {Gvaramadze}, {Langer}, {Mackey},
  {Boumis} \& {Mohamed}}{{Meyer} et~al.}{2014}]{meyer_mnras_2013}
{Meyer} D.~M.-A.,  {Gvaramadze} V.~V.,  {Langer} N.,  {Mackey} J.,  {Boumis}
  P.,    {Mohamed} S.,  2014, \mnras, 439, L41

\bibitem[\protect\citeauthoryear{{Meyer}, {Mackey}, {Langer}, {Gvaramadze},
  {Mignone}, {Izzard} \& {Kaper}}{{Meyer} et~al.}{2014}]{meyer}
{Meyer} D.~M.-A.,  {Mackey} J.,  {Langer} N.,  {Gvaramadze} V.~V.,  {Mignone}
  A.,  {Izzard} R.~G.,    {Kaper} L.,  2014, \mnras, 444, 2754

\bibitem[\protect\citeauthoryear{{Mignone}, {Bodo}, {Massaglia}, {Matsakos},
  {Tesileanu}, {Zanni} \& {Ferrari}}{{Mignone}
  et~al.}{2007}]{mignone_apj_170_2007}
{Mignone} A.,  {Bodo} G.,  {Massaglia} S.,  {Matsakos} T.,  {Tesileanu} O.,
  {Zanni} C.,    {Ferrari} A.,  2007, \apjs, 170, 228

\bibitem[\protect\citeauthoryear{{Mignone}, {Zanni}, {Tzeferacos}, {van
  Straalen}, {Colella} \& {Bodo}}{{Mignone}
  et~al.}{2012}]{migmone_apjs_198_2012}
{Mignone} A.,  {Zanni} C.,  {Tzeferacos} P.,  {van Straalen} B.,  {Colella} P.,
     {Bodo} G.,  2012, \apjs, 198, 7

\bibitem[\protect\citeauthoryear{{Mohamed}, {Mackey} \& {Langer}}{{Mohamed}
  et~al.}{2012}]{mohamed_aa_541_2012}
{Mohamed} S.,  {Mackey} J.,    {Langer} N.,  2012, \aap, 541, A1

\bibitem[\protect\citeauthoryear{{Neufeld}, {Gusdorf}, {G{\"u}sten}, {Herczeg},
  {Kristensen}, {Melnick}, {Nisini}, {Ossenkopf}, {Tafalla} \& {van
  Dishoeck}}{{Neufeld} et~al.}{2014}]{neufeld_apj_781_2014}
{Neufeld} D.~A.,  {Gusdorf} A.,  {G{\"u}sten} R.,  {Herczeg} G.~J.,
  {Kristensen} L.,  {Melnick} G.~J.,  {Nisini} B.,  {Ossenkopf} V.,  {Tafalla}
  M.,    {van Dishoeck} E.~F.,  2014, \apj, 781, 102

\bibitem[\protect\citeauthoryear{{Noriega-Crespo}, {van Buren}, {Cao} \&
  {Dgani}}{{Noriega-Crespo} et~al.}{1997}]{noriegacrespo_aj_114_1997}
{Noriega-Crespo} A.,  {van Buren} D.,  {Cao} Y.,    {Dgani} R.,  1997, \aj,
  114, 837

\bibitem[\protect\citeauthoryear{{Orlando}, {Bocchino}, {Miceli}, {Petruk} \&
  {Pumo}}{{Orlando} et~al.}{2012}]{orlando_apj_749_2012}
{Orlando} S.,  {Bocchino} F.,  {Miceli} M.,  {Petruk} O.,    {Pumo} M.~L.,
  2012, \apj, 749, 156

\bibitem[\protect\citeauthoryear{{Orlando}, {Bocchino}, {Reale}, {Peres} \&
  {Pagano}}{{Orlando} et~al.}{2008}]{orlando_apj_678_2008}
{Orlando} S.,  {Bocchino} F.,  {Reale} F.,  {Peres} G.,    {Pagano} P.,  2008,
  \apj, 678, 274

\bibitem[\protect\citeauthoryear{{Orlando}, {Bocchino}, {Reale}, {Peres} \&
  {Petruk}}{{Orlando} et~al.}{2007}]{orlando_aa_470_2007}
{Orlando} S.,  {Bocchino} F.,  {Reale} F.,  {Peres} G.,    {Petruk} O.,  2007,
  \aap, 470, 927

\bibitem[\protect\citeauthoryear{{Osterbrock} \& {Bochkarev}}{{Osterbrock} \&
  {Bochkarev}}{1989}]{osterbrock_1989}
{Osterbrock} D.~E.,  {Bochkarev} N.~G.,  1989, \sovast, 33, 694

\bibitem[\protect\citeauthoryear{{Pannuti}, {Rho}, {Heinke} \&
  {Moffitt}}{{Pannuti} et~al.}{2014}]{pannuti_147_aj_2014}
{Pannuti} T.~G.,  {Rho} J.,  {Heinke} C.~O.,    {Moffitt} W.~P.,  2014, \aj,
  147, 55

\bibitem[\protect\citeauthoryear{{Parker}, {Phillipps}, {Pierce}, {Hartley},
  {Hambly}, {Read} \& {MacGillivray}}{{Parker}
  et~al.}{2005}]{parker_mnras_362_2005}
{Parker} Q.~A.,  {Phillipps} S.,  {Pierce} M.~J.,  {Hartley} M.,  {Hambly}
  N.~C.,  {Read} M.~A.,    {MacGillivray} 2005, \mnras, 362, 689

\bibitem[\protect\citeauthoryear{{P{\'e}rez-Rend{\'o}n}, {Garc{\'{\i}}a-Segura}
  \& {Langer}}{{P{\'e}rez-Rend{\'o}n} et~al.}{2009}]{peresrendon_aa_506_2009}
{P{\'e}rez-Rend{\'o}n} B.,  {Garc{\'{\i}}a-Segura} G.,    {Langer} N.,  2009,
  \aap, 506, 1249

\bibitem[\protect\citeauthoryear{{Petruk}, {Dubner}, {Castelletti}, {Bocchino},
  {Iakubovskyi}, {Kirsch}, {Miceli}, {Orlando} \& {Telezhinsky}}{{Petruk}
  et~al.}{2009}]{petruk_393_mnras_2009}
{Petruk} O.,  {Dubner} G.,  {Castelletti} G.,  {Bocchino} F.,  {Iakubovskyi}
  D.,  {Kirsch} M.~G.~F.,  {Miceli} M.,  {Orlando} S.,    {Telezhinsky} I.,
  2009, \mnras, 393, 1034

\bibitem[\protect\citeauthoryear{{Reach} \& {Rho}}{{Reach} \&
  {Rho}}{1999}]{reach_apj_511_1999}
{Reach} W.~T.,  {Rho} J.,  1999, \apj, 511, 836

\bibitem[\protect\citeauthoryear{{Reach}, {Rho}, {Tappe}, {Pannuti}, {Brogan},
  {Churchwell}, {Meade}, {Babler}, {Indebetouw} \& {Whitney}}{{Reach}
  et~al.}{2006}]{reach_aj_131_2006}
{Reach} W.~T.,  {Rho} J.,  {Tappe} A.,  {Pannuti} T.~G.,  {Brogan} C.~L.,
  {Churchwell} E.~B.,  {Meade} M.~R.,  {Babler} B.,  {Indebetouw} R.,
  {Whitney} B.~A.,  2006, \aj, 131, 1479

\bibitem[\protect\citeauthoryear{{Rozyczka} \& {Tenorio-Tagle}}{{Rozyczka} \&
  {Tenorio-Tagle}}{1995}]{rozyczka_274_MNRAS_1995}
{Rozyczka} M.,  {Tenorio-Tagle} G.,  1995, \mnras, 274, 1157

\bibitem[\protect\citeauthoryear{{Rozyczka}, {Tenorio-Tagle}, {Franco} \&
  {Bodenheimer}}{{Rozyczka} et~al.}{1993}]{rozyczka_mnras_261_1993}
{Rozyczka} M.,  {Tenorio-Tagle} G.,  {Franco} J.,    {Bodenheimer} P.,  1993,
  \mnras, 261, 674

\bibitem[\protect\citeauthoryear{{Schlegel}}{{Schlegel}}{1990}]{schlegel_mnras%
_244_1990}
{Schlegel} E.~M.,  1990, \mnras, 244, 269

\bibitem[\protect\citeauthoryear{{Schneiter}, {Vel{\'a}zquez}, {Reynoso} \& {de
  Colle}}{{Schneiter} et~al.}{2010}]{schneiter_mnras_408_2010}
{Schneiter} E.~M.,  {Vel{\'a}zquez} P.~F.,  {Reynoso} E.~M.,    {de Colle} F.,
  2010, \mnras, 408, 430

\bibitem[\protect\citeauthoryear{{Schure} \& {Bell}}{{Schure} \&
  {Bell}}{2013}]{schure_mnras_435_2013}
{Schure} K.~M.,  {Bell} A.~R.,  2013, \mnras, 435, 1174

\bibitem[\protect\citeauthoryear{{Spitzer}}{{Spitzer}}{1962}]{spitzer_1962}
{Spitzer} L.,  1962, {Physics of Fully Ionized Gases}

\bibitem[\protect\citeauthoryear{{Stevens}, {Blondin} \& {Pollock}}{{Stevens}
  et~al.}{1992}]{stevens_apj_386_1992}
{Stevens} I.~R.,  {Blondin} J.~M.,    {Pollock} A.~M.~T.,  1992, \apj, 386, 265

\bibitem[\protect\citeauthoryear{{Stone} \& {Norman}}{{Stone} \&
  {Norman}}{1992}]{stone_apjs_80_1992}
{Stone} J.~M.,  {Norman} M.~L.,  1992, \apjs, 80, 753

\bibitem[\protect\citeauthoryear{{Sutherland} \& {Dopita}}{{Sutherland} \&
  {Dopita}}{1993}]{sutherland_apjs_88_1993}
{Sutherland} R.~S.,  {Dopita} M.~A.,  1993, \apjs, 88, 253

\bibitem[\protect\citeauthoryear{{Tenorio-Tagle}, {Bodenheimer}, {Franco} \&
  {Rozyczka}}{{Tenorio-Tagle} et~al.}{1990}]{tenoriotagle_mnras_244_1990}
{Tenorio-Tagle} G.,  {Bodenheimer} P.,  {Franco} J.,    {Rozyczka} M.,  1990,
  \mnras, 244, 563

\bibitem[\protect\citeauthoryear{{Tenorio-Tagle}, {Rozyczka}, {Franco} \&
  {Bodenheimer}}{{Tenorio-Tagle} et~al.}{1991}]{tenoriotagle_mnras_251_1991}
{Tenorio-Tagle} G.,  {Rozyczka} M.,  {Franco} J.,    {Bodenheimer} P.,  1991,
  \mnras, 251, 318

\bibitem[\protect\citeauthoryear{{Tenorio-Tagle}, {Rozyczka} \&
  {Yorke}}{{Tenorio-Tagle} et~al.}{1985}]{tenoriotagle_aa_148_1985}
{Tenorio-Tagle} G.,  {Rozyczka} M.,    {Yorke} H.~W.,  1985, \aap, 148, 52

\bibitem[\protect\citeauthoryear{{Toledo-Roy}, {Esquivel}, {Vel{\'a}zquez} \&
  {Reynoso}}{{Toledo-Roy} et~al.}{2014}]{toledo_mnras_442_2014}
{Toledo-Roy} J.~C.,  {Esquivel} A.,  {Vel{\'a}zquez} P.~F.,    {Reynoso} E.~M.,
   2014, \mnras, 442, 229

\bibitem[\protect\citeauthoryear{{Truelove} \& {McKee}}{{Truelove} \&
  {McKee}}{1999}]{truelove_apjs_120_1999}
{Truelove} J.~K.,  {McKee} C.~F.,  1999, \apjs, 120, 299

\bibitem[\protect\citeauthoryear{{Uchida}, {Tsunemi}, {Katsuda}, {Kimura} \&
  {Kosugi}}{{Uchida} et~al.}{2009}]{uchida_pasj_61_2009}
{Uchida} H.,  {Tsunemi} H.,  {Katsuda} S.,  {Kimura} M.,    {Kosugi} H.,  2009,
  \pasj, 61, 301

\bibitem[\protect\citeauthoryear{{van Dishoeck}, {Jansen} \& {Phillips}}{{van
  Dishoeck} et~al.}{1993}]{vandishoeck_aa_279_1993}
{van Dishoeck} E.~F.,  {Jansen} D.~J.,    {Phillips} T.~G.,  1993, \aap, 279,
  541

\bibitem[\protect\citeauthoryear{{van Marle}, {Decin} \& {Meliani}}{{van Marle}
  et~al.}{2014}]{vanmarle_aa_561_2014}
{van Marle} A.~J.,  {Decin} L.,    {Meliani} Z.,  2014, \aap, 561, A152

\bibitem[\protect\citeauthoryear{{van Marle}, {Langer}, {Yoon} \&
  {Garc{\'{\i}}a-Segura}}{{van Marle} et~al.}{2008}]{vanmarle_aa_478_2008}
{van Marle} A.~J.,  {Langer} N.,  {Yoon} S.-C.,    {Garc{\'{\i}}a-Segura} G.,
  2008, \aap, 478, 769

\bibitem[\protect\citeauthoryear{{van Marle}, {Meliani}, {Keppens} \&
  {Decin}}{{van Marle} et~al.}{2011}]{vanmarle_apj_734_2011}
{van Marle} A.~J.,  {Meliani} Z.,  {Keppens} R.,    {Decin} L.,  2011, \apjl,
  734, L26

\bibitem[\protect\citeauthoryear{{van Marle}, {Smith}, {Owocki} \& {van
  Veelen}}{{van Marle} et~al.}{2010}]{vanmarle_mnras_407_2010}
{van Marle} A.~J.,  {Smith} N.,  {Owocki} S.~P.,    {van Veelen} B.,  2010,
  \mnras, 407, 2305

\bibitem[\protect\citeauthoryear{{van Veelen}, {Langer}, {Vink},
  {Garc{\'{\i}}a-Segura} \& {van Marle}}{{van Veelen}
  et~al.}{2009}]{vanveelen_aa_50._2009}
{van Veelen} B.,  {Langer} N.,  {Vink} J.,  {Garc{\'{\i}}a-Segura} G.,    {van
  Marle} A.~J.,  2009, \aap, 503, 495

\bibitem[\protect\citeauthoryear{{Vel{\'a}zquez}, {Martinell}, {Raga} \&
  {Giacani}}{{Vel{\'a}zquez} et~al.}{2004}]{velazquez_apj_601_2004}
{Vel{\'a}zquez} P.~F.,  {Martinell} J.~J.,  {Raga} A.~C.,    {Giacani} E.~B.,
  2004, \apj, 601, 885

\bibitem[\protect\citeauthoryear{{Vel{\'a}zquez}, {Vigh}, {Reynoso},
  {G{\'o}mez} \& {Schneiter}}{{Vel{\'a}zquez}
  et~al.}{2006}]{velazquez_apj_649_2006}
{Vel{\'a}zquez} P.~F.,  {Vigh} C.~D.,  {Reynoso} E.~M.,  {G{\'o}mez} D.~O.,
  {Schneiter} E.~M.,  2006, \apj, 649, 779

\bibitem[\protect\citeauthoryear{{Vigh}, {Vel{\'a}zquez}, {G{\'o}mez},
  {Reynoso}, {Esquivel} \& {Matias Schneiter}}{{Vigh}
  et~al.}{2011}]{vigh_apj_727_2011}
{Vigh} C.~D.,  {Vel{\'a}zquez} P.~F.,  {G{\'o}mez} D.~O.,  {Reynoso} E.~M.,
  {Esquivel} A.,    {Matias Schneiter} E.,  2011, \apj, 727, 32

\bibitem[\protect\citeauthoryear{{Vink}}{{Vink}}{2012}]{vink_aarv_20_2012}
{Vink} J.,  2012, \aapr, 20, 49

\bibitem[\protect\citeauthoryear{{Vink}, {Kaastra} \& {Bleeker}}{{Vink}
  et~al.}{1996}]{vink_aa_307_1996}
{Vink} J.,  {Kaastra} J.~S.,    {Bleeker} J.~A.~M.,  1996, \aap, 307, L41

\bibitem[\protect\citeauthoryear{{Vink}, {Kaastra} \& {Bleeker}}{{Vink}
  et~al.}{1997}]{vink_aa_328_1997}
{Vink} J.,  {Kaastra} J.~S.,    {Bleeker} J.~A.~M.,  1997, \aap, 328, 628

\bibitem[\protect\citeauthoryear{{Vink}}{{Vink}}{2006}]{vink_asp_353_2006}
{Vink} J.~S.,  2006, in {Lamers} H.~J.~G.~L.~M.,  {Langer} N.,  {Nugis} T.,
  {Annuk} K.,  eds, Stellar Evolution at Low Metallicity: Mass Loss,
  Explosions, Cosmology Vol.~353 of Astronomical Society of the Pacific
  Conference Series, {Massive star feedback -- from the first stars to the
  present}.
p.~113

\bibitem[\protect\citeauthoryear{{Vishniac}}{{Vishniac}}{1994}]{vishniac_apj_4%
28_1994}
{Vishniac} E.~T.,  1994, \apj, 428, 186

\bibitem[\protect\citeauthoryear{{Wang}, {Dyson} \& {Kahn}}{{Wang}
  et~al.}{1993}]{wang_MNRAS_261_1993}
{Wang} L.,  {Dyson} J.~E.,    {Kahn} F.~D.,  1993, \mnras, 261, 391

\bibitem[\protect\citeauthoryear{{Weaver}, {McCray}, {Castor}, {Shapiro} \&
  {Moore}}{{Weaver} et~al.}{1977}]{weaver_apj_218_1977}
{Weaver} R.,  {McCray} R.,  {Castor} J.,  {Shapiro} P.,    {Moore} R.,  1977,
  \apj, 218, 377

\bibitem[\protect\citeauthoryear{{Whalen}, {van Veelen}, {O'Shea} \&
  {Norman}}{{Whalen} et~al.}{2008}]{whalen_apj_682_2008}
{Whalen} D.,  {van Veelen} B.,  {O'Shea} B.~W.,    {Norman} M.~L.,  2008, \apj,
  682, 49

\bibitem[\protect\citeauthoryear{{Whiteoak} \& {Green}}{{Whiteoak} \&
  {Green}}{1996}]{whiteoak_aas_118_1996}
{Whiteoak} J.~B.~Z.,  {Green} A.~J.,  1996, \aaps, 118, 329

\bibitem[\protect\citeauthoryear{{Wolfire}, {McKee}, {Hollenbach} \&
  {Tielens}}{{Wolfire} et~al.}{2003}]{wolfire_apj_587_2003}
{Wolfire} M.~G.,  {McKee} C.~F.,  {Hollenbach} D.,    {Tielens} A.~G.~G.~M.,
  2003, \apj, 587, 278

\bibitem[\protect\citeauthoryear{{Woosley}, {Heger} \& {Weaver}}{{Woosley}
  et~al.}{2002}]{woosley_rvmp_74_2002}
{Woosley} S.~E.,  {Heger} A.,    {Weaver} T.~A.,  2002, Reviews of Modern
  Physics, 74, 1015

\bibitem[\protect\citeauthoryear{{Yusef-Zadeh}, {Wardle}, {Rho} \&
  {Sakano}}{{Yusef-Zadeh} et~al.}{2003}]{yusefzadeh_apj_585_2003}
{Yusef-Zadeh} F.,  {Wardle} M.,  {Rho} J.,    {Sakano} M.,  2003, \apj, 585,
  319

\bibitem[\protect\citeauthoryear{{Zavlin}, {Pavlov} \& {Trumper}}{{Zavlin}
  et~al.}{1998}]{zavlin_aa_331_1998}
{Zavlin} V.~E.,  {Pavlov} G.~G.,    {Trumper} J.,  1998, \aap, 331, 821

\end{thebibliography}


\begin{thebibliography}{}

\bibitem[\protect\citeauthoryear{{Adams}, {Ruden} \& {Shu}}{{Adams}
  et~al.}{1989}]{adam_apj_347_1989}
{Adams} F.~C.,  {Ruden} S.~P.,    {Shu} F.~H.,  1989, \apj, 347, 959

\bibitem[\protect\citeauthoryear{{Arcos}, {Kanaan}, {Ch{\'a}vez}, {Vanzi},
  {Araya} \& {Cur{\'e}}}{{Arcos} et~al.}{2018}]{arcos_mnras_474_2018}
{Arcos} C.,  {Kanaan} S.,  {Ch{\'a}vez} J.,  {Vanzi} L.,  {Araya} I.,
  {Cur{\'e}} M.,  2018, \mnras, 474, 5287

\bibitem[\protect\citeauthoryear{{Armitage}, {Livio} \& {Pringle}}{{Armitage}
  et~al.}{2001}]{armitage_mnras_324_2001}
{Armitage} P.~J.,  {Livio} M.,    {Pringle} J.~E.,  2001, \mnras, 324, 705

\bibitem[\protect\citeauthoryear{{Asplund}, {Grevesse} \& {Sauval}}{{Asplund}
  et~al.}{2005}]{asplund_ASPC_2005}
{Asplund} M.,  {Grevesse} N.,    {Sauval} A.~J.,  2005, in {Barnes} III T.~G.,
  {Bash} F.~N.,  eds, Cosmic Abundances as Records of Stellar Evolution and
  Nucleosynthesis Vol.~336 of Astronomical Society of the Pacific Conference
  Series, {The Solar Chemical Composition}.
p.~25

\bibitem[\protect\citeauthoryear{{Banerjee}, {Pudritz} \&
  {Anderson}}{{Banerjee} et~al.}{2006}]{banerjee_mnras_373_2006}
{Banerjee} R.,  {Pudritz} R.~E.,    {Anderson} D.~W.,  2006, \mnras, 373, 1091

\bibitem[\protect\citeauthoryear{{Baraffe}, {Chabrier} \& {Gallardo}}{{Baraffe}
  et~al.}{2009}]{baraffe_apj_702_2009}
{Baraffe} I.,  {Chabrier} G.,    {Gallardo} J.,  2009, \apjl, 702, L27

\bibitem[\protect\citeauthoryear{{Baraffe}, {Elbakyan}, {Vorobyov} \&
  {Chabrier}}{{Baraffe} et~al.}{2017}]{baraffe_aa_597_2017}
{Baraffe} I.,  {Elbakyan} V.~G.,  {Vorobyov} E.~I.,    {Chabrier} G.,  2017,
  \aap, 597, A19

\bibitem[\protect\citeauthoryear{{Beech} \& {Mitalas}}{{Beech} \&
  {Mitalas}}{1994}]{beech_apjs_95_1994}
{Beech} M.,  {Mitalas} R.,  1994, \apjs, 95, 517

\bibitem[\protect\citeauthoryear{{Behrend} \& {Maeder}}{{Behrend} \&
  {Maeder}}{2001}]{behrend_aa_373_2001}
{Behrend} R.,  {Maeder} A.,  2001, \aap, 373, 190

\bibitem[\protect\citeauthoryear{{Bell} \& {Lin}}{{Bell} \&
  {Lin}}{1994}]{bell_apj_427_1994}
{Bell} K.~R.,  {Lin} D.~N.~C.,  1994, \apj, 427, 987

\bibitem[\protect\citeauthoryear{{Bell}, {Lin}, {Hartmann} \& {Kenyon}}{{Bell}
  et~al.}{1995}]{bell_apj_444_1995}
{Bell} K.~R.,  {Lin} D.~N.~C.,  {Hartmann} L.~W.,    {Kenyon} S.~J.,  1995,
  \apj, 444, 376

\bibitem[\protect\citeauthoryear{{Bernasconi}}{{Bernasconi}}{1996}]{bernasconi_aa_120_1996}
{Bernasconi} P.~A.,  1996, \aaps, 120, 57

\bibitem[\protect\citeauthoryear{{Bernasconi} \& {Maeder}}{{Bernasconi} \&
  {Maeder}}{1996}]{bernasconi_aa_307_1996}
{Bernasconi} P.~A.,  {Maeder} A.,  1996, \aap, 307, 829

\bibitem[\protect\citeauthoryear{{Beuther}, {Ragan}, {Johnston}, {Henning},
  {Hacar} \& {Kainulainen}}{{Beuther} et~al.}{2015}]{beuther_aa_584_2015}
{Beuther} H.,  {Ragan} S.~E.,  {Johnston} K.,  {Henning} T.,  {Hacar} A.,
  {Kainulainen} J.~T.,  2015, \aap, 584, A67

\bibitem[\protect\citeauthoryear{{Bhandare}, {Kuiper}, {Henning}, {Fendt},
  {Marleau} \& {K{\"o}lligan}}{{Bhandare} et~al.}{2018}]{bhandare_aa_618_2018}
{Bhandare} A.,  {Kuiper} R.,  {Henning} T.,  {Fendt} C.,  {Marleau} G.-D.,
  {K{\"o}lligan} A.,  2018, \aap, 618, A95

\bibitem[\protect\citeauthoryear{{Bitsch}, {Morbidelli}, {Lega} \&
  {Crida}}{{Bitsch} et~al.}{2014}]{bitsch_aa_564_2014}
{Bitsch} B.,  {Morbidelli} A.,  {Lega} E.,    {Crida} A.,  2014, \aap, 564,
  A135

\bibitem[\protect\citeauthoryear{{Black} \& {Bodenheimer}}{{Black} \&
  {Bodenheimer}}{1975}]{black_apj_199_1975}
{Black} D.~C.,  {Bodenheimer} P.,  1975, \apj, 199, 619

\bibitem[\protect\citeauthoryear{{Bonnell} \& {Bastien}}{{Bonnell} \&
  {Bastien}}{1992}]{bonnell_apj_401_1992}
{Bonnell} I.,  {Bastien} P.,  1992, \apjl, 401, L31

\bibitem[\protect\citeauthoryear{{Bonnell}, {Bate} \& {Zinnecker}}{{Bonnell}
  et~al.}{1998}]{1998MNRAS.298...93B}
{Bonnell} I.~A.,  {Bate} M.~R.,    {Zinnecker} H.,  1998, \mnras, 298, 93

\bibitem[\protect\citeauthoryear{{Brown}, {Johnston}, {Hoare} \&
  {Lumsden}}{{Brown} et~al.}{2016}]{brown_mnras_463_2016}
{Brown} G.~M.,  {Johnston} K.~G.,  {Hoare} M.~G.,    {Lumsden} S.~L.,  2016,
  \mnras, 463, 2839

\bibitem[\protect\citeauthoryear{{Burns}}{{Burns}}{2018}]{arXiv180102211B}
{Burns} R.~A.,  2018, ArXiv e-prints

\bibitem[\protect\citeauthoryear{{Burns}, {Handa}, {Imai}, {Nagayama},
  {Omodaka}, {Hirota}, {Motogi}, {van Langevelde} \& {Baan}}{{Burns}
  et~al.}{2017}]{burns_mnras_467_2017}
{Burns} R.~A.,  {Handa} T.,  {Imai} H.,  {Nagayama} T.,  {Omodaka} T.,
  {Hirota} T.,  {Motogi} K.,  {van Langevelde} H.~J.,    {Baan} W.~A.,  2017,
  \mnras, 467, 2367

\bibitem[\protect\citeauthoryear{{Butler} \& {Tan}}{{Butler} \&
  {Tan}}{2012}]{butler_apj_754_2012}
{Butler} M.~J.,  {Tan} J.~C.,  2012, \apj, 754, 5

\bibitem[\protect\citeauthoryear{{Butler}, {Tan} \& {Kainulainen}}{{Butler}
  et~al.}{2014}]{butler_apj_782_2014}
{Butler} M.~J.,  {Tan} J.~C.,    {Kainulainen} J.,  2014, \apjl, 782, L30

\bibitem[\protect\citeauthoryear{{Caratti o Garatti}, {Stecklum}, {Garcia
  Lopez}, {Eisloffel}, {Ray}, {Sanna}, {Cesaroni}, {Walmsley}, {Oudmaijer}, {de
  Wit}, {Moscadelli}, {Greiner}, {Krabbe}, {Fischer}, {Klein} \&
  {Ibanez}}{{Caratti o Garatti} et~al.}{2016}]{caratti_nature_2016}
{Caratti o Garatti} A.,  {Stecklum} B.,  {Garcia Lopez} R.,  {Eisloffel} J.,
  {Ray} T.~P.,  {Sanna} A.,  {Cesaroni} R.,  {Walmsley} C.~M.,  {Oudmaijer}
  R.~D.,  {de Wit} W.~J.,  {Moscadelli} L.,  {Greiner} J.,  {Krabbe} A.,
  {Fischer} C.,  {Klein} R.,    {Ibanez} J.~M.,  2016, Nature, pp 1745--2481

\bibitem[\protect\citeauthoryear{{Caratti o Garatti}, {Stecklum}, {Linz},
  {Garcia Lopez} \& {Sanna}}{{Caratti o Garatti}
  et~al.}{2015}]{caratti_aa_573_2015}
{Caratti o Garatti} A.,  {Stecklum} B.,  {Linz} H.,  {Garcia Lopez} R.,
  {Sanna} A.,  2015, \aap, 573, A82

\bibitem[\protect\citeauthoryear{{Cesaroni}, {Galli}, {Lodato}, {Walmsley} \&
  {Zhang}}{{Cesaroni} et~al.}{2006}]{cesaroni_natur_444_2006}
{Cesaroni} R.,  {Galli} D.,  {Lodato} G.,  {Walmsley} M.,    {Zhang} Q.,  2006,
  \nat, 444, 703

\bibitem[\protect\citeauthoryear{{Cesaroni}, {Hofner}, {Araya} \&
  {Kurtz}}{{Cesaroni} et~al.}{2010}]{cesaroni_aa_509_2010}
{Cesaroni} R.,  {Hofner} P.,  {Araya} E.,    {Kurtz} S.,  2010, \aap, 509, A50

\bibitem[\protect\citeauthoryear{{Cesaroni}, {Moscadelli}, {Neri}, {Sanna},
  {Garatti}, {Eisloeffel}, {Stecklum}, {Ray} \& {Walmsley}}{{Cesaroni}
  et~al.}{2018}]{cesaroni_2018}
{Cesaroni} R.,  {Moscadelli} L.,  {Neri} R.,  {Sanna} A.,  {Garatti} A.~C.~o.,
  {Eisloeffel} J.,  {Stecklum} B.,  {Ray} T.,    {Walmsley} C.~M.,  2018, ArXiv
  e-prints

\bibitem[\protect\citeauthoryear{{Chen}, {Ren}, {Zhang}, {Shen} \&
  {Qiu}}{{Chen} et~al.}{2017}]{chen_apj_835_2017}
{Chen} X.,  {Ren} Z.,  {Zhang} Q.,  {Shen} Z.,    {Qiu} K.,  2017, \apj, 835,
  227

\bibitem[\protect\citeauthoryear{{Chini}, {Hoffmeister}, {Nasseri}, {Stahl} \&
  {Zinnecker}}{{Chini} et~al.}{2012}]{chini_424_mnras_2012}
{Chini} R.,  {Hoffmeister} V.~H.,  {Nasseri} A.,  {Stahl} O.,    {Zinnecker}
  H.,  2012, \mnras, 424, 1925

\bibitem[\protect\citeauthoryear{{Cieza}, {Ru{\'{\i}}z-Rodr{\'{\i}}guez},
  {Perez}, {Casassus}, {Williams}, {Zurlo}, {Principe}, {Hales}, {Prieto},
  {Tobin}, {Zhu} \& {Marino}}{{Cieza} et~al.}{2018}]{cieza_mnras_474_2018}
{Cieza} L.~A.,  {Ru{\'{\i}}z-Rodr{\'{\i}}guez} D.,  {Perez} S.,  {Casassus} S.,
   {Williams} J.~P.,  {Zurlo} A.,  {Principe} D.~A.,  {Hales} A.,  {Prieto}
  J.~L.,  {Tobin} J.~J.,  {Zhu} Z.,    {Marino} S.,  2018, \mnras, 474, 4347

\bibitem[\protect\citeauthoryear{{Clarke}, {Lin} \& {Pringle}}{{Clarke}
  et~al.}{1990}]{clarke_mnras_442_1990}
{Clarke} C.~J.,  {Lin} D.~N.~C.,    {Pringle} J.~E.,  1990, \mnras, 242, 439

\bibitem[\protect\citeauthoryear{{Commer{\c c}on}, {Teyssier}, {Audit},
  {Hennebelle} \& {Chabrier}}{{Commer{\c c}on}
  et~al.}{2011}]{commercon_aa_529_2011}
{Commer{\c c}on} B.,  {Teyssier} R.,  {Audit} E.,  {Hennebelle} P.,
  {Chabrier} G.,  2011, \aap, 529, A35

\bibitem[\protect\citeauthoryear{{Cunha}, {Hubeny} \& {Lanz}}{{Cunha}
  et~al.}{2006}]{cunha_apj_647_2006}
{Cunha} K.,  {Hubeny} I.,    {Lanz} T.,  2006, \apjl, 647, L143

\bibitem[\protect\citeauthoryear{{Cunningham}, {Moeckel} \&
  {Bally}}{{Cunningham} et~al.}{2009}]{Cunningham_apj_692_2009}
{Cunningham} N.~J.,  {Moeckel} N.,    {Bally} J.,  2009, \apj, 692, 943

\bibitem[\protect\citeauthoryear{{de Jager}}{{de
  Jager}}{1998}]{jaeger_arv_8_1998}
{de Jager} C.,  1998, \aapr, 8, 145

\bibitem[\protect\citeauthoryear{{Eggenberger}, {Meynet}, {Maeder}, {Hirschi},
  {Charbonnel}, {Talon} \& {Ekstr{\"o}m}}{{Eggenberger}
  et~al.}{2008}]{eggenberger_apss_316_2008}
{Eggenberger} P.,  {Meynet} G.,  {Maeder} A.,  {Hirschi} R.,  {Charbonnel} C.,
  {Talon} S.,    {Ekstr{\"o}m} S.,  2008, \apss, 316, 43

\bibitem[\protect\citeauthoryear{{Ekstr{\"o}m}, {Georgy}, {Eggenberger},
  {Meynet}, {Mowlavi}, {Wyttenbach}, {Granada}, {Decressin}, {Hirschi},
  {Frischknecht}, {Charbonnel} \& {Maeder}}{{Ekstr{\"o}m}
  et~al.}{2012}]{ekstroem_aa_537_2012}
{Ekstr{\"o}m} S.,  {Georgy} C.,  {Eggenberger} P.,  {Meynet} G.,  {Mowlavi} N.,
   {Wyttenbach} A.,  {Granada} A.,  {Decressin} T.,  {Hirschi} R.,
  {Frischknecht} U.,  {Charbonnel} C.,    {Maeder} A.,  2012, \aap, 537, A146

\bibitem[\protect\citeauthoryear{{Federrath}, {Banerjee}, {Clark} \&
  {Klessen}}{{Federrath} et~al.}{2010}]{federrath_apj_713_2010}
{Federrath} C.,  {Banerjee} R.,  {Clark} P.~C.,    {Klessen} R.~S.,  2010,
  \apj, 713, 269

\bibitem[\protect\citeauthoryear{{Flock}, {Fromang}, {Gonz{\'a}lez} \&
  {Commer{\c c}on}}{{Flock} et~al.}{2013}]{flock_aa_560_2013}
{Flock} M.,  {Fromang} S.,  {Gonz{\'a}lez} M.,    {Commer{\c c}on} B.,  2013,
  \aap, 560, A43

\bibitem[\protect\citeauthoryear{{Fontani}, {Palau}, {Busquet}, {Isella},
  {Estalella}, {Sanchez-Monge}, {Caselli} \& {Zhang}}{{Fontani}
  et~al.}{2012}]{fontani_mnras_423_2012}
{Fontani} F.,  {Palau} A.,  {Busquet} G.,  {Isella} A.,  {Estalella} R.,
  {Sanchez-Monge} {\'A}.,  {Caselli} P.,    {Zhang} Q.,  2012, \mnras, 423,
  1691

\bibitem[\protect\citeauthoryear{{Forgan}, {Ilee}, {Cyganowski}, {Brogan} \&
  {Hunter}}{{Forgan} et~al.}{2016}]{forgan_mnras_463_2016}
{Forgan} D.~H.,  {Ilee} J.~D.,  {Cyganowski} C.~J.,  {Brogan} C.~L.,
  {Hunter} T.~R.,  2016, \mnras, 463, 957

\bibitem[\protect\citeauthoryear{{Fossati}, {Castro}, {Sch{\"o}ller}, {Hubrig},
  {Langer}, {Morel}, {Briquet}, {Herrero}, {Przybilla}, {Sana}, {Schneider},
  {de Koter} \& {BOB Collaboration}}{{Fossati}
  et~al.}{2015}]{fossati_aa_582_2015}
{Fossati} L.,  {Castro} N.,  {Sch{\"o}ller} M.,  {Hubrig} S.,  {Langer} N.,
  {Morel} T.,  {Briquet} M.,  {Herrero} A.,  {Przybilla} N.,  {Sana} H.,
  {Schneider} F.~R.~N.,  {de Koter} A.,    {BOB Collaboration} 2015, \aap, 582,
  A45

\bibitem[\protect\citeauthoryear{{Fossati}, {Schneider}, {Castro}, {Langer},
  {Sim{\'o}n-D{\'{\i}}az}, {M{\"u}ller}, {de Koter}, {Morel}, {Petit}, {Sana}
  \& {Wade}}{{Fossati} et~al.}{2016}]{fossati_aa_592_2016}
{Fossati} L.,  {Schneider} F.~R.~N.,  {Castro} N.,  {Langer} N.,
  {Sim{\'o}n-D{\'{\i}}az} S.,  {M{\"u}ller} A.,  {de Koter} A.,  {Morel} T.,
  {Petit} V.,  {Sana} H.,    {Wade} G.~A.,  2016, \aap, 592, A84

\bibitem[\protect\citeauthoryear{{Fuente}, {Neri}, {Mart{\'{\i}}n-Pintado},
  {Bachiller}, {Rodr{\'{\i}}guez-Franco} \& {Palla}}{{Fuente}
  et~al.}{2001}]{fuente_aa_366_2001}
{Fuente} A.,  {Neri} R.,  {Mart{\'{\i}}n-Pintado} J.,  {Bachiller} R.,
  {Rodr{\'{\i}}guez-Franco} A.,    {Palla} F.,  2001, \aap, 366, 873

\bibitem[\protect\citeauthoryear{{Fujisawa}, {Yonekura}, {Sugiyama},
  {Horiuchi}, {Hayashi}, {Hachisuka}, {Matsumoto} \& {Niinuma}}{{Fujisawa}
  et~al.}{2015}]{fujisawa_atel_2015}
{Fujisawa} K.,  {Yonekura} Y.,  {Sugiyama} K.,  {Horiuchi} H.,  {Hayashi} T.,
  {Hachisuka} K.,  {Matsumoto} N.,    {Niinuma} K.,  2015, The Astronomer's
  Telegram, 8286

\bibitem[\protect\citeauthoryear{{Ginsburg}, {Bally}, {Goddi}, {Plambeck} \&
  {Wright}}{{Ginsburg} et~al.}{2018}]{2018arXiv180410622G}
{Ginsburg} A.,  {Bally} J.,  {Goddi} C.,  {Plambeck} R.,    {Wright} M.,  2018,
  ArXiv e-prints

\bibitem[\protect\citeauthoryear{{Groh}, {Meynet}, {Georgy} \&
  {Ekstr{\"o}m}}{{Groh} et~al.}{2013}]{groh_aa_558_2013}
{Groh} J.~H.,  {Meynet} G.,  {Georgy} C.,    {Ekstr{\"o}m} S.,  2013, \aap,
  558, A131

\bibitem[\protect\citeauthoryear{{Haemmerl{\'e}}}{{Haemmerl{\'e}}}{2014}]{haemmerle_phd_2014}
{Haemmerl{\'e}} L.,  2014, PhD thesis, University of Geneva

\bibitem[\protect\citeauthoryear{{Haemmerl{\'e}}, {Eggenberger}, {Meynet},
  {Maeder} \& {Charbonnel}}{{Haemmerl{\'e}}
  et~al.}{2016}]{haemmerle_585_aa_2016}
{Haemmerl{\'e}} L.,  {Eggenberger} P.,  {Meynet} G.,  {Maeder} A.,
  {Charbonnel} C.,  2016, \aap, 585, A65

\bibitem[\protect\citeauthoryear{{Haemmerl{\'e}}, {Eggenberger}, {Meynet},
  {Maeder}, {Charbonnel} \& {Klessen}}{{Haemmerl{\'e}}
  et~al.}{2017}]{haemmerle_602_aap_2017}
{Haemmerl{\'e}} L.,  {Eggenberger} P.,  {Meynet} G.,  {Maeder} A.,
  {Charbonnel} C.,    {Klessen} R.~S.,  2017, \aap, 602, A17

\bibitem[\protect\citeauthoryear{{Haemmerl{\'e}} \& {Peters}}{{Haemmerl{\'e}}
  \& {Peters}}{2016}]{haemmerle_458_mnras_2016}
{Haemmerl{\'e}} L.,  {Peters} T.,  2016, \mnras, 458, 3299

\bibitem[\protect\citeauthoryear{{Harries}}{{Harries}}{2015}]{harries_mnras_448_2015}
{Harries} T.~J.,  2015, \mnras, 448, 3156

\bibitem[\protect\citeauthoryear{{Harries}, {Douglas} \& {Ali}}{{Harries}
  et~al.}{2017}]{harries_2017}
{Harries} T.~J.,  {Douglas} T.~A.,    {Ali} A.,  2017, \mnras, 471, 4111

\bibitem[\protect\citeauthoryear{{Hirano}, {Hosokawa}, {Yoshida} \&
  {Kuiper}}{{Hirano} et~al.}{2017}]{hirano_sci_357_2017}
{Hirano} S.,  {Hosokawa} T.,  {Yoshida} N.,    {Kuiper} R.,  2017, Science,
  357, 1375

\bibitem[\protect\citeauthoryear{{Hosokawa}, {Hirano}, {Kuiper}, {Yorke},
  {Omukai} \& {Yoshida}}{{Hosokawa} et~al.}{2016a}]{hosokawa_2015}
{Hosokawa} T.,  {Hirano} S.,  {Kuiper} R.,  {Yorke} H.~W.,  {Omukai} K.,
  {Yoshida} N.,  2016a, \apj, 824, 119

\bibitem[\protect\citeauthoryear{{Hosokawa}, {Hirano}, {Kuiper}, {Yorke},
  {Omukai} \& {Yoshida}}{{Hosokawa} et~al.}{2016b}]{2016ApJ...824..119H}
{Hosokawa} T.,  {Hirano} S.,  {Kuiper} R.,  {Yorke} H.~W.,  {Omukai} K.,
  {Yoshida} N.,  2016b, \apj, 824, 119

\bibitem[\protect\citeauthoryear{{Hosokawa}, {Offner} \& {Krumholz}}{{Hosokawa}
  et~al.}{2011}]{hosokawa_apj_738_2011}
{Hosokawa} T.,  {Offner} S.~S.~R.,    {Krumholz} M.~R.,  2011, \apj, 738, 140

\bibitem[\protect\citeauthoryear{{Hosokawa} \& {Omukai}}{{Hosokawa} \&
  {Omukai}}{2009}]{hosokawa_apj_691_2009}
{Hosokawa} T.,  {Omukai} K.,  2009, \apj, 691, 823

\bibitem[\protect\citeauthoryear{{Hosokawa}, {Yorke} \& {Omukai}}{{Hosokawa}
  et~al.}{2010}]{hosokawa_apj_721_2010}
{Hosokawa} T.,  {Yorke} H.~W.,    {Omukai} K.,  2010, \apj, 721, 478

\bibitem[\protect\citeauthoryear{{Hunter}, {Brogan}, {MacLeod}, {Cyganowski},
  {Chandler}, {Chibueze}, {Friesen}, {Indebetouw}, {Thesner} \&
  {Young}}{{Hunter} et~al.}{2017}]{2017arXiv170108637H}
{Hunter} T.~R.,  {Brogan} C.~L.,  {MacLeod} G.,  {Cyganowski} C.~J.,
  {Chandler} C.~J.,  {Chibueze} J.~O.,  {Friesen} R.,  {Indebetouw} R.,
  {Thesner} C.,    {Young} K.~H.,  2017, \apjl, 837, L29

\bibitem[\protect\citeauthoryear{{Ilee}, {Cyganowski}, {Nazari}, {Hunter},
  {Brogan}, {Forgan} \& {Zhang}}{{Ilee} et~al.}{2016}]{ilee_mnras_462_2016}
{Ilee} J.~D.,  {Cyganowski} C.~J.,  {Nazari} P.,  {Hunter} T.~R.,  {Brogan}
  C.~L.,  {Forgan} D.~H.,    {Zhang} Q.,  2016, \mnras, 462, 4386

\bibitem[\protect\citeauthoryear{{Johnston}, {Robitaille}, {Beuther}, {Linz},
  {Boley}, {Kuiper}, {Keto}, {Hoare} \& {van Boekel}}{{Johnston}
  et~al.}{2015}]{johnston_apj_813_2015}
{Johnston} K.~G.,  {Robitaille} T.~P.,  {Beuther} H.,  {Linz} H.,  {Boley} P.,
  {Kuiper} R.,  {Keto} E.,  {Hoare} M.~G.,    {van Boekel} R.,  2015, \apjl,
  813, L19

\bibitem[\protect\citeauthoryear{{Kee}, {Owocki} \& {Kuiper}}{{Kee}
  et~al.}{2018}]{kee_mnras_479_2018}
{Kee} N.~D.,  {Owocki} S.,    {Kuiper} R.,  2018, \mnras, 479, 4633

\bibitem[\protect\citeauthoryear{{Kessel}, {Yorke} \& {Richling}}{{Kessel}
  et~al.}{1998}]{kessel_aa_337_1998}
{Kessel} O.,  {Yorke} H.~W.,    {Richling} S.,  1998, \aap, 337, 832

\bibitem[\protect\citeauthoryear{{Keto} \& {Wood}}{{Keto} \&
  {Wood}}{2006}]{keto_apj_637_2006}
{Keto} E.,  {Wood} K.,  2006, \apj, 637, 850

\bibitem[\protect\citeauthoryear{{Klassen}, {Pudritz}, {Kuiper}, {Peters} \&
  {Banerjee}}{{Klassen} et~al.}{2016}]{klassen_apj_823_2016}
{Klassen} M.,  {Pudritz} R.~E.,  {Kuiper} R.,  {Peters} T.,    {Banerjee} R.,
  2016, \apj, 823, 28

\bibitem[\protect\citeauthoryear{{Kolb}, {Stute}, {Kley} \& {Mignone}}{{Kolb}
  et~al.}{2013}]{kolb_aa_559_2013}
{Kolb} S.~M.,  {Stute} M.,  {Kley} W.,    {Mignone} A.,  2013, \aap, 559, A80

\bibitem[\protect\citeauthoryear{{Kratter} \& {Lodato}}{{Kratter} \&
  {Lodato}}{2016}]{kratter_araa_54_2016}
{Kratter} K.,  {Lodato} G.,  2016, \araa, 54, 271

\bibitem[\protect\citeauthoryear{{Kratter} \& {Matzner}}{{Kratter} \&
  {Matzner}}{2006}]{kratter_mnras_373_2006}
{Kratter} K.~M.,  {Matzner} C.~D.,  2006, \mnras, 373, 1563

\bibitem[\protect\citeauthoryear{{Kratter}, {Matzner} \& {Krumholz}}{{Kratter}
  et~al.}{2008}]{kratter_apj_681_2008}
{Kratter} K.~M.,  {Matzner} C.~D.,    {Krumholz} M.~R.,  2008, \apj, 681, 375

\bibitem[\protect\citeauthoryear{{Kraus}, {Kluska}, {Kreplin}, {Bate},
  {Harries}, {Hofmann}, {Hone}, {Monnier}, {Weigelt}, {Anugu}, {de Wit} \&
  {Wittkowski}}{{Kraus} et~al.}{2017}]{kraus_apj_835_2017}
{Kraus} S.,  {Kluska} J.,  {Kreplin} A.,  {Bate} M.,  {Harries} T.~J.,
  {Hofmann} K.-H.,  {Hone} E.,  {Monnier} J.~D.,  {Weigelt} G.,  {Anugu} A.,
  {de Wit} W.~J.,    {Wittkowski} M.,  2017, \apjl, 835, L5

\bibitem[\protect\citeauthoryear{{Krumholz}, {Klein} \& {McKee}}{{Krumholz}
  et~al.}{2007}]{krumholz_apj_656_2007}
{Krumholz} M.~R.,  {Klein} R.~I.,    {McKee} C.~F.,  2007, \apj, 656, 959

\bibitem[\protect\citeauthoryear{{Krumholz}, {Klein}, {McKee}, {Offner} \&
  {Cunningham}}{{Krumholz} et~al.}{2009}]{krumholz_sci_323_2009}
{Krumholz} M.~R.,  {Klein} R.~I.,  {McKee} C.~F.,  {Offner} S.~S.~R.,
  {Cunningham} A.~J.,  2009, Science, 323, 754

\bibitem[\protect\citeauthoryear{{Kuiper} \& {Yorke}}{{Kuiper} \&
  {Yorke}}{2013}]{kuiper_apj_772_2013}
{Kuiper} R.,  {Yorke} H.~W.,  2013, \apj, 772, 61

\bibitem[\protect\citeauthoryear{{Larson}}{{Larson}}{1969}]{larson_mnras_145_1969}
{Larson} R.~B.,  1969, \mnras, 145, 271

\bibitem[\protect\citeauthoryear{{Larson}}{{Larson}}{1972}]{larson_mnras_157_1972}
{Larson} R.~B.,  1972, \mnras, 157, 121

\bibitem[\protect\citeauthoryear{{Levesque}, {Massey}, {Olsen}, {Plez},
  {Josselin}, {Maeder} \& {Meynet}}{{Levesque}
  et~al.}{2005}]{levesque_apj_628_2005}
{Levesque} E.~M.,  {Massey} P.,  {Olsen} K.~A.~G.,  {Plez} B.,  {Josselin} E.,
  {Maeder} A.,    {Meynet} G.,  2005, \apj, 628, 973

\bibitem[\protect\citeauthoryear{{Lodato} \& {Clarke}}{{Lodato} \&
  {Clarke}}{2004}]{lodato_mnras_353_2004}
{Lodato} G.,  {Clarke} C.~J.,  2004, \mnras, 353, 841

\bibitem[\protect\citeauthoryear{{Lumsden}, {Wheelwright}, {Hoare}, {Oudmaijer}
  \& {Drew}}{{Lumsden} et~al.}{2012}]{lumsen_mnras_424_2012}
{Lumsden} S.~L.,  {Wheelwright} H.~E.,  {Hoare} M.~G.,  {Oudmaijer} R.~D.,
  {Drew} J.~E.,  2012, \mnras, 424, 1088

\bibitem[\protect\citeauthoryear{{Machida} \& {Matsumoto}}{{Machida} \&
  {Matsumoto}}{2011}]{machida_mnras_413_2011}
{Machida} M.~N.,  {Matsumoto} T.,  2011, \mnras, 413, 2767

\bibitem[\protect\citeauthoryear{{Mahy}, {Rauw}, {De Becker}, {Eenens} \&
  {Flores}}{{Mahy} et~al.}{2013}]{2013A&A...550A..27M}
{Mahy} L.,  {Rauw} G.,  {De Becker} M.,  {Eenens} P.,    {Flores} C.~A.,  2013,
  \aap, 550, A27

\bibitem[\protect\citeauthoryear{{Maud}, {Hoare}, {Galv{\'a}n-Madrid}, {Zhang},
  {de Wit}, {Keto}, {Johnston} \& {Pineda}}{{Maud}
  et~al.}{2017}]{maud_467_mnras_2017}
{Maud} L.~T.,  {Hoare} M.~G.,  {Galv{\'a}n-Madrid} R.,  {Zhang} Q.,  {de Wit}
  W.~J.,  {Keto} E.,  {Johnston} K.~G.,    {Pineda} J.~E.,  2017, \mnras, 467,
  L120

\bibitem[\protect\citeauthoryear{{Menu}, {van Boekel}, {Henning}, {Leinert},
  {Waelkens} \& {Waters}}{{Menu} et~al.}{2015}]{menu_aa_581_2015}
{Menu} J.,  {van Boekel} R.,  {Henning} T.,  {Leinert} C.,  {Waelkens} C.,
  {Waters} L.~B.~F.~M.,  2015, \aap, 581, A107

\bibitem[\protect\citeauthoryear{{Meyer}, {Kuiper}, {Kley}, {Johnston} \&
  {Vorobyov}}{{Meyer} et~al.}{2018}]{meyer_mnras_473_2018}
{Meyer} D.~M.-A.,  {Kuiper} R.,  {Kley} W.,  {Johnston} K.~G.,    {Vorobyov}
  E.,  2018, \mnras, 473, 3615

\bibitem[\protect\citeauthoryear{{Meyer}, {Vorobyov}, {Kuiper} \&
  {Kley}}{{Meyer} et~al.}{2017}]{meyer_mnras_464_2017}
{Meyer} D.~M.-A.,  {Vorobyov} E.~I.,  {Kuiper} R.,    {Kley} W.,  2017, \mnras,
  464, L90

\bibitem[\protect\citeauthoryear{{Meyer}, {Vorobyov E.}, {Elbakyan V.},
  {Stecklum}, {Eisl{\"o}ffel} \& {Sobolev A.}}{{Meyer}
  et~al.}{2018}]{2018arXiv181100574A}
{Meyer} D.~M.-A.,  {Vorobyov E.} I.,  {Elbakyan V.} G.,  {Stecklum} B.,
  {Eisl{\"o}ffel} J.,    {Sobolev A.} M.,  2018, ArXiv e-prints 1811.00574

\bibitem[\protect\citeauthoryear{{Mignone}, {Bodo}, {Massaglia}, {Matsakos},
  {Tesileanu}, {Zanni} \& {Ferrari}}{{Mignone}
  et~al.}{2007}]{mignone_apj_170_2007}
{Mignone} A.,  {Bodo} G.,  {Massaglia} S.,  {Matsakos} T.,  {Tesileanu} O.,
  {Zanni} C.,    {Ferrari} A.,  2007, \apjs, 170, 228

\bibitem[\protect\citeauthoryear{{Mignone}, {Zanni}, {Tzeferacos}, {van
  Straalen}, {Colella} \& {Bodo}}{{Mignone}
  et~al.}{2012}]{migmone_apjs_198_2012}
{Mignone} A.,  {Zanni} C.,  {Tzeferacos} P.,  {van Straalen} B.,  {Colella} P.,
     {Bodo} G.,  2012, \apjs, 198, 7

\bibitem[\protect\citeauthoryear{{Moscadelli}, {Sanna}, {Goddi}, {Walmsley},
  {Cesaroni}, {Caratti o Garatti}, {Stecklum}, {Menten} \&
  {Kraus}}{{Moscadelli} et~al.}{2017}]{moscadelli_aa_600_2017}
{Moscadelli} L.,  {Sanna} A.,  {Goddi} C.,  {Walmsley} M.~C.,  {Cesaroni} R.,
  {Caratti o Garatti} A.,  {Stecklum} B.,  {Menten} K.~M.,    {Kraus} A.,
  2017, \aap, 600, L8

\bibitem[\protect\citeauthoryear{{Nayakshin} \& {Lodato}}{{Nayakshin} \&
  {Lodato}}{2012}]{nayakshin_mnras_426_2012}
{Nayakshin} S.,  {Lodato} G.,  2012, \mnras, 426, 70

\bibitem[\protect\citeauthoryear{{Nikoghosyan}, {Azatyan} \&
  {Khachatryan}}{{Nikoghosyan} et~al.}{2017}]{nikoghosyan_aa_603_2017}
{Nikoghosyan} E.~H.,  {Azatyan} N.~M.,    {Khachatryan} K.~G.,  2017, \aap,
  603, A26

\bibitem[\protect\citeauthoryear{{Norberg} \& {Maeder}}{{Norberg} \&
  {Maeder}}{2000}]{norberg_aa_159_2000}
{Norberg} P.,  {Maeder} A.,  2000, \aap, 359, 1025

\bibitem[\protect\citeauthoryear{{Padoan}, {Haugb{\o}lle} \&
  {Nordlund}}{{Padoan} et~al.}{2014}]{padoan_apj_797_2014}
{Padoan} P.,  {Haugb{\o}lle} T.,    {Nordlund} {\AA}.,  2014, \apj, 797, 32

\bibitem[\protect\citeauthoryear{{Palla} \& {Stahler}}{{Palla} \&
  {Stahler}}{1991}]{palla_apj_375_1991}
{Palla} F.,  {Stahler} S.~W.,  1991, \apj, 375, 288

\bibitem[\protect\citeauthoryear{{Palla} \& {Stahler}}{{Palla} \&
  {Stahler}}{1992}]{palla_apj_392_1992}
{Palla} F.,  {Stahler} S.~W.,  1992, \apj, 392, 667

\bibitem[\protect\citeauthoryear{{Palla} \& {Stahler}}{{Palla} \&
  {Stahler}}{1993}]{palla_1993}
{Palla} F.,  {Stahler} S.~W.,  1993, \apj, 418, 414

\bibitem[\protect\citeauthoryear{{Peters}, {Banerjee}, {Klessen}, {Mac Low},
  {Galv{\'a}n-Madrid} \& {Keto}}{{Peters} et~al.}{2010}]{peters_apj_711_2010}
{Peters} T.,  {Banerjee} R.,  {Klessen} R.~S.,  {Mac Low} M.-M.,
  {Galv{\'a}n-Madrid} R.,    {Keto} E.~R.,  2010, \apj, 711, 1017

\bibitem[\protect\citeauthoryear{{Peters}, {Klessen}, {Mac Low} \&
  {Banerjee}}{{Peters} et~al.}{2010}]{peters_apj_725_2010}
{Peters} T.,  {Klessen} R.~S.,  {Mac Low} M.-M.,    {Banerjee} R.,  2010, \apj,
  725, 134

\bibitem[\protect\citeauthoryear{{Purser}, {Lumsden}, {Hoare} \&
  {Cunningham}}{{Purser} et~al.}{2018}]{purser_mnras_475_2018}
{Purser} S.~J.~D.,  {Lumsden} S.~L.,  {Hoare} M.~G.,    {Cunningham} N.,  2018,
  \mnras, 475, 2

\bibitem[\protect\citeauthoryear{{Purser}, {Lumsden}, {Hoare}, {Urquhart},
  {Cunningham}, {Purcell}, {Brooks}, {Garay}, {G{\'u}zman} \&
  {Voronkov}}{{Purser} et~al.}{2016}]{purser_mnras_460_2016}
{Purser} S.~J.~D.,  {Lumsden} S.~L.,  {Hoare} M.~G.,  {Urquhart} J.~S.,
  {Cunningham} N.,  {Purcell} C.~R.,  {Brooks} K.~J.,  {Garay} G.,
  {G{\'u}zman} A.~E.,    {Voronkov} M.~A.,  2016, \mnras, 460, 1039

\bibitem[\protect\citeauthoryear{{Reg{\'a}ly} \& {Vorobyov}}{{Reg{\'a}ly} \&
  {Vorobyov}}{2017}]{regaly_aa_601_2017}
{Reg{\'a}ly} Z.,  {Vorobyov} E.,  2017, \aap, 601, A24

\bibitem[\protect\citeauthoryear{{Reiter}, {Kiminki}, {Smith} \&
  {Bally}}{{Reiter} et~al.}{2017a}]{reiter_mnras_470_2017}
{Reiter} M.,  {Kiminki} M.~M.,  {Smith} N.,    {Bally} J.,  2017a, \mnras, 470,
  4671

\bibitem[\protect\citeauthoryear{{Reiter}, {Kiminki}, {Smith} \&
  {Bally}}{{Reiter} et~al.}{2017b}]{2017arXiv170604657R}
{Reiter} M.,  {Kiminki} M.~M.,  {Smith} N.,    {Bally} J.,  2017b, ArXiv
  e-prints

\bibitem[\protect\citeauthoryear{{Romanova}, {Ustyugova}, {Koldoba} \&
  {Lovelace}}{{Romanova} et~al.}{2012}]{romanova_mnras_421_2012}
{Romanova} M.~M.,  {Ustyugova} G.~V.,  {Koldoba} A.~V.,    {Lovelace} R.~V.~E.,
   2012, \mnras, 421, 63

\bibitem[\protect\citeauthoryear{{Rosen}, {Krumholz}, {McKee} \&
  {Klein}}{{Rosen} et~al.}{2016}]{rosen_mnras_463_2016}
{Rosen} A.~L.,  {Krumholz} M.~R.,  {McKee} C.~F.,    {Klein} R.~I.,  2016,
  \mnras, 463, 2553

\bibitem[\protect\citeauthoryear{{Samal}, {Chen}, {Takami}, {Jose} \&
  {Froebrich}}{{Samal} et~al.}{2018}]{2018arXiv180311413S}
{Samal} M.~R.,  {Chen} W.-P.,  {Takami} M.,  {Jose} J.,    {Froebrich} D.,
  2018, ArXiv e-prints

\bibitem[\protect\citeauthoryear{{Sanna}, {Koelligan}, {Moscadelli}, {Kuiper},
  {Cesaroni}, {Pillai}, {Menten}, {Zhang}, {Garatti}, {Goddi}, {Leurini} \&
  {Carrasco-Gonzalez}}{{Sanna} et~al.}{2018}]{2018arXiv180509842S}
{Sanna} A.,  {Koelligan} A.,  {Moscadelli} L.,  {Kuiper} R.,  {Cesaroni} R.,
  {Pillai} T.,  {Menten} K.~M.,  {Zhang} Q.,  {Garatti} A.~C.~o.,  {Goddi} C.,
  {Leurini} S.,    {Carrasco-Gonzalez} C.,  2018, ArXiv e-prints

\bibitem[\protect\citeauthoryear{{Seifried}, {Banerjee}, {Klessen}, {Duffin} \&
  {Pudritz}}{{Seifried} et~al.}{2011}]{seifried_mnras_417_2011}
{Seifried} D.,  {Banerjee} R.,  {Klessen} R.~S.,  {Duffin} D.,    {Pudritz}
  R.~E.,  2011, \mnras, 417, 1054

\bibitem[\protect\citeauthoryear{{Shu}}{{Shu}}{1977}]{shu_apj_214_1977}
{Shu} F.~H.,  1977, \apj, 214, 488

\bibitem[\protect\citeauthoryear{{Stecklum}, {Caratti o Garatti}, {Cardenas},
  {Greiner}, {Kruehler}, {Klose} \& {Eisloeffel}}{{Stecklum}
  et~al.}{2016}]{stecklum_ATel_2016}
{Stecklum} B.,  {Caratti o Garatti} A.,  {Cardenas} M.~C.,  {Greiner} J.,
  {Kruehler} T.,  {Klose} S.,    {Eisloeffel} J.,  2016, The Astronomer's
  Telegram, 8732

\bibitem[\protect\citeauthoryear{{Stecklum}, {Garatti}, {Hodapp}, {Linz},
  {Moscadelli} \& {Sanna}}{{Stecklum} et~al.}{2017}]{stecklum_2017b}
{Stecklum} B.,  {Garatti} A.~C.~o.,  {Hodapp} K.,  {Linz} H.,  {Moscadelli} L.,
     {Sanna} A.,  2017, ArXiv e-prints

\bibitem[\protect\citeauthoryear{{Stecklum}, {Heese}, {Wolf}, {Garatti},
  {Ibanez} \& {Linz}}{{Stecklum} et~al.}{2017}]{stecklum_2017a}
{Stecklum} B.,  {Heese} S.,  {Wolf} S.,  {Garatti} A.~C.~o.,  {Ibanez} J.~M.,
   {Linz} H.,  2017, ArXiv e-prints

\bibitem[\protect\citeauthoryear{{Tanaka}, {Tan}, {Zhang} \&
  {Hosokawa}}{{Tanaka} et~al.}{2018}]{2018arXiv180401132T}
{Tanaka} K.~E.~I.,  {Tan} J.~C.,  {Zhang} Y.,    {Hosokawa} T.,  2018, ArXiv
  e-prints

\bibitem[\protect\citeauthoryear{{Testi}}{{Testi}}{2003}]{testi_2003}
{Testi} L.,  2003, in {De Buizer} J.~M.,  {van der Bliek} N.~S.,  eds, Galactic
  Star Formation Across the Stellar Mass Spectrum Vol.~287 of Astronomical
  Society of the Pacific Conference Series, {Intermediate Mass Stars (Invited
  Review)}.
pp 163--173

\bibitem[\protect\citeauthoryear{{Torrelles}, {Curiel}, {Estalella}, {Anglada},
  {G{\'o}mez}, {Cant{\'o}}, {Patel}, {Trinidad}, {Girart},
  {Carrasco-Gonz{\'a}lez} \& {Rodr{\'{\i}}guez}}{{Torrelles}
  et~al.}{2014}]{torrelles_mnras_442_2014}
{Torrelles} J.~M.,  {Curiel} S.,  {Estalella} R.,  {Anglada} G.,  {G{\'o}mez}
  J.~F.,  {Cant{\'o}} J.,  {Patel} N.~A.,  {Trinidad} M.~A.,  {Girart} J.~M.,
  {Carrasco-Gonz{\'a}lez} C.,    {Rodr{\'{\i}}guez} L.~F.,  2014, \mnras, 442,
  148

\bibitem[\protect\citeauthoryear{{Vaidya}, {Fendt}, {Beuther} \&
  {Porth}}{{Vaidya} et~al.}{2011}]{vaidya_apj_742_2011}
{Vaidya} B.,  {Fendt} C.,  {Beuther} H.,    {Porth} O.,  2011, \apj, 742, 56

\bibitem[\protect\citeauthoryear{{Vaytet} \& {Haugb{\o}lle}}{{Vaytet} \&
  {Haugb{\o}lle}}{2017}]{vaytet_aa_598_2017}
{Vaytet} N.,  {Haugb{\o}lle} T.,  2017, \aap, 598, A116

\bibitem[\protect\citeauthoryear{{Vorobyov}}{{Vorobyov}}{2009}]{vorobyov_apj_704_2009}
{Vorobyov} E.~I.,  2009, \apj, 704, 715

\bibitem[\protect\citeauthoryear{{Vorobyov} \& {Basu}}{{Vorobyov} \&
  {Basu}}{2005}]{vorobyov_apj_633_2005}
{Vorobyov} E.~I.,  {Basu} S.,  2005, \apjl, 633, L137

\bibitem[\protect\citeauthoryear{{Vorobyov} \& {Basu}}{{Vorobyov} \&
  {Basu}}{2010}]{vorobyov_apj_719_2010}
{Vorobyov} E.~I.,  {Basu} S.,  2010, \apj, 719, 1896

\bibitem[\protect\citeauthoryear{{Vorobyov} \& {Basu}}{{Vorobyov} \&
  {Basu}}{2015}]{vorobyov_apj_805_2015}
{Vorobyov} E.~I.,  {Basu} S.,  2015, \apj, 805, 115

\bibitem[\protect\citeauthoryear{{Vorobyov} \& {Elbakyan}}{{Vorobyov} \&
  {Elbakyan}}{2018}]{2018arXiv180607675V}
{Vorobyov} E.~I.,  {Elbakyan} V.,  2018, ArXiv e-prints

\bibitem[\protect\citeauthoryear{{Vorobyov}, {Elbakyan}, {Dunham} \&
  {Guedel}}{{Vorobyov} et~al.}{2017}]{2017A&A...600A..36V}
{Vorobyov} E.~I.,  {Elbakyan} V.,  {Dunham} M.~M.,    {Guedel} M.,  2017, \aap,
  600, A36

\bibitem[\protect\citeauthoryear{{Vorobyov}, {Elbakyan}, {Hosokawa}, {Sakurai},
  {Guedel} \& {Yorke}}{{Vorobyov} et~al.}{2017}]{vorobyov_aa_605_2017}
{Vorobyov} E.~I.,  {Elbakyan} V.,  {Hosokawa} T.,  {Sakurai} Y.,  {Guedel} M.,
    {Yorke} H.,  2017, \aap, 605, A77

\bibitem[\protect\citeauthoryear{{Yorke} \& {Sonnhalter}}{{Yorke} \&
  {Sonnhalter}}{2002}]{2002ApJ...569..846Y}
{Yorke} H.~W.,  {Sonnhalter} C.,  2002, \apj, 569, 846

\bibitem[\protect\citeauthoryear{{Zakhozhay}, {Miroshnichenko}, {Kuratov},
  {Zakhozhay}, {Khokhlov}, {Zharikov} \& {Manset}}{{Zakhozhay}
  et~al.}{2018}]{zakhozhay_mnras_477_2018}
{Zakhozhay} O.~V.,  {Miroshnichenko} A.~S.,  {Kuratov} K.~S.,  {Zakhozhay}
  V.~A.,  {Khokhlov} S.~A.,  {Zharikov} S.~V.,    {Manset} N.,  2018, \mnras,
  477, 977

\bibitem[\protect\citeauthoryear{{Zhao}, {Caselli}, {Li} \&
  {Krasnopolsky}}{{Zhao} et~al.}{2018}]{zhao_mnras_473_2018}
{Zhao} B.,  {Caselli} P.,  {Li} Z.-Y.,    {Krasnopolsky} R.,  2018, \mnras,
  473, 4868

\bibitem[\protect\citeauthoryear{{Zhu}, {Hartmann} \& {Gammie}}{{Zhu}
  et~al.}{2009}]{zhu_apj_694_2009}
{Zhu} Z.,  {Hartmann} L.,    {Gammie} C.,  2009, \apj, 694, 1045

\bibitem[\protect\citeauthoryear{{Zhu}, {Hartmann}, {Gammie} \&
  {McKinney}}{{Zhu} et~al.}{2009}]{zhu_apj_701_2009}
{Zhu} Z.,  {Hartmann} L.,  {Gammie} C.,    {McKinney} J.~C.,  2009, \apj, 701,
  620

\end{thebibliography}
}

%%%%%%%%%%%%%%%%%%%%%%%%%%%%%%%%%%%%%%%%%%%%%%%%%%%%%%%%%%%%%%%%%%%%%%%%%%%%%%%%%%%%%%%%%%%
%%%%%%%%%%%%%%%%%%%%%%%%%%%%%%%%%%%%%%%%%%%%%%%%%%%%%%%%%%%%%%%%%%%%%%%%%%%%%%%%%%%%%%%%%%%
%%%%%%%%%%%%%%%%%%%%%%%%%%%%%%%%%%%%%%%%%%%%%%%%%%%%%%%%%%%%%%%%%%%%%%%%%%%%%%%%%%%%%%%%%%%

\end{document}